\documentclass[prd,amsmath,superscriptaddress,twocolumn]{revtex4}

\usepackage{hyperref}
\usepackage{ifthen}
\usepackage{amsmath}
\usepackage{amssymb}
\usepackage{graphicx}
\usepackage{color,bm}

\newcommand{\Om}{\Omega_{\rm m}}

\newcommand{\rhom}{\rho_{\rm m}}

\newcommand{\Msun}{M_\odot}

\definecolor{darkgreen}{cmyk}{0.85,0.2,1.00,0.2}
\definecolor{purple}{cmyk}{0.5,1.0,0,0}

\newcommand{\hMs}{h^{-1}~M_\odot}
\newcommand{\hMpc}{h^{-1}~\text{Mpc}}
\newcommand{\hGpc}{h^{-1}~\text{Gpc}}
\newcommand{\hMpcc}{h^{-3}~\text{Mpc}^3}
\newcommand{\sgL}{\sigma_8^{\Lambda{\rm CDM}}}
\newcommand{\absfR}{|f_{R0}|}
\newcommand{\arr}{\mathcal{Q}}
\newcommand{\be}{\begin{equation}}
\newcommand{\ee}{\end{equation}}

\newcommand{\rmd}{\ensuremath{\mathrm{d}}}
\newcommand{\ds}{\ensuremath{\Delta\Sigma}}

\begin{document}

\title{Cluster Density Profiles as a Test of Modified Gravity}

\author{Lucas~Lombriser}
\affiliation{Institute of Cosmology \& Gravitation, University of Portsmouth, Portsmouth, PO1 3FX, UK}
\affiliation{Institute for Theoretical Physics, University of Zurich, Winterthurerstrasse 190, CH-8057 Z\"{u}rich, Switzerland}
\author{Fabian~Schmidt}
\affiliation{Theoretical Astrophysics, California Institute of Technology M/C 350-17, Pasadena, CA 91125, USA}
\author{Tobias~Baldauf}
\affiliation{Institute for Theoretical Physics, University of Zurich, Winterthurerstrasse 190, CH-8057 Z\"{u}rich, Switzerland}
\author{Rachel~Mandelbaum}
\affiliation{Princeton University Observatory, Peyton Hall, Princeton,
  NJ 08544, USA}
\affiliation{Department of Physics, Carnegie Mellon University, Pittsburgh,
PA 15213, USA}
\author{Uro\v{s}~Seljak}
\affiliation{Institute for Theoretical Physics, University of Zurich, Winterthurerstrasse 190, CH-8057 Z\"{u}rich, Switzerland}
\affiliation{Physics and Astronomy Department, University of California, and Lawrence Berkeley National Laboratory, Berkeley, California 94720, USA}
\affiliation{Ewha University, Seoul 120-750, Korea}
\author{Robert~E.~Smith}
\affiliation{Institute for Theoretical Physics, University of Zurich, Winterthurerstrasse 190, CH-8057 Z\"{u}rich, Switzerland}
\affiliation{Argelander-Institute for Astronomy, Auf dem H\"ugel 71, D-53121 Bonn, Germany}

\date{\today}

\begin{abstract}

We present a new test of gravitational interactions at the $r\simeq(0.2-20)~{\rm Mpc}$ scale, around the virial radius of dark matter halos measured through cluster-galaxy lensing of maxBCG clusters from the Sloan Digital Sky Survey (SDSS).  We employ predictions from self-consistent simulations of $f(R)$ gravity to
find an upper bound on the background field amplitude of $\absfR<3.5\times10^{-3}$ at the 1D-marginalized 95\% confidence level.
As a model-independent assessment of the constraining power of cluster profiles measured through weak gravitational lensing, we also constrain the amplitude $F_0$ of a phenomenological modification based on the profile enhancement induced by $f(R)$ gravity when not including effects from the increased cluster abundance in $f(R)$.
In both scenarios, dark-matter-only simulations of the concordance model corresponding to $\absfR = 0$ and $F_0=0$ are consistent with the lensing
measurements, i.e., at the $68\%$ and $95\%$ confidence level, respectively.

\end{abstract}

\maketitle

\section{Introduction}

Modifications of gravity can serve as an alternative explanation to the dark energy paradigm for the late-time accelerated expansion of our Universe. Such modifications have extensively been tested on solar-system scales (see, e.g.,~\cite{will:05}) and to a lesser degree at large cosmological scales using specific alternative theories of gravity (see, e.g.,~\cite{fang:08a, lombriser:09, reyes:10, song:07, giannantonio:09, lombriser:10, hojjati:11, schmidt:09}), as well as generic modifications to general relativity (GR) while adopting a $\Lambda$CDM background (see, e.g.,~\cite{diporto:07, rapetti:08, daniel:09, rapetti:09, bean:10, daniel:10, daniel:10b, dossett:10, tereno:10, zhao:10, dossett:11, dossett:11b}) or simultaneously allowing a dynamic effective dark energy equation of state~\cite{lombriser:11, zhao:11}. However, gravity may also be tested by the structure observed at intermediate scales~\cite{smith:09t, wojtak:11}. In this regime, nonlinear gravitational interactions gain in importance and need to be modeled correctly to obtain reliable predictions for both GR and its competitors, which in turn can be compared with observations to infer constraints on modified gravity theories.

To study nonlinear effects in structure formation, we need to specialize to a particular gravitational modification. In our case, this is $f(R)$ gravity. Within this model, the Einstein-Hilbert action is supplemented with a free function $f(R)$ of the Ricci scalar $R$. It has been shown that such models can reproduce the late-time accelerated expansion of the Universe without invoking dark energy~\cite{carroll:03, nojiri:03, capozziello:03}. However, they also produce a stronger gravitational coupling and enhance the growth of structure. $f(R)$ gravity is formally equivalent to a scalar-tensor theory where the additional degree of freedom is described by the \emph{scalaron} field $f_R \equiv \rmd f/\rmd R$~\cite{starobinsky:79, starobinsky:80}. We parametrize our models by the background value of the scalaron field today, $\absfR$. The $f_R$ field is massive, and below its Compton wavelength, it enhances gravitational forces by a factor of $4/3$. Due to the density dependence of the scalaron's mass, viable $f(R)$ gravity models experience a mechanism dubbed the chameleon effect~\cite{khoury:03, navarro:06, faulkner:06}, which returns gravitational forces to the standard relations in high-density regions, making them compatible with solar-system tests~\cite{hu:07a} at $r\lesssim 20~{\rm AU}$.
The transition required to interpolate between the low curvature of the large-scale structure and the high curvature of the galactic halo sets the currently strongest bound on the background field, $\absfR<|\Psi|\sim (10^{-6} - 10^{-5})$~\cite{hu:07a}, i.e., the typical depth of cosmological potential wells.  
A bound of the same order is obtained from galaxies serving as strong gravitational lenses~\cite{smith:09t} at $r \sim (1 - 10)~\textrm{kpc}$.  
Independently, strong constraints can also be inferred from the large-scale structure $(r\gtrsim10~\textrm{Mpc})$. The enhanced growth of structure observed in $f(R)$ gravity models manifests itself on the largest scales of the cosmic microwave background (CMB) temperature anisotropy power spectrum~\cite{song:06}, where compatibility with CMB data places an upper bound on $\absfR$ of order unity~\cite{song:07}. Cross correlations of the CMB temperature field with foreground galaxies tighten this constraint by an order of magnitude~\cite{song:06, song:07, giannantonio:09, lombriser:10, hojjati:11}. However, the currently strongest constraints on $f(R)$ gravity models from large-scale structures are inferred from the analysis of the abundance of clusters, yielding an improvement over the CMB constraints of nearly four orders of magnitude~\cite{schmidt:09, lombriser:10}.

In this paper, we present a new test of gravity at the $r \sim (0.2 - 20)~\textrm{Mpc}$ scale, i.e., around the virial radius of dark matter halos measured through the excess surface mass density from cluster-galaxy lensing.
$N$-body simulations of modified gravity scenarios have shown that halo-density
profiles exhibit a characteristic enhancement at a few virial radii when 
compared to halo profiles in GR simulations with the same expansion history
\cite{schmidt:08,schmidt:09b}.  In models which attempt to explain
the accelerated expansion of the Universe without dark energy, the 
modifications to the gravitational force generally increase towards 
late times, leading to a
pileup of matter in the infall regions of massive halos.  In contrast, the
inner profiles of halos are less affected since they formed earlier, when
the force modifications were weak or absent.  

Here, we use this effect to constrain the field amplitude $\absfR$ of the Hu-Sawicki~\cite{hu:07a} $f(R)$ gravity model with measurements of weak lensing around the maxBCG galaxy cluster sample~\cite{koester:07} from the Sloan Digital Sky Survey (SDSS)~\cite{SDSS:09}.
Through matching clusters by abundance, we consistently take
into account the modified gravity effects on the halo-mass function as well as the 
profiles.  
In addition, we consider a phenomenological approach modeled on the
$f(R)$ effects on the halo profiles at fixed mass.  
While this approach is not entirely consistent (since it does not include the
effects on the halo-mass function), the constraints are largely \emph{independent} of 
halo number counts, and, moreover, are given directly in terms
of the observable, rather than a model parameter.  They can thus be used to assess the 
constraining power of halo profiles measured through weak lensing on a wider range 
of modified gravity models, including, for example, models where gravity is 
weakened and profiles are consequently suppressed with respect to GR.
In both cases, we perform a Markov chain Monte Carlo (MCMC) likelihood analysis on the underlying parameter spaces.
 
The outline of the paper is as follows.  In~\textsection\ref{sec:theory}, we review the $f(R)$ gravity model and weak gravitational lensing.  We then describe the $N$-body simulations employed to derive the dark matter halo properties (\textsection\ref{sec:sims}), and the procedure used to predict weak-lensing observables in $f(R)$ and $\Lambda$CDM cosmologies (\textsection\ref{sec:pheno}).  
\textsection\ref{sec:observations} then introduces the observational data as well as external priors used in this study.  
The constraints on the alternative gravity models are presented in \textsection\ref{sec:constraints}, along with a discussion of systematic effects that may contaminate the data or complicate its interpretation.  We conclude in~\textsection\ref{sec:discussion}.  The appendices give further details about the halo model and interpolation used in \textsection\ref{sec:pheno}.

\section{Modified Gravity \& Gravitational Lensing}\label{sec:theory}

When gravitational interactions are modified, the growth of structure and thus the distribution of mass, as well as the relation between light deflection and mass distribution change \cite{KnoxSongTyson,Schmidt:08a,JainZhang,Tsujikawa2008}.
Effects of modified gravity on halo properties were studied in the case of $f(R)$ gravity in, e.g.,~\cite{schmidt:08, borisov:11} (cf.~\cite{martino:08}) and the DGP braneworld scenario in, e.g.,~\cite{schmidt:09a, schmidt:09b} (cf.~\cite{narikawa:12}). 

We concentrate on Hu-Sawicki~\cite{hu:07a} $f(R)$ gravity and rely on the nonlinear behavior measured in $N$-body simulations of this model~\cite{oyaizu:08a, oyaizu:08b, schmidt:08} (cf.~\cite{zhao:10b, li:11}).  
We shall first review the details of the Hu-Sawicki model and 
how to relate lensing observables to the underlying matter distribution.  
We then briefly review how stacked weak-lensing observables measure the mass
distribution around halos.

\subsection{$f(R)$ gravity}\label{sec:theoryfR}

In $f(R)$ gravity, the Einstein-Hilbert action is supplemented by a free function of the Ricci scalar $R$,
\begin{equation}
S = \frac{1}{16 \pi \, G} \int \rmd^4 x \sqrt{-g} \left[ R + f(R) \right] + \int \rmd^4x \sqrt{-g} \mathcal{L}_{\rm m}.
\end{equation}
Here, $\mathcal{L}_{\rm m}$ is the matter Lagrangian and we have set $c=1$. Variation with respect to the metric $g_{\mu\nu}$ yields the modified Einstein equations for metric $f(R)$ gravity,
\begin{equation}
G_{\mu\nu} + f_R R_{\mu\nu} - \left( \frac{f}{2} - \Box f_R \right) g_{\mu\nu} - \nabla_{\mu} \nabla_{\nu} f_R = 8 \pi \, G \, T_{\mu\nu},
\end{equation}
where the connection is of Levi-Civita type and $f_R \equiv \rmd f/\rmd R$ is the additional scalar degree of freedom of the model, characterizing the force modifications.

We specialize our considerations to the functional form~\cite{hu:07a}
\begin{equation}
f(R) = -m^2 \frac{c_1 \left( R/m^2 \right)^n}{c_2 \left( R/m^2 \right)^n + 1},
\label{eq:husawicki}
\end{equation}
where $m^2 \equiv 8 \pi \, G \, \bar{\rho}_{\rm m} / 3$. The free parameters of the model $c_1$, $c_2$, and $n$ can be chosen to reproduce the $\Lambda$CDM expansion history and satisfy solar-system tests~\cite{hu:07a} through the chameleon mechanism~\cite{khoury:03, navarro:06, faulkner:06}. In the high-curvature regime, $c_2^{1/n} R \gg m^2$, Eq.~(\ref{eq:husawicki}) simplifies to
\begin{equation}
f(R) = -\frac{c_1}{c_2} m^2 - \frac{f_{R0}}{n} \frac{\bar{R}_0^{n+1}}{R^n},
\label{eq:backgroundmimick}
\end{equation}
where $\bar{R}_0$ denotes the background curvature today, $\bar{R}_0 = \bar{R}|_{z=0}$ , and $f_{R0} \equiv f_R(\bar{R}_0)$. We further infer
\begin{equation}
 \frac{c_1}{c_2} m^2 = 16 \pi \, G \, \bar{\rho}_{\Lambda}
\end{equation}
from requiring equivalence with $\Lambda$CDM when $\absfR \rightarrow 0$ and restrict to models with $n=1$.  Varying $n$ changes the evolution of the
Compton wavelength of the $f_R$ field with redshift.  Generally, constraints
on $f_{R0}$ become weaker (stronger) for $n > 1$ ($n < 1$) (see \cite{FerraroEtal} for a study of the mass function of halos in $f(R)$ with varying $n$).  In 
the following, we will further assume that $|f_{R0}| \ll 1$, and drop
terms that are higher order in $f_R$.  

In the quasistatic limit, the trace and time-time component of the modified Einstein equations yield the $f_R$ field equation and Poisson equation for the Newtonian potential $\Psi = \delta g_{00} / (2g_{00})$ in the longitudinal gauge.  
Specifically,
\begin{eqnarray}
\nabla^2 \delta f_R & = & \frac{a^2}{3} \left[ \delta R (f_R) - 8 \pi \, G \, \delta \rho_{\rm m} \right], \label{eq:fR} \\
\nabla^2 \Psi & = & \frac{16 \pi \, G}{3} a^2 \delta\rho_{\rm m} - \frac{a^2}{6} \delta R (f_R). \label{eq:pot} 
\end{eqnarray}
Here, coordinates are comoving, $\delta f_R = f_R(R) - f_R(\bar{R})$, $\delta R = R - \bar{R}$, $\delta \rho_{\rm m} = \rho_{\rm m} - \bar{\rho}_{\rm m}$.  
In contrast, the potential $\Psi_- \equiv (\Psi-\Phi)/2$, where $\Phi = \delta g_{ii} / (2g_{ii})$, governing the
propagation of light and hence lensing is only affected at order 
$f_R$, which is of order $10^{-2}$ or less for the models we consider here.  
Hence, the modifications to cluster-galaxy lensing studied in
this paper are caused by modifications in the distribution of matter, which 
arise from the enhanced gravitational forces.

If the background field $|f_{R0}|$ is large compared to typical gravitational potentials ($\sim 10^{-5}$), we may linearize the field equations via the approximation
\begin{equation}
\delta R \approx \left. \frac{\rmd R}{\rmd f_R} \right|_{R=\bar{R}} \delta f_{R} = 3 \lambda_{\rm C}^{-2} \delta f_R,
\end{equation}
where $\lambda_{\rm C} = 1 / m_{f_R}$ is the Compton wavelength of the field at the background.
In Fourier space, the solution to Eqs.~(\ref{eq:fR}) and (\ref{eq:pot}) within the linearized approximation is
\begin{equation}
k^2 \Psi({\bf k}) = - 4 \pi \, G \left\{ \frac{4}{3} - \frac{1}{3} \left[ \left( \lambda_{\rm C} \frac{k}{a} \right)^2 + 1 \right]^{-1} \right\} a^2 \delta \rhom({\bf k}),
\label{eq:linearized}
\end{equation}
where $k=|{\bf k}|$.
For scales $k \gg 2\pi \, \lambda_{\rm C}^{-1} a$, this leads to an enhancement of gravitational forces by a factor of $4/3$. Computations using Eq.~(\ref{eq:linearized}) are referred to as the \emph{no-chameleon} or \emph{linearized} $f(R)$ case~\cite{oyaizu:08b}.  

If the background field becomes small compared to the depth of the gravitational potential of the object considered ($\absfR \lesssim 10^{-5}$, small-field limit), the chameleon mechanism becomes active, suppressing non-Newtonian forces. More precisely, $\delta f_R \approx -\overline{f_R}$ and from Eq.~(\ref{eq:fR}), $\delta R \simeq 8 \pi \, G \, \delta\rho_{\rm m}$, which restores the standard Poisson equation in Eq.~(\ref{eq:pot}). Given that the constraints on $\absfR$ expected from our lensing data are well within the large-field regime ($\absfR \gg 10^{-5}$), we can apply the approximation Eq.~(\ref{eq:linearized}) in the simulations.

\subsection{Weak Gravitational Lensing} \label{sec:weak_lensing}

Weak gravitational lensing serves as a powerful probe of the total matter
distribution (baryonic + dark matter) within our Universe.
Here, we focus on stacked cluster-galaxy lensing, which measures the average deformation of
background galaxy images around foreground maxBCG galaxy clusters.  
By averaging over many lenses, the contribution of unassociated
large-scale structure is suppressed.  

We use the tangential shear $\gamma_t$ , measured using the ellipticities 
of galaxy shapes, as a function of the comoving transverse separation from the
lens $r_{\perp,{\rm l}} \approx
\theta \left(1+z_{\rm l}\right) D_{\rm l}$.
Here, $D_{\rm l}$ is the angular diameter distance to the lens and $z_{\rm l}$
is the lens redshift. After stacking many clusters, the mass distribution
becomes symmetric around the line of sight.  Then, the shear is related to
the excess surface mass density around the dark matter halos hosting the
clusters, $\Delta\Sigma(r_\perp)$, through~\cite{squires:95}
\begin{equation}
\gamma_t(r_\perp) = \frac{\Delta\Sigma(r_\perp)}{\Sigma_{\rm crit}}.
 \label{eq:gammat}
\end{equation}
The excess surface mass density is related to the projected
surface density $\Sigma(r_\perp)$ through
\begin{align}
\Delta \Sigma(r_\perp) =\:& \bar{\Sigma}(r_\perp) - \Sigma(r_\perp)\nonumber\\
\bar{\Sigma}(r_\perp) =\:& \frac{2}{r_\perp^2} \int_0^{r_\perp}\Sigma(r_\perp')r_\perp' \,\rmd r_\perp'.
 \label{eq:DSigma}
\end{align}
Here and throughout the paper, $r$ denotes a three-dimensional separation, while $r_{\perp}$ refers to a projected two-dimensional separation.  
The comoving critical surface mass density is given by
\begin{equation}
 \Sigma_{\rm crit} = \frac{c^2}{4 \pi \, G} \frac{D_{\rm s}}{D_{\rm ls}D_{\rm l}(1+z_{\rm l})^2},
\end{equation}
where $D_{\rm s}$ and $D_{\rm ls}$ denote the angular diameter
distance to the source
and between the lens and the source, respectively. Note that both $\Sigma_{\rm crit}$ and the conversion between $\theta$ and $r_\perp$ are dependent on the specific cosmological model (see~\textsection\ref{sec:systematics}).

Assuming perfect centering of the lenses,
the projected surface mass density is related to the halo profiles by
\begin{eqnarray}
 \Sigma(r_\perp) & = & \frac{H^2 \Omega_{\rm m}}{4 \pi \, G} \int_{-\chi_l}^{+\infty}
 g_{\rm l}(\chi_l+y) \nonumber \\
 & & \times \left[ 1 + \xi_{\rm hm} \left( \sqrt{r_\perp^2+y^2} \right) \right] \rmd y,
 \label{eq:projsmd}
\end{eqnarray}
where $H$ indicates the Hubble parameter, $\chi_l$ denotes the comoving
distance to the lens, and $y$ denotes the distance from the lens along the line 
of sight.  $\xi_{\rm hm}(r)$
is the halo-matter correlation function which quantifies the total mass distribution
around halo centers (see \S \ref{sec:halomodelpred}).  The lensing window $g_{\rm l}(\chi)$ depends on the source redshift distribution $p_{\rm s}(\chi)$ as
\begin{equation}
g_{\rm l}(\chi) = 2 \int_{\chi}^\infty p_{\rm s}(\chi') \frac{D_{\rm l}(\chi)D_{\rm ls}(\chi,\chi')}{a(\chi) D_{\rm s}(\chi')} \rmd\chi',
\end{equation}
assuming that $p_{\rm s}(\chi)$ is normalized to integrate to unity. 
The halo-matter correlation function decays strongly with increasing separation, so for 
the transverse scales considered in this study, the lensing strength $g_{\rm
l}(\chi)$ is effectively a constant.

\section{Simulations}\label{sec:sims}

In our study, we consider gravitational lensing measurements on scales of
$0.5 \hMpc \leq r_\perp \leq 25 \hMpc $. These scales are affected by nonlinear
clustering such that numerical simulations are required to obtain reliable
predictions for the mass distribution.   
We utilize $f(R)$ gravity simulations to obtain the deviations induced by the modified forces in the halo profiles with respect to the $\Lambda$CDM predictions, i.e., $\absfR=0$, from the same initial conditions and simulation setup.  
We then employ the Z\"urich Horizon ({\sc zhorizon}) simulations~\cite{smith:08},
which provide $\Lambda$CDM predictions of better resolution and larger volume, and scale
these results with the deviations from the $f(R)$ gravity simulations.  
Note that we use simulations where the matter density field consists exclusively of dark matter particles, hereafter dark-matter-only (DMO) simulations.

\subsection{$f(R)$ gravity simulations} \label{sec:fRsimulations}

\begin{table}
\begin{tabular}{lcc}
\hline
& $L_{\rm box}~[\hMpc]$ & Number of runs \\
\hline
$\absfR = 0$ & 128 & 30 \\
             & 64 & 28 \\
$\absfR = 10^{-4}$ & 128 & 6 \\
             & 64 & 6 \\
$\absfR = 10^{-3}$ & 128 & 30 \\
             & 64 & 28 \\
$\absfR = 10^{-2}$ & 128 & 30 \\
             & 64 & 28 \\
\hline
\end{tabular}
\caption{Summary of $f(R)$ simulation runs.  All simulations use the
linearized $f_R$ field equation, Eq.~(\ref{eq:linearized}). The cosmological parameters of the simulations are given in~\textsection\ref{sec:fRsimulations}.}
\label{tab:fR_par}
\end{table}

Since our constraints lie in a regime where the chameleon mechanism
is not active and we require sufficient halo statistics, we employ
no-chameleon $f(R)$ gravity simulations, which solve the linearized $f_R$ field 
equation, Eq.~(\ref{eq:linearized}) \cite{oyaizu:08a, oyaizu:08b, schmidt:08}.  
Simulations are conducted for $\absfR = 10^{-2},\ 10^{-3},\ 10^{-4},\
0$ and $n=1$.  Note that $\absfR=0$ corresponds to $\Lambda$CDM.
Other cosmological parameters are fixed to values following the
WMAP 3-year results,
$\Omega_{\Lambda}=0.76$, $\Omega_{\rm b}=0.04181$, $h=0.73$, $n_{\rm
  s}=0.958$, and the initial power in curvature fluctuations $A_{\rm
  s}=(4.89\times 10^{-5})^2$ at $k=0.05~\textrm{Mpc}^{-1}$, corresponding
to $\sigma_8=0.82$ at $z=0$.  
The simulations are carried out on $512^3$ grid cells with a total 
of $N_{\rm p} = 256^3$ particles.  Due to the limited volume and resolution
of the $f(R)$ simulations, we combine results from two different box sizes, 
$L_{\rm box} = 64\hMpc,\ 128\hMpc$.  Only the smaller boxes contribute 
for $r < 0.75\hMpc$, corresponding to 3 grid cells for $L_{\rm box} = 128\hMpc$.    
The box sizes and number of runs for each value of $\absfR$ are summarized
in Table~\ref{tab:fR_par}. 

Halos within the simulation and their associated masses are identified via a spherical overdensity (SO) algorithm (cf.~\cite{jenkins:00}). The particles are placed on the grid by a cloud-in-cell interpolation and counted within a growing sphere around the center of mass until the required overdensity is reached. The mass of the halo is then defined by the sum of the particle masses contained in the sphere. This process is started at the highest overdensity grid point and hierarchically continued to lower overdensity grid points until all halos are identified.  The halos employed for this analysis ($\log_{10} M \gtrsim 10^{14}h^{-1}~\Msun$) 
generally contain more than $10^3$ particles.

\subsection{Concordance model simulations}\label{sec:LCDMsimulations}

\begin{table}
\begin{tabular}{ccccccc}
\hline
$\Omega_{\rm m}$ & $\Omega_{\rm b}$ & $h$ & $\sigma_8$ & $n_{\rm s}$ & $L_{\rm box}~[\hGpc]$ & Number of runs \\
\hline
0.25 & 0.04 & 0.7 & 0.8 & 1.00 & 1.5 & 30 \\
0.20 & 0.04 & 0.7 & 0.8 & 1.00 & 1.5 & 4 \\
0.30 & 0.04 & 0.7 & 0.8 & 1.00 & 1.5 & 4 \\
0.25 & 0.04 & 0.7 & 0.7 & 1.00 & 1.5 & 4 \\
0.25 & 0.04 & 0.7 & 0.9 & 1.00 & 1.5 & 4 \\
0.25 & 0.04 & 0.7 & 0.8 & 0.95 & 1.5 & 4 \\
0.25 & 0.04 & 0.7 & 0.8 & 1.05 & 1.5 & 4 \\
\hline
\end{tabular}
\caption{Parameter values for the {\sc zhorizon} simulations: total and baryonic matter density parameters
$\Omega_{\rm m}$ and $\Omega_{\rm b}$, respectively, the dimensionless Hubble parameter $h$,
the power spectrum normalization $\sigma_8=\sgL$, and the primordial spectral index $n_{\rm s}$. The first
row indicates the fiducial cosmological parameters inspired by the three-year WMAP best-fit
values~\cite{spergel:03, spergel:06}.}
\label{tab:cos_par}
\end{table}

The {\sc zhorizon} simulations comprise $30+24$ pure dissipationless
dark matter $N$-body simulations of different $\Lambda$CDM cosmologies (see
Table~\ref{tab:cos_par}),
designed for high-precision studies of cosmological structures on scales of up
to a few $100\hMpc$~\cite{smith:08, smith:09}.

The matter density field is sampled by $N_{\rm p} = 750^3$ dark matter particles of mass
$M_{\rm dm} = 5.55 \times 10^{11}\hMs$, in the fiducial case, with a box size of $1.5\hGpc$.
For the nonlinear gravitational evolution of the equal-mass particles, the publicly available
{\sc gadget-2} code~\cite{springel:05} is used. In order to avoid two-particle collisions,
a force softening length of $60h^{-1}~\textrm{kpc}$ is employed.
The transfer function at redshift $z = 0$ is generated using {\sc
cmbfast}~\cite{seljak:96} and then rescaled to the initial redshift $z_{\rm i} =
50$, where a realization of the potential on the grid is calculated. The
particles are placed on a Cartesian grid of spacing
$\Delta x = 2\hMpc$ and then displaced according to second-order 
Lagrangian perturbation theory using the {\sc 2lpt} code~\cite{scoccimarro:97,
crocce:06}.

For each cosmology, we use four boxes from the {\sc zhorizon} simulations,
yielding an effective volume of $13.5h^{-3}~\textrm{Gpc}^3$.
For all snapshots of each simulation, gravitationally bound structures are
identified by a Friends-of-Friends (FoF) algorithm~\cite{davis:85} with linking
length of 0.2 times the mean interparticle spacing~\footnote{The FoF code used
({\sc b-fof}) was kindly provided by V. Springel.}. The halo center is
associated with the minimum of the
potential of the particle distribution. Halos with fewer than 20 particles are
rejected, resulting in a halo-mass
resolution of $M>1.2\times10^{13}~\hMs$, corresponding to a halo-number
density $\bar{n}=3.7 \times10^{-4}\hMpcc$.

\subsection{Cluster density profiles and sample selection}\label{sec:halomodelpred}

Cluster-galaxy lensing measures a projection of the halo-matter cross
correlation $\xi_{\rm hm}(r)$.  We measure $\xi_{\rm hm}$ by
averaging the spherically averaged density distribution around halos
in the $\Lambda$CDM and $f(R)$ simulations:
\begin{equation}
\xi_{\rm hm}(r) = \left\langle\frac{\rho(r)}{\bar\rho_{\rm m}} -1\right\rangle.
\end{equation}

In observations, clusters are selected according to their optical richness. The true mass can, however,
deviate from the mass inferred from the mass-richness relation \cite{rozo:08}.
Thus, it is important to take into account the
scatter in the mass-richness relation.  In the simulations, we model the scatter by
a log-normal distribution, assigning a new mass to each halo in the simulations
by 
\begin{equation}
 M=\exp\left[\ln(M_0)+\mathcal{N}(0,\sigma) - \frac12 \sigma^2\right],
\end{equation}
where $M_0$ is the true mass and $\mathcal{N}$ is the normal distribution 
with zero mean and variance $\sigma^2$.  The scatter $\sigma$ is left as a free
parameter in the likelihood analysis.  
We apply this scatter, with $\sigma=0,0.4,0.6,0.8$, to the halo masses in the 
simulations, then we mass order the halos according to the simulated mass with scatter, and finally select the $N_\text{h}$
most massive ones until the required cluster abundance $\bar n$ is achieved.  
Here, $\bar n$ is the estimated true average number density of the maxBCG sample (see \textsection\ref{sec:observations}).  
Specifically, we require $N_{\rm h} = \bar{n} \, V_{\rm tot}$ at 
$z=0.23$, the mean redshift of the lens sample (see~\textsection\ref{sec:observations}), where $V_{\rm tot}$ is the simulation volume.  
In the following, we will
denote the corresponding mass profile as $\xi_{\rm hm}^{\Lambda\rm CDM}$.  
The same procedure is applied to the $f(R)$ simulations at $z=0.22$, but 
only for values of $\sigma=0,0.6$ for the scatter.  

As discussed in~\textsection\ref{sec:observations}, a cylindrical cut is applied to the
observational data in order to remove satellite galaxies. In the data this procedure
removes both true clusters and satellite galaxies. To account for the removal of
clusters, we mimic this approach in the {\sc zhorizon} simulation analysis, following the same 
algorithm and using $\delta \chi=\pm 100 \hMpc$ for the length of the cylinder.
This reduces the number density by 20\% from $\bar{n}=1.8\times10^{-5}\hMpcc$ to $\bar{n}=1.45\times10^{-5}\hMpcc$ for zero scatter and to $\bar{n}=1.43\times10^{-5}\hMpcc$
for scatter $\sigma=0.4$.
Since the simulations contain only true halo centers, we conclude that $2/3$ of the
30\% of the maxBCG sample removed from the data were true clusters and $1/3$ were contaminating satellite galaxies.  
After applying the cylindrical cuts, the abundances of halos in the simulation
and maxBCGs in the data sample agree very well.

\section{From Simulations to Observables}\label{sec:pheno}

In this section, we describe how we obtain cluster-galaxy lensing predictions 
for $f(R)$ gravity from the simulations described in the previous section.  
We also introduce our phenomenological approach
modeled on the effects on the halo profile from $f(R)$ modifications
when averaging halos with the same lower mass threshold as in the concordance model. 
The intention of this approach, being largely unaffected by differences in halo number counts, is to yield a model-independent assessment of the constraining power of cluster density profiles measured through weak gravitational lensing.
For this purpose, it is essential to not only study the $f(R)$ modification on the abundance-matched halo profile but also its counterpart in a fixed mass range scenario as described in detail in~\textsection\ref{sec:fR_halomodel}.

\subsection{$f(R)$ gravity halo profile predictions} \label{sec:fR_halomodel}

\begin{figure*}
 \resizebox{\hsize}{!}{\includegraphics{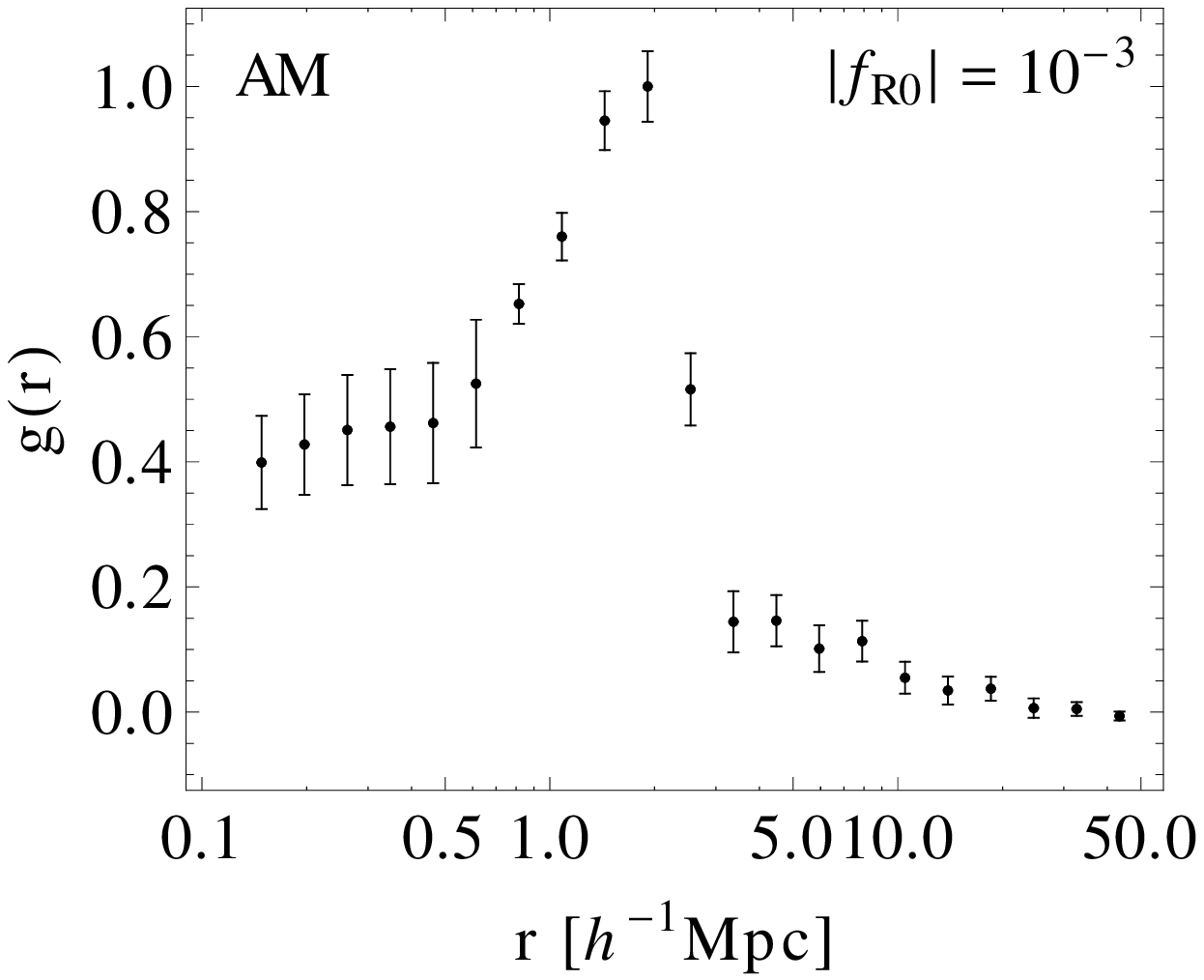}\includegraphics{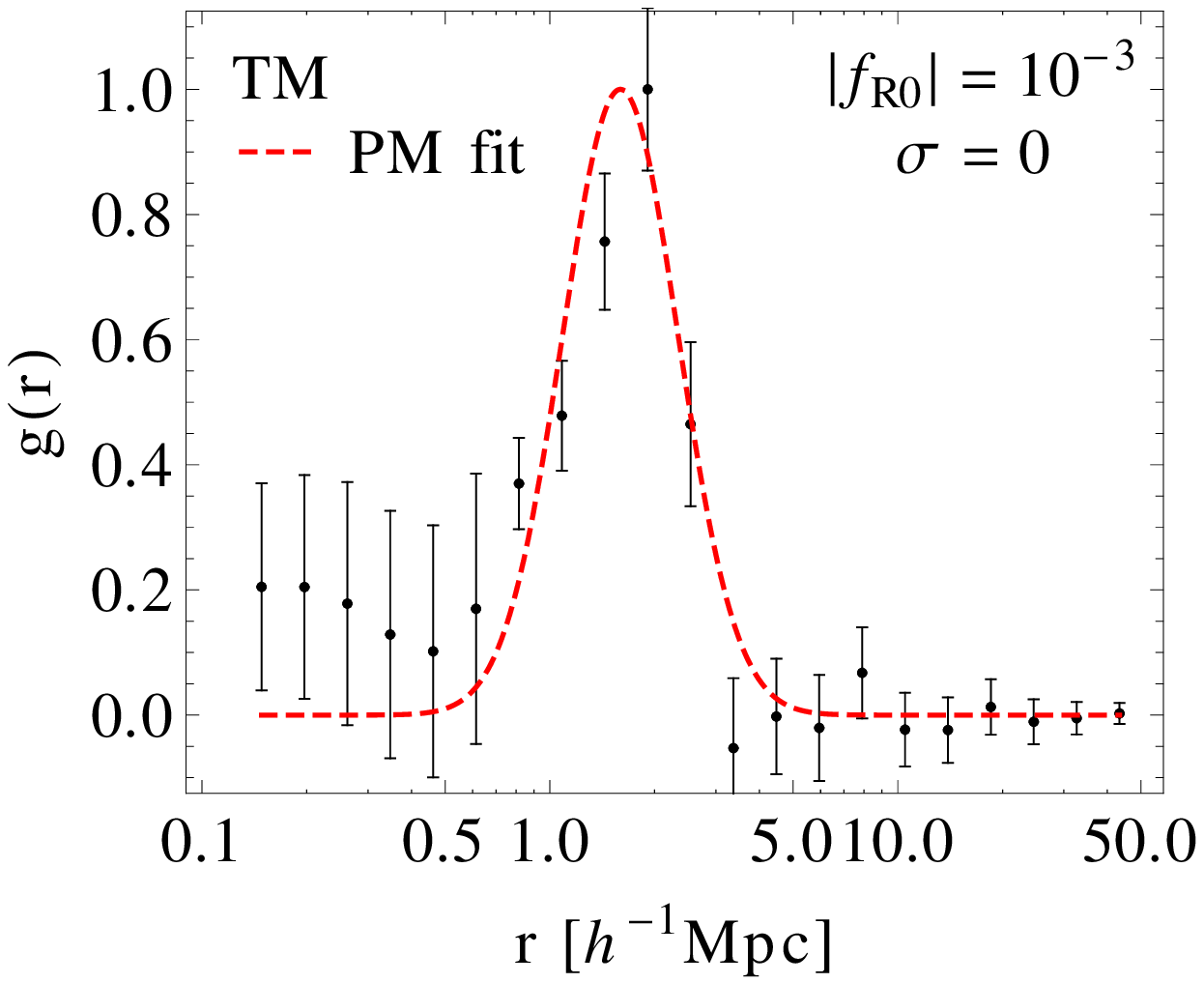}\includegraphics{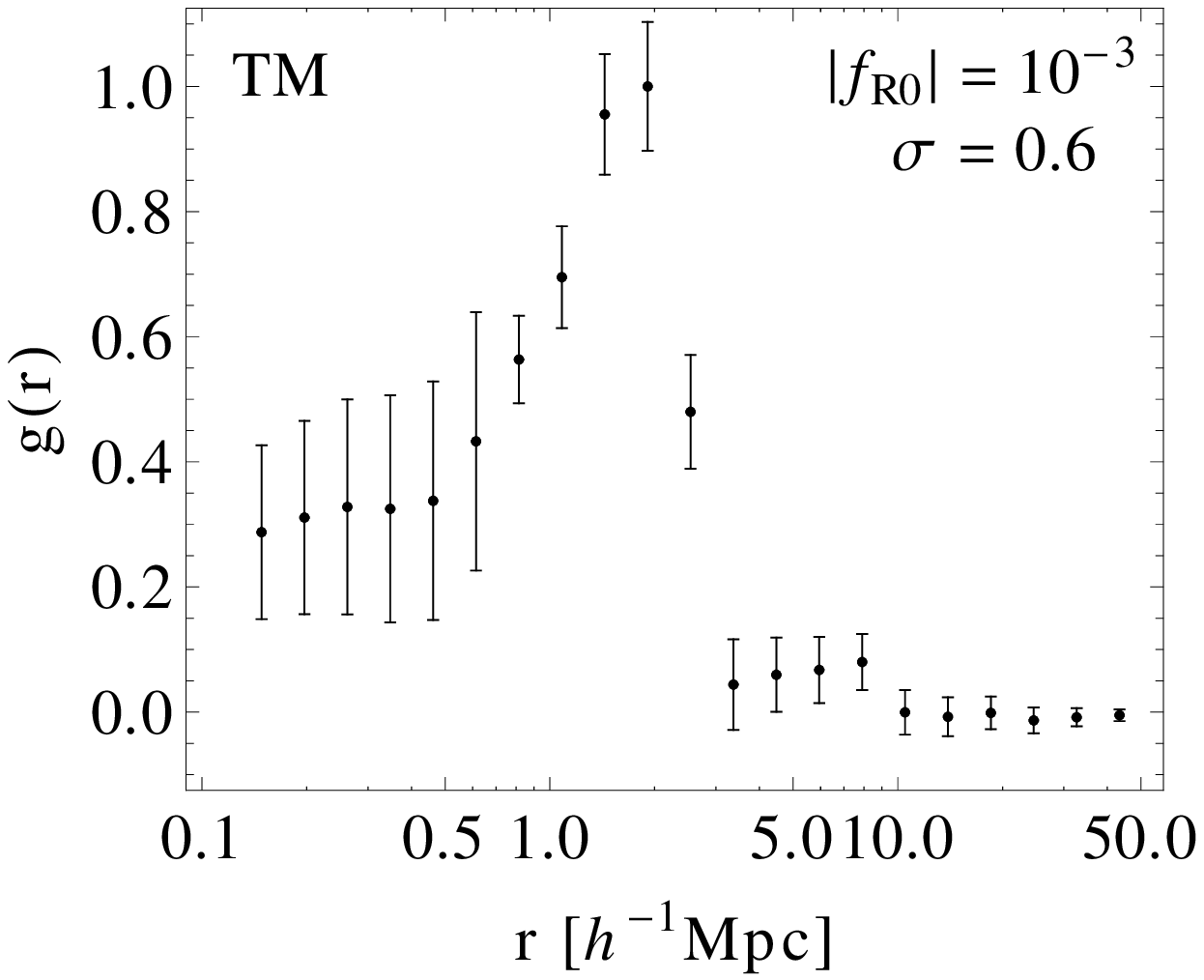}}
\caption{The shape $g(r)$ of the relative enhancement of $\xi_{\rm hm}(r)$ in $f(R)$ gravity simulations, for $\absfR=10^{-3}$ in the abundance- (left panel) and threshold-matched case with scatter $\sigma=0$ (middle panel) and $\sigma=0.6$ (right panel), respectively. The middle panel shows the best-fit Gaussian function, Eq.~(\ref{eq:xipheno}), to the simulation output.}
\label{fig:shape}
\end{figure*}

\begin{figure*}
  \includegraphics[width=0.326\textwidth]{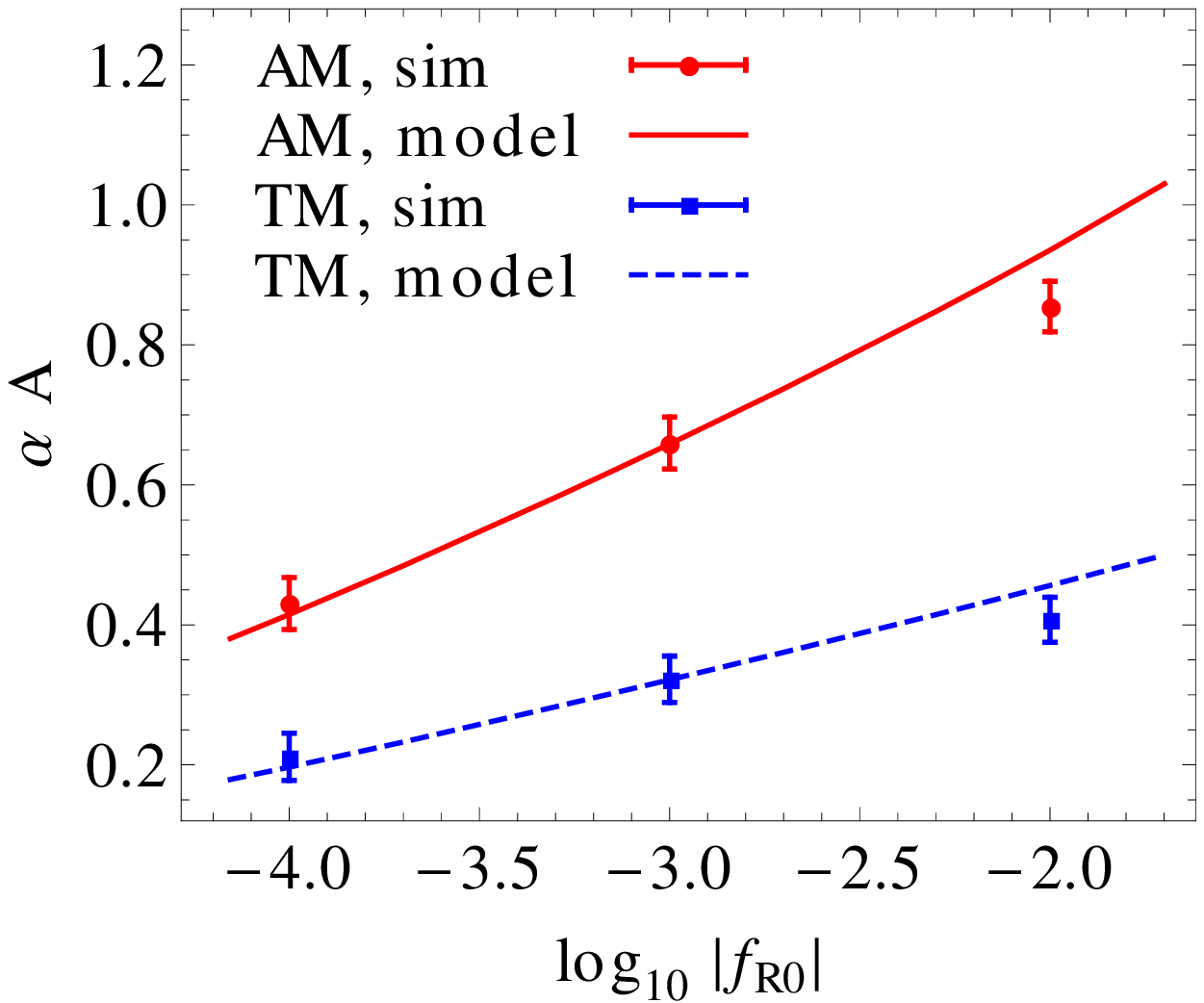}
  \includegraphics[width=0.326\textwidth]{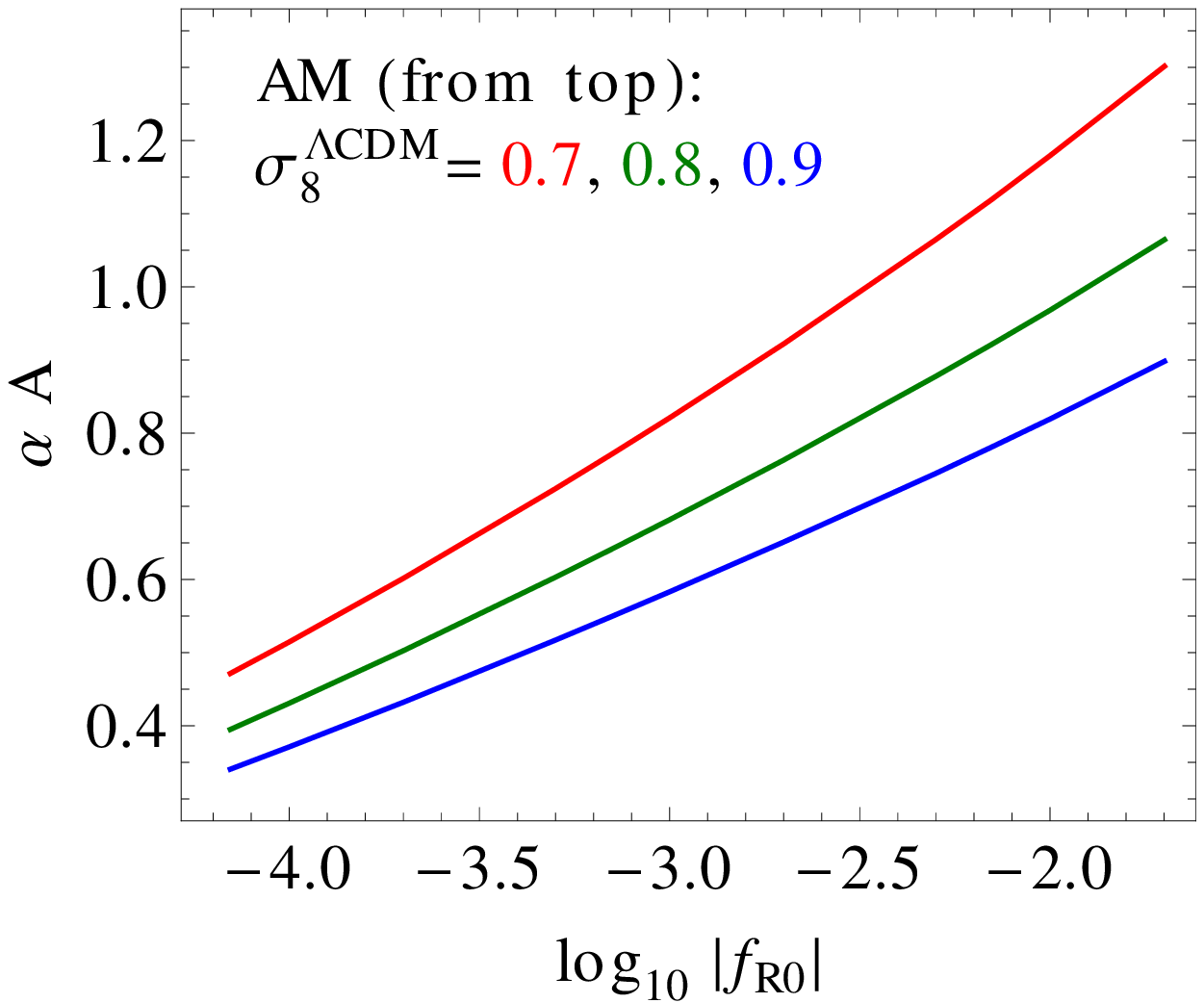}
  \includegraphics[width=0.336\textwidth]{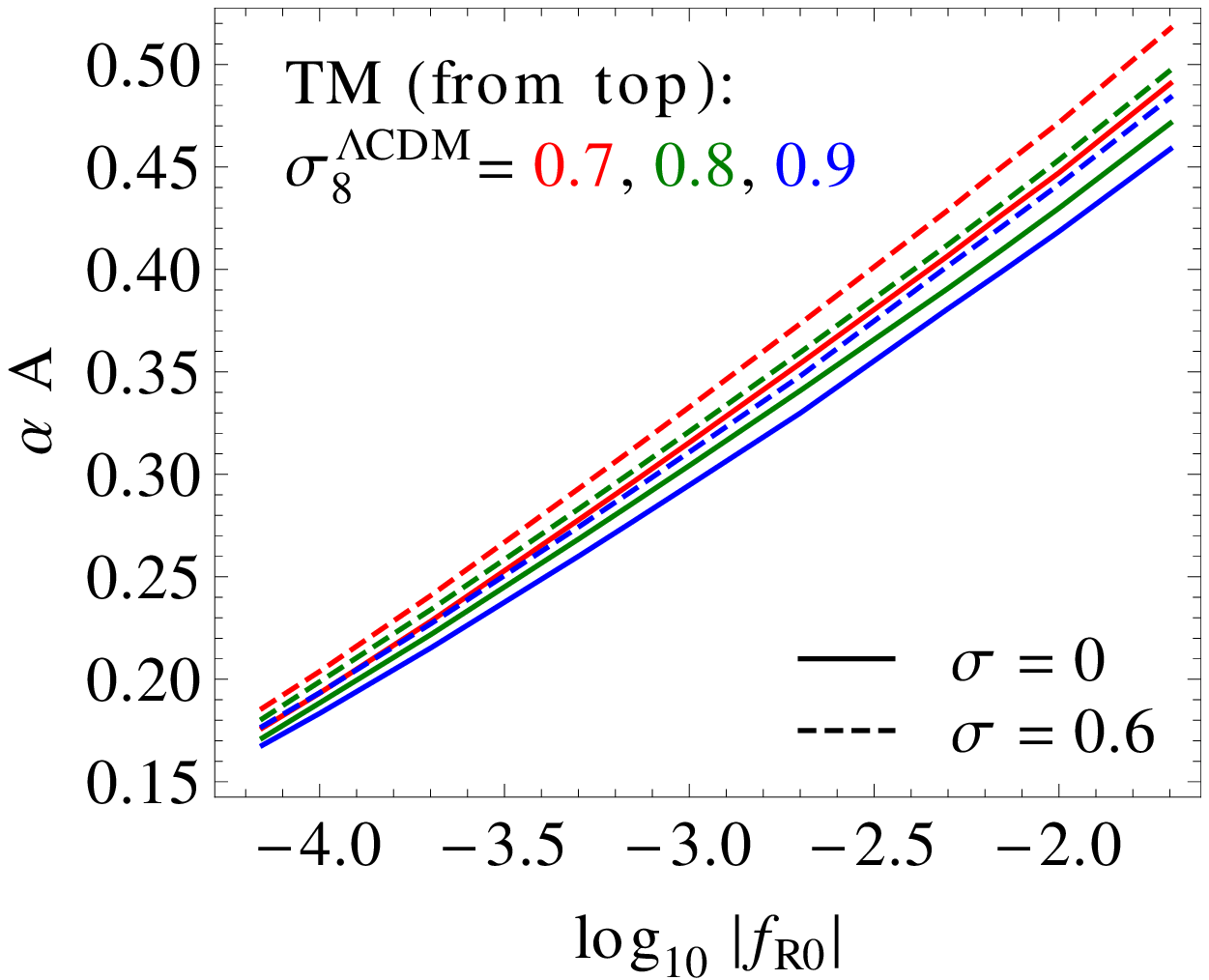}
 \caption{
  \textit{Left:}  Simulation measurements (points) and model 
predictions (lines) for the peak enhancement of $\xi_{\rm hm}(r)$, i.e., $\alpha \, A\left(\absfR,\sgL\right)$, as a function
 of $\absfR$.  Note the approximately logarithmic dependence of $\alpha\,A$ 
on $\absfR$.  
  \textit{Middle:} The peak enhancement of $\xi_{\rm hm}(r)$ in the abundance-matched case as a function of $\absfR$ for different values of the power spectrum normalization $\sigma_8^{\Lambda\textrm{CDM}}$.
  \textit{Right:}  Same for the threshold-matched case, including the dependence on the scatter $\sigma$.  Effects from scatter are negligible in the abundance-matched case.
 } 
 \label{fig:amplitude}
\end{figure*}

Since the $f(R)$ simulations are of worse resolution and smaller volume
compared to the $\Lambda$CDM simulations, we parametrize the relative
effect on the halo-matter cross correlation $\xi_{\rm hm}$, rather
than $\xi_{\rm hm}$ itself.  That is, we measure
\begin{equation}
\arr_{\rm sim}(r, \absfR) \equiv \frac{\xi_{\rm hm,sim}(r, \absfR)}
{\xi_{\rm hm,sim}
(r, \absfR=0)} - 1
\end{equation}
from the simulation outputs at $z=0.22$ with $\absfR > 0$ and $\absfR = 0$.  We apply the
scatter in mass as is done in the $\Lambda$CDM simulations (but only
for $\sigma =0, \, 0.6$).  
In order to compare the $f(R)$ gravity profiles to their $\Lambda$CDM counterparts, 
we consider two cases: 
a fixed common lower mass limit $M_0$, derived from the
$\Lambda$CDM concordance cosmology (threshold-matched case, TM); and a 
lower mass limit for $f(R)$ adjusted to match the abundance of 
tracers $\bar{n}$ (abundance-matched case, AM).  Since
the mass function of halos is enhanced in $f(R)$ gravity, the $f(R)$ mass
threshold is higher in the second case.  
The AM case is a consistent approach 
for comparing $f(R)$ gravity to $\Lambda$CDM;  on the other hand,
in the TM approach, we purely rely on the modified gravity effects on halo
profiles, without explicitly using the information from the mass function that has
been used to place constraints on $f(R)$ in \cite{lombriser:10}.  

The effects of a modification of gravity are significantly less severe 
in the TM case as compared to the abundance-matched case, i.e.,
when taking into account that massive halos are more abundant in $f(R)$.  
This effect is illustrated in Fig.~\ref{fig:shape},
which shows $\arr_{\rm sim}(r)$ normalized to unity at the peak, i.e., $g(r)$ [see Eq.~(\ref{eq:xiratio})], and 
Fig.~\ref{fig:amplitude}, which shows the peak amplitude as function of
$\absfR$.  
The profile enhancements peak at a few virial radii, corresponding to the
infall region onto massive clusters.  This effect has also been found in
simulations of other modified gravity models \cite{schmidt:09b}, and is a generic
result of modified gravitational forces increasing towards late times
(which typically is the case for models linked to the late-time acceleration
of the Universe).

Since the $f(R)$ simulations have only been run for one cosmology and
a small set of values of $\absfR$, we use the halo model to interpolate
between the simulation predictions.  We have found that the shape $g(r)$
of the profile enhancement (see Fig.~\ref{fig:shape}),
when normalized to unity at the peak of the enhancement,
is independent of $\absfR$ to within a few percent for the simulated values of $\absfR$.  
In the following, we will adopt $g(r)$ measured for
$\absfR = 10^{-3}$.  Hence, we write
\begin{equation}
\arr(r) = \alpha \,A\left(\absfR, \sigma_8^{\Lambda\rm CDM}, \sigma\right)\,g(r, \sigma),
\label{eq:xiratio}
\end{equation}
where $A$ is the peak height predicted in the halo model as function
of $\absfR$ and $\sigma_8^{\Lambda\rm CDM}$, the $\sigma_8$ a
$\Lambda$CDM universe would have for a given primordial power spectrum
amplitude, and the scatter $\sigma$.
The halo model predictions are described in Appendix~\ref{app:HM}.  
$\alpha$ is a fudge factor, which is determined by matching to $\arr_{\rm sim}(r)$ at
the peak;
in other words, we are only using the halo model to predict
the scaling with $f_{R0}$ and $\sigma_8$, while the simulations are used to
match the precise amplitude.
$\alpha$ depends on whether we are
considering the AM or TM case.
In the AM case, scatter effects on $\alpha$ 
and $g$ can be neglected, i.e., $g$ is only a function of $r$, and
$\alpha=0.52$.  In the TM case, we have
$\alpha(\sigma=0)=0.73$, $\alpha(\sigma=0.6)=0.77$,
and interpolate $\alpha$ and $g(r)$ linearly in $\sigma$. 

For $\absfR$ and $\sgL$, we use an interpolation based on the halo model for
$\absfR \leq 2\times10^{-2}$ and $\sgL \in [0.7,0.9]$.
In order for the MCMC runs to converge, however, we need to
cover a larger parameter space in $\absfR$ and $\sgL$ than can reasonably be covered
by the halo model.  Thus, when
$\absfR > 2\times10^{-2}$ and $\sgL \notin [0.7,0.9]$,
we use an extrapolation fitted to the halo model predictions for
$\absfR \leq 2\times10^{-2}$ and $\sgL \in [0.7,0.9]$, as described in
Appendix~\ref{sec:extrapolation}.  However, the details of 
this extrapolation are not important for the final parameter constraints 
since they lie well within the region that is covered by the simulations and the halo model inter- and extrapolation
(see \textsection\ref{sec:constraints}).

Finally, the prediction for the halo-mass correlation function in $f(R)$ gravity is
given by
\begin{equation}
\xi_{\rm hm}(r) =  \left[ \arr(r) + 1 \right] \xi^{\Lambda\rm CDM}_{\rm hm}(r),
\label{eq:xifR}
\end{equation}
where here and throughout $\xi^{\Lambda\rm CDM}_{\rm hm}(r)$ is 
the $\Lambda$CDM prediction interpolated from
the measurements in the {\sc zhorizon} simulations.

\begin{figure*}
 \resizebox{\hsize}{!}{\includegraphics{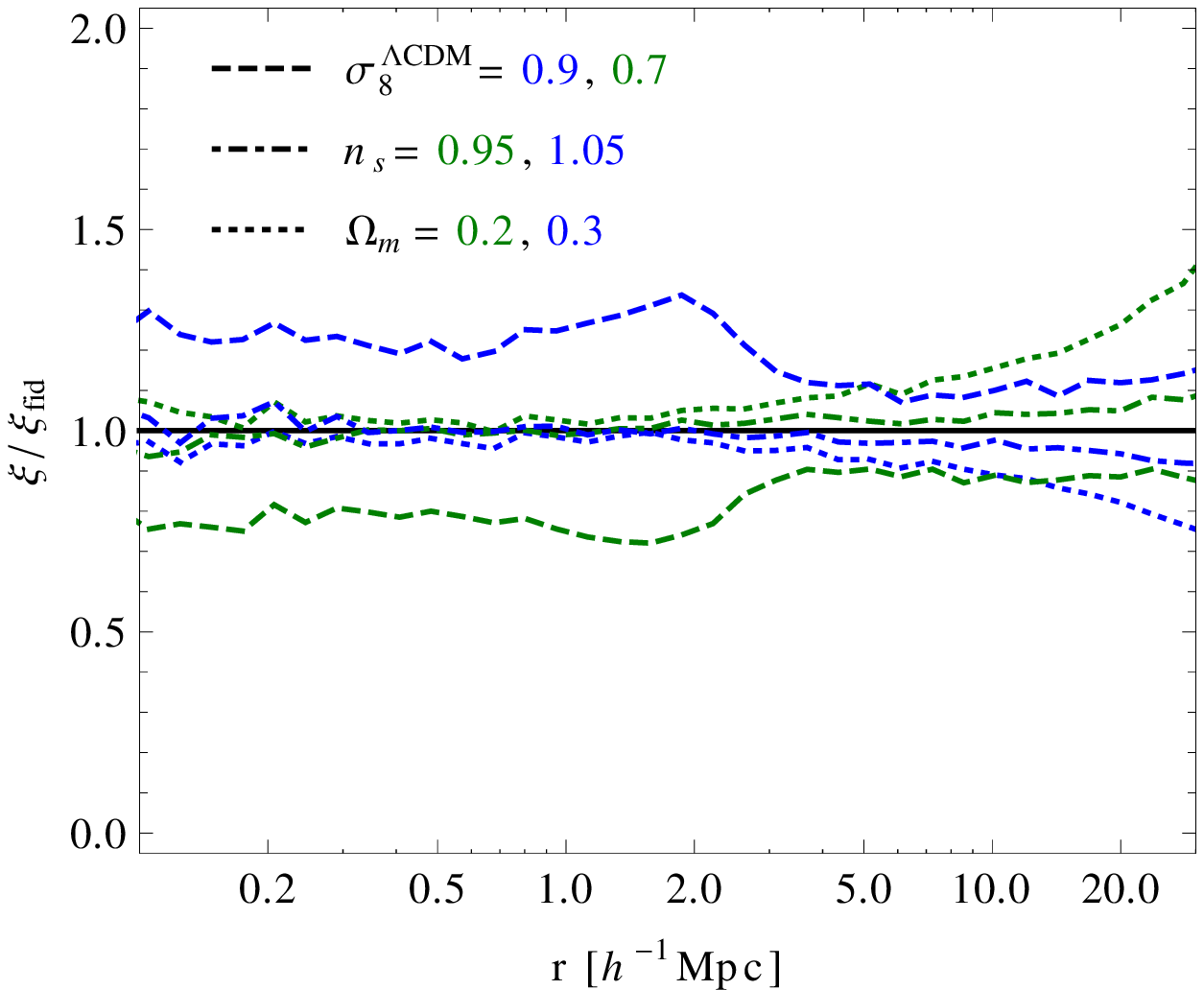}\includegraphics{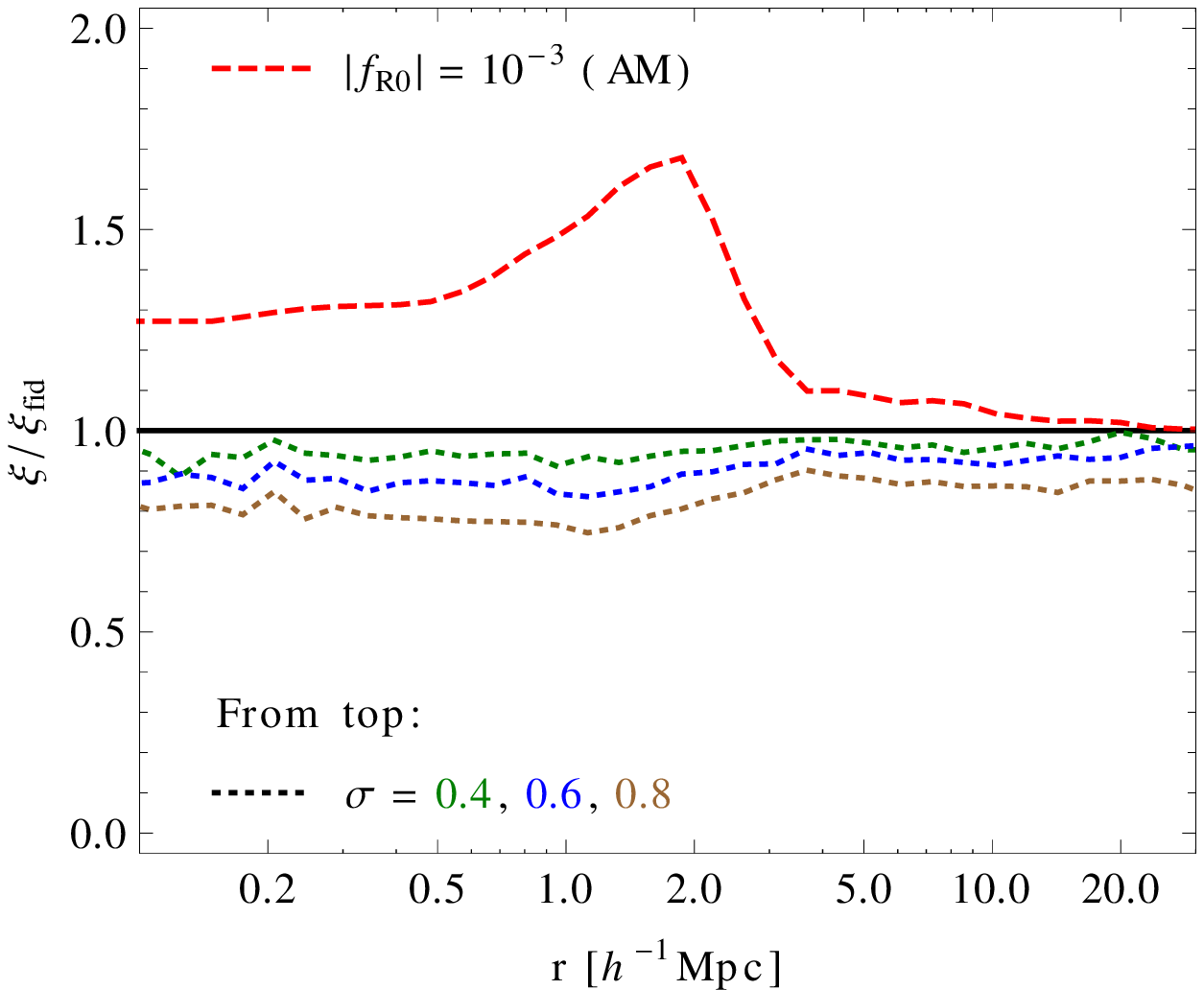}}
 \resizebox{\hsize}{!}{\includegraphics{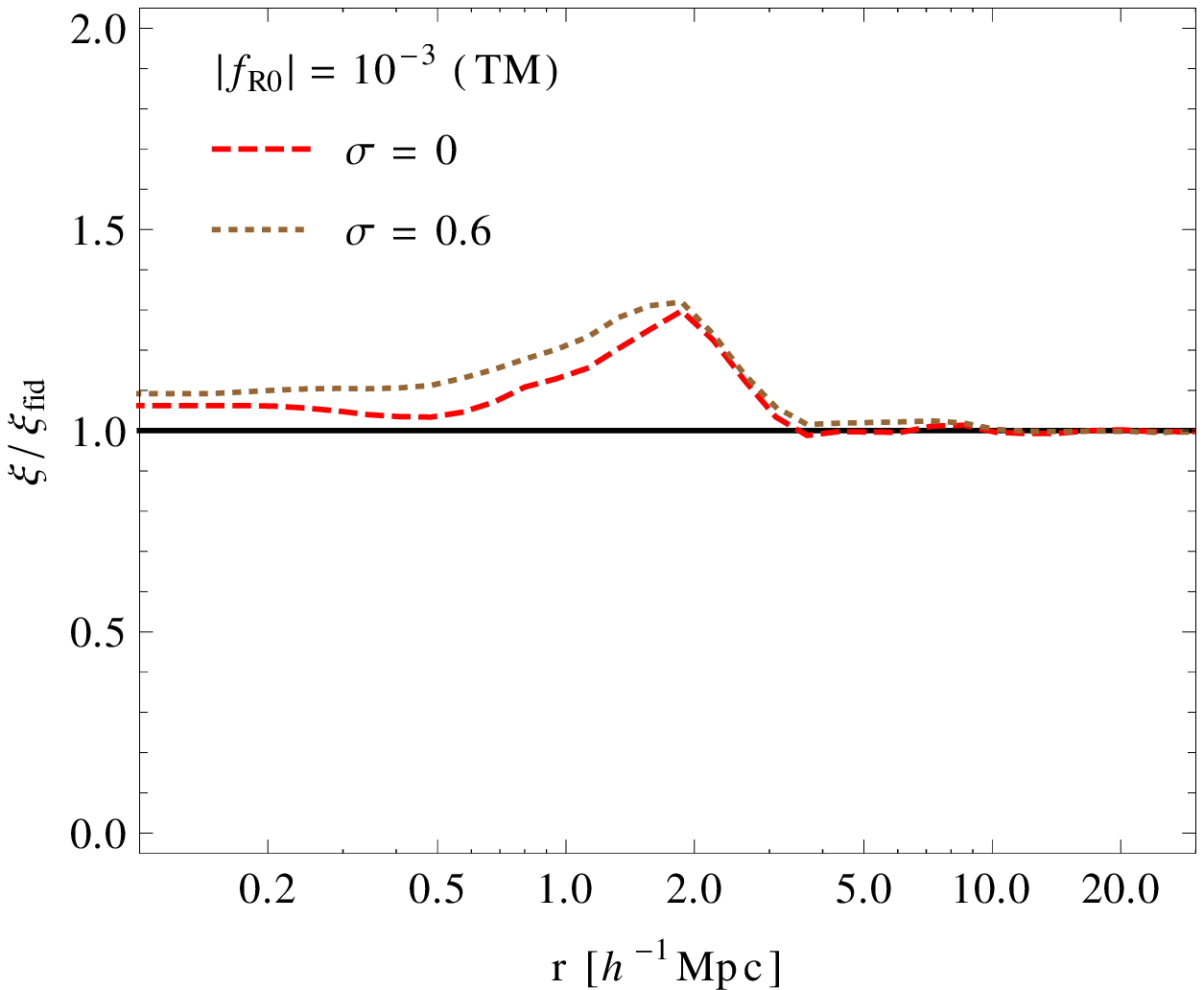}\includegraphics{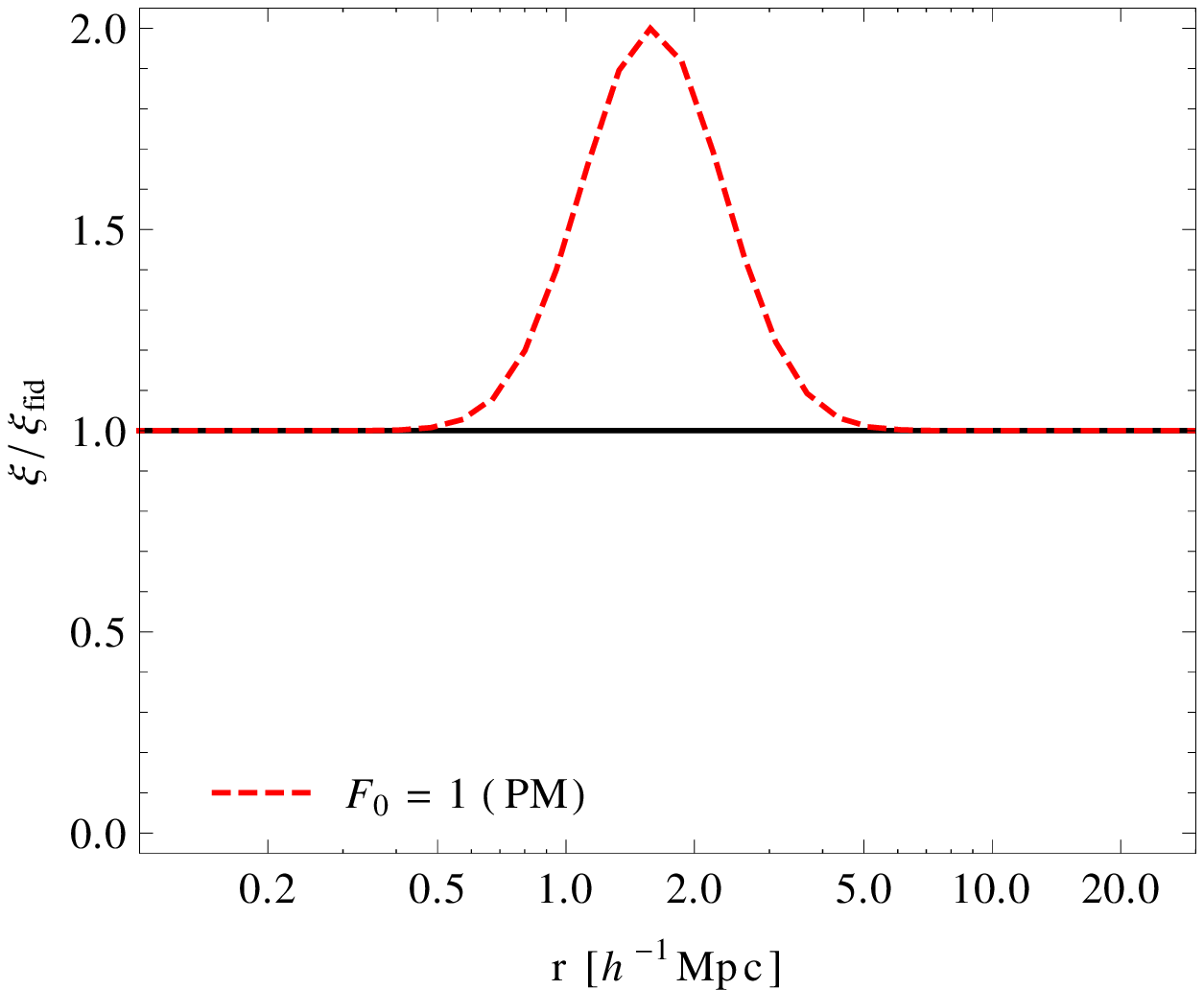}}
 \caption{Effects on the halo density profile $\xi_{\rm hm}$ from varying the cosmological parameters with respect to the fiducial case. \textit{Upper left}: Different parameter values for $\sgL$ (dashed), $n_{\rm s}$ (dot-dashed), and $\Om$ (dotted). \textit{Upper right}: $\absfR=10^{-3}$ for the abundance-matched case (dashed) and for the fiducial $\Lambda$CDM cosmology with different values of scatter $\sigma$ (dotted). \textit{Lower left}: $\absfR=10^{-3}$ for the threshold-matched case with $\sigma=0$ (dashed) and $\sigma=0.6$ (dotted) (with the fiducial case corrected for scatter). \textit{Lower right}: $F_0=1$ for the phenomenological scenario (see~\textsection\ref{sec:xigauss}).}
\label{fig:xi}
\end{figure*}

\subsection{Phenomenology with a Gaussian fit} \label{sec:xigauss}

In addition to the consistent, abundance-matched constraints on $f(R)$ 
gravity, we also consider a phenomenological approach modeled on the
profile enhancement in $f(R)$ at fixed halo mass (TM case).  This case
serves to illustrate the ability of halo profiles to probe gravity, independently
of halo abundances and the specific $f(R)$ model.  To do this,
we fit $\arr_{\rm sim}(r)$ for the $\absfR=10^{-3}$ threshold-matched case 
without scatter
for the amplitude, width, and position of a Gaussian function in $\ln r$ and then take the amplitude $F_0$ to be the free parameter controlling the modification, i.e.,
\begin{equation}
\arr^{\rm PM}(r, F_0) = F_0 \exp \left[ - \frac{1}{2} \left( \frac{\ln r - \mu}{\varsigma} \right)^2 \right].
\label{eq:xipheno}
\end{equation}
The minimum $\chi^2$ for the fit of the fixed mass simulation 
(see Fig.~\ref{fig:shape}) is obtained for $e^{\varsigma}=1.47\hMpc$ and 
$e^{\mu}=1.59 \hMpc$.  
Note that $\arr_{\rm sim}(r)$ in the AM case is not simply described by a 
Gaussian enhancement.  

In the middle panel of Fig.~\ref{fig:shape}, we show the enhancement of the modified relative to the $\Lambda$CDM ($|f_{R0}|=0$) simulated density profile for $\absfR=10^{-3}$ and the corresponding Gaussian function.  
In the following, we refer to this approach as the phenomenological model (PM) case.

For comparison, $F_0$ matches the peak height of the enhancement in the threshold-matched scenario for
\begin{equation}
 F_0= \alpha\, A\left(\absfR=10^{-3},\sgL=0.8, \sigma=0\right) \simeq 0.306.
\end{equation}
In general, one can map $F_0$ to the corresponding value of $\absfR$ in the
TM case through the right-hand panel of Fig.~\ref{fig:amplitude}.
We shall, however, not restrict the likelihood analysis to only non-negative values of $F_0$,
in correspondence with $\absfR \geq 0$ but extend it to cases where $F_0<0$, i.e., models where gravity
is weakened and profiles are consequently suppressed. A suppression of this kind may, for instance, be
observed in self-accelerating DGP braneworlds \cite{schmidt:09b}.

\subsection{Lensing predictions} \label{sec:DSigma_pre}

\begin{figure*}
 \resizebox{\hsize}{!}{\includegraphics{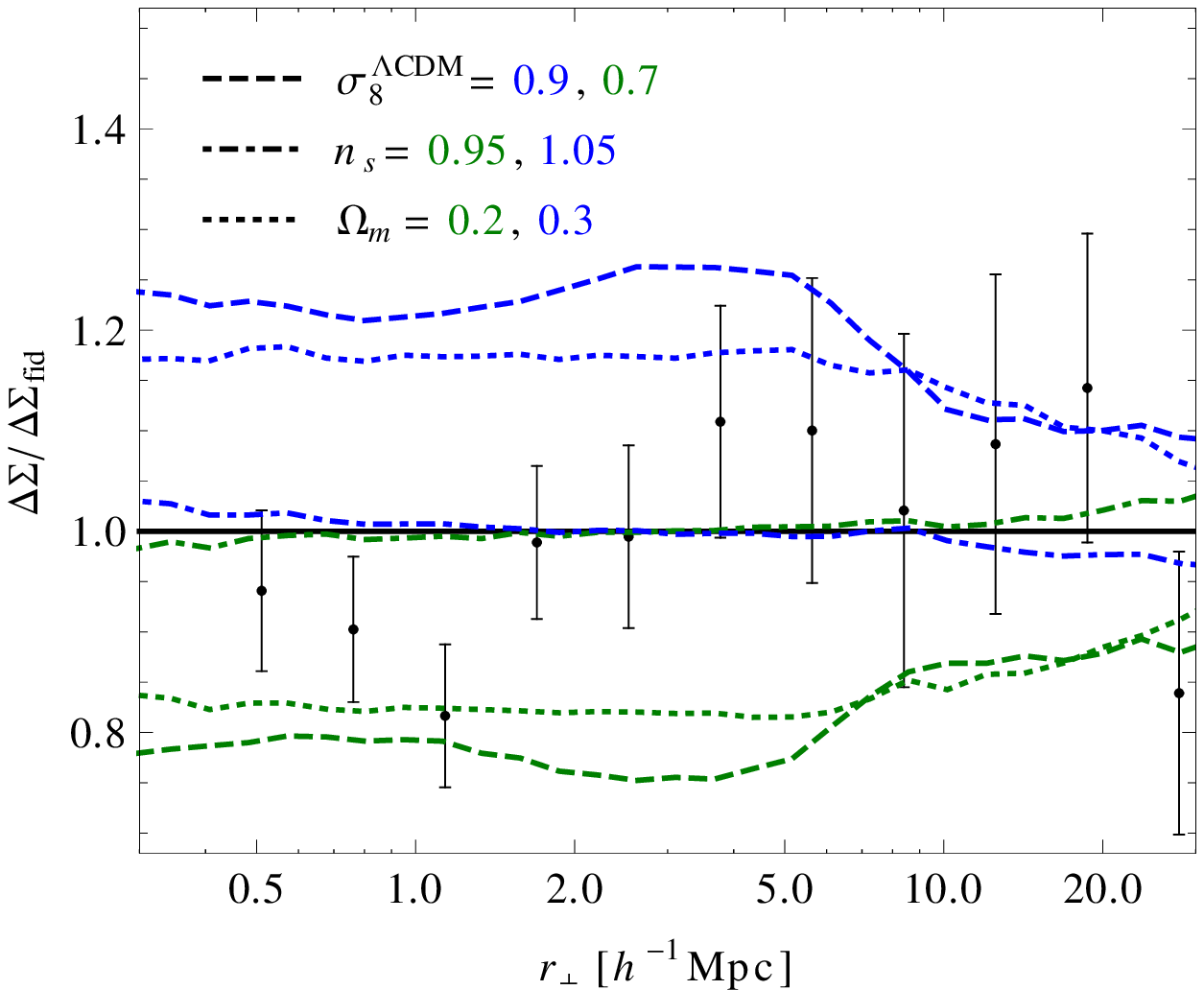}\includegraphics{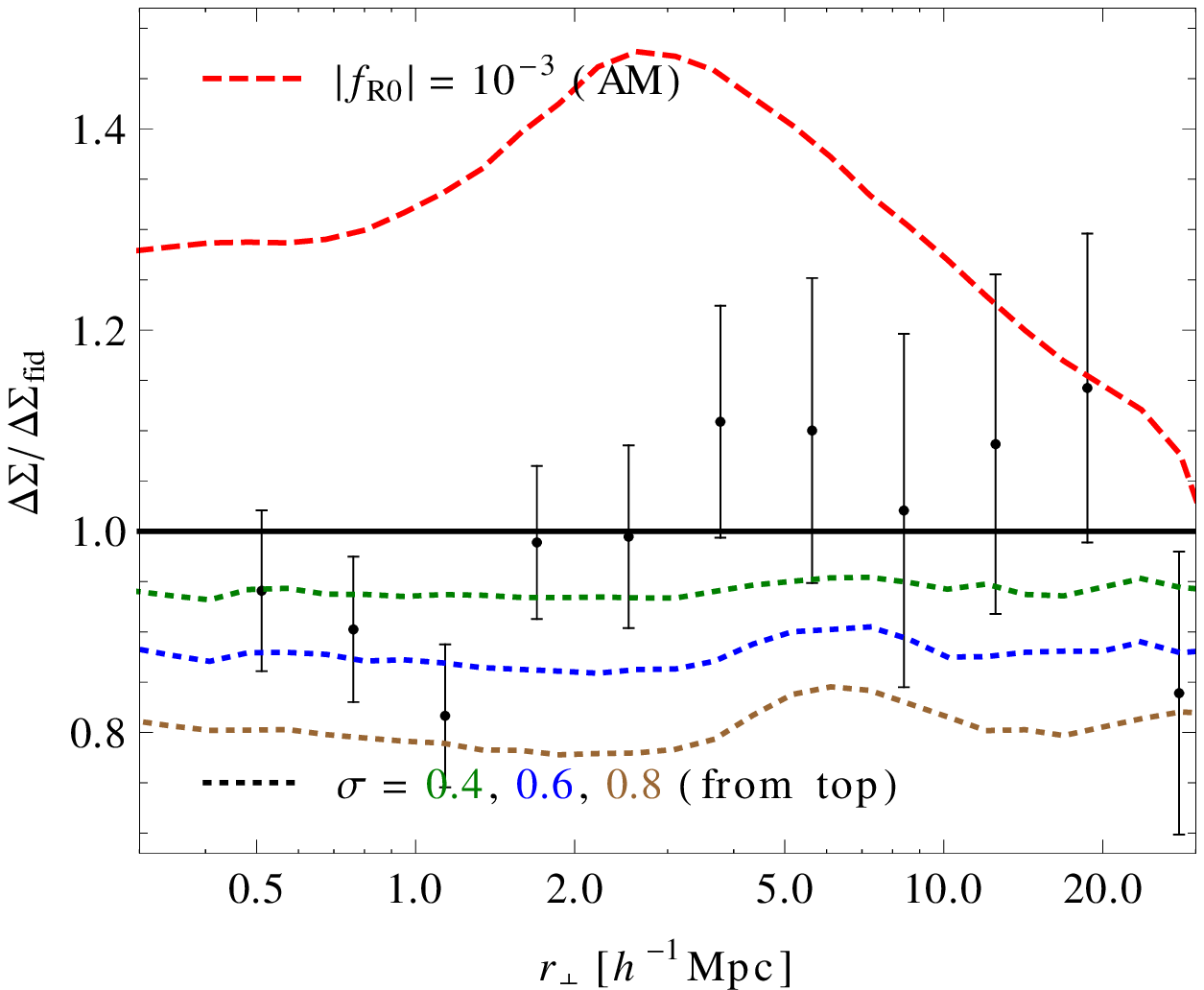}}
 \resizebox{\hsize}{!}{\includegraphics{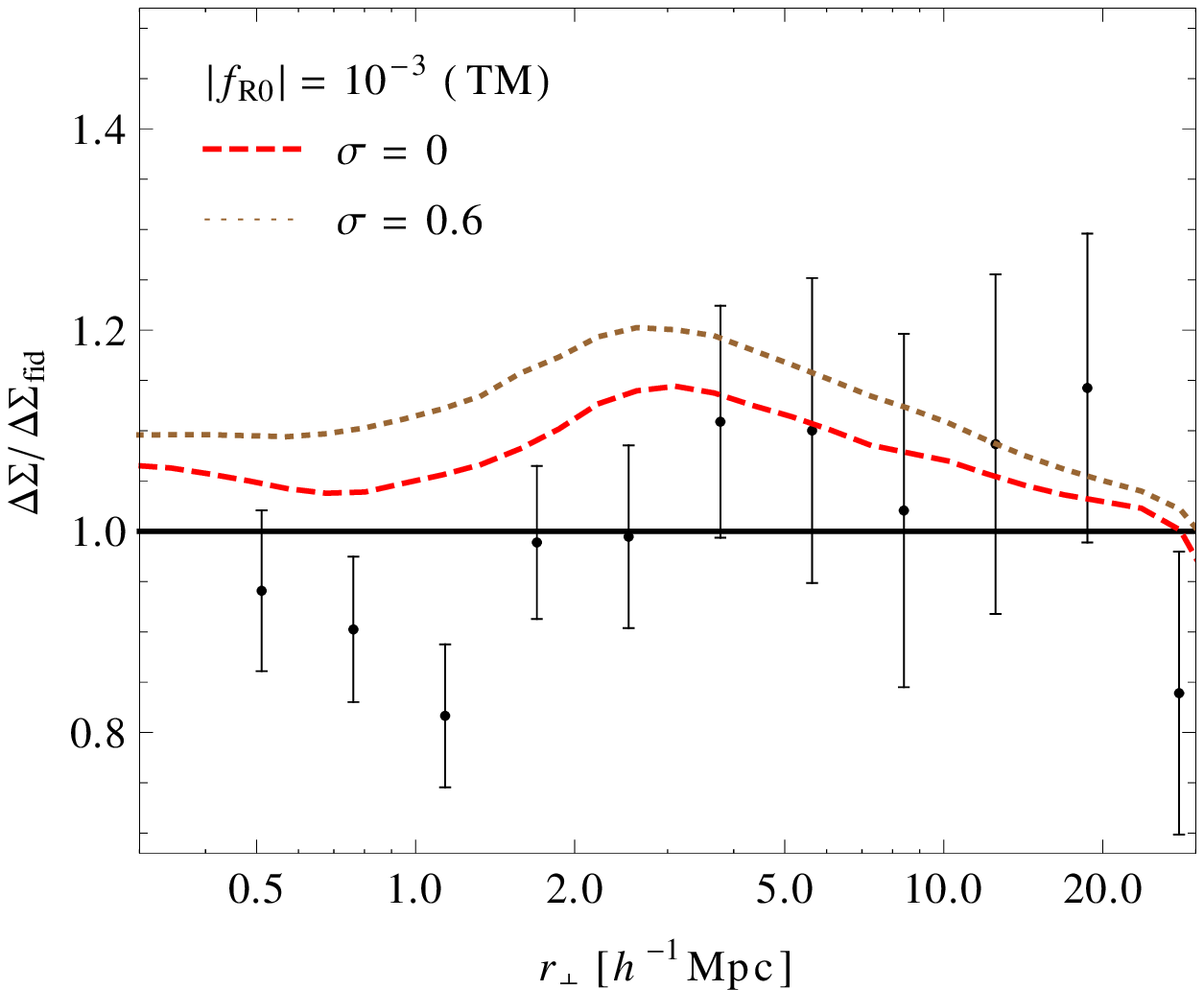}\includegraphics{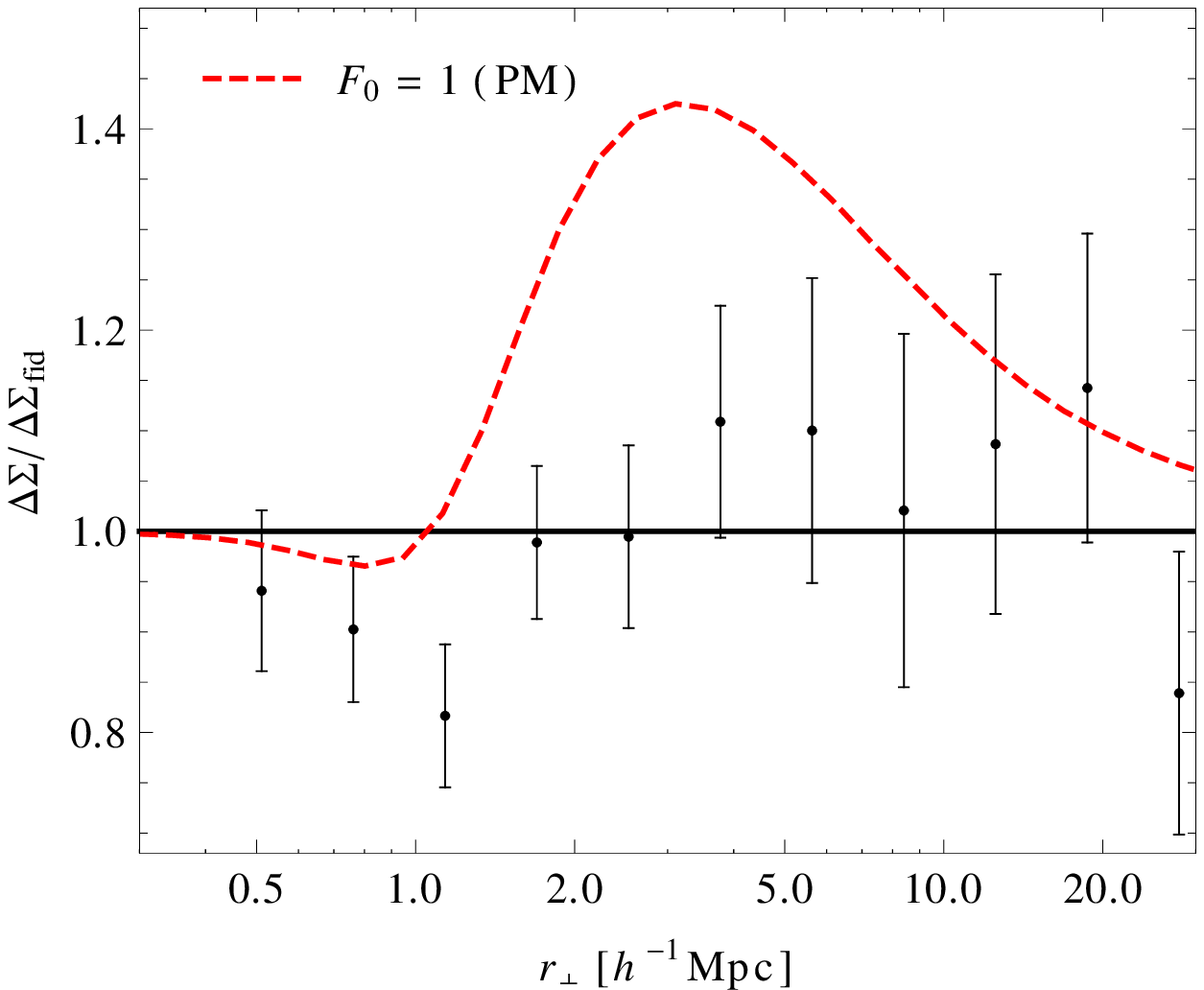}}
\caption{Effects on the excess surface mass density $\Delta\Sigma$ from varying the cosmological parameters with respect to the fiducial case.
The lensing data has been rebinned for illustrative purposes (cf. Fig.~\ref{fig:DSigma_obs})
\textit{Top left}: $\sgL$ (dashed), $n_{\rm s}$ (dot-dashed), $\Om$ (dotted). 
\textit{Top right}: $\absfR=10^{-3}$ for the abundance-matched case (dashed) and for the fiducial $\Lambda$CDM cosmology with different values of scatter $\sigma$ (dotted). 
\textit{Bottom left}: $\absfR=10^{-3}$ for the threshold-matched case with $\sigma=0$ (dashed) and $\sigma=0.6$ (dotted) (with corresponding values of $\sigma$ used in the 
fiducial $\ds$ for each case).
\textit{Bottom right}: $F_0=1$ for the phenomenological scenario.
\label{fig:DSigma}}
\end{figure*}

Fig.~\ref{fig:xi} illustrates the effect of varying the cosmological
parameters and mass scatter $\sigma$ on $\xi_{\rm hm}$.  It is apparent that
the $f(R)$ field strength $\absfR$ and $\sigma$ have the largest impact
on the profiles; this shows that halo density profiles in the region
of one to a few virial radii are useful as probes of gravity, in particular,
if external information on the scatter is available.  Note also that the
profile enhancement is significantly smaller in the TM case when compared
to AM at a fixed value of $\absfR$.

We first determine $\Delta\Sigma^{\Lambda{\rm CDM}}$ from 
$\xi_{\rm hm}^{\Lambda{\rm CDM}}$, i.e., without including 
modified gravity effects, using Eqs.~(\ref{eq:DSigma}) through (\ref{eq:projsmd}) for each concordance model cosmology in Table~\ref{tab:cos_par}.  
At each $r_\perp$, we use a four dimensional paraboloid to interpolate $\Delta\Sigma^{\Lambda{\rm CDM}}$ in the parameters $\left\{ \Omega_{\rm m}, \sgL, n_{\rm s}, \sigma \right\}$.
The paraboloid is defined by three simulations in each parameter direction.
We then interpolate linearly in $\log r_{\perp}$.  

In order to include the modified gravity effects for the AM/TM case, we write
\begin{eqnarray}
\Delta\Sigma(r_\perp, \absfR) & = & \left[ 1 + \frac{A(\absfR,\sgL)}{A_{\rm fid}} \Delta\arr_{\rm fid}(r_\perp) \right] \nonumber \\
& & \times \Delta\Sigma^{\Lambda\textrm{CDM}}(r_\perp),
\label{eq:DSigma_app}
\end{eqnarray}
where $\Delta\Sigma^{\Lambda\textrm{CDM}}$ contains the dependency on the cosmological parameters, $A_{\rm fid} = A(\absfR=10^{-3},\sgL=0.8)$, and
\begin{equation}
\Delta\arr_{\rm fid}(r_\perp) = \frac{\Delta\Sigma_{\rm fid}(r_\perp, \absfR=10^{-3})}{\Delta\Sigma_{\rm fid}^{\Lambda{\rm CDM}}(r_\perp)} - 1
\end{equation}
is obtained by inserting Eq.~(\ref{eq:xiratio}) into Eq.~(\ref{eq:xifR}) when performing the projection, Eqs.~(\ref{eq:DSigma}) through (\ref{eq:projsmd}),
using the fiducial values for the cosmological parameters defined in Table~\ref{tab:cos_par} with $\sigma=0$.

Similarly, for the PM case we write
\begin{equation}
\Delta\Sigma(r_\perp, F_0) = \left[ 1 + F_0 \Delta\arr^{\rm PM}_{\rm fid}(r_\perp) \right] 
\Delta\Sigma^{\Lambda\textrm{CDM}}(r_\perp)
\label{eq:DSigma_app2}
\end{equation}
where
\begin{equation}
\Delta\arr^{\rm PM}_{\rm fid}(r_\perp) = \frac{\Delta\Sigma_{\rm fid}(r_\perp, F_0=1)}{\Delta\Sigma_{\rm fid}^{\Lambda{\rm CDM}}(r_\perp)} - 1
\end{equation}
is obtained by inserting Eq.~(\ref{eq:xipheno}) 
into Eq.~(\ref{eq:xifR}) when performing the projection, Eqs.~(\ref{eq:DSigma}) through (\ref{eq:projsmd}), with fiducial values for the cosmological parameters.
Eqs.~(\ref{eq:DSigma_app}) and (\ref{eq:DSigma_app2}) are approximate and assume
that the $r$-dependence of the modified gravity effects does not depend
on the cosmological parameters.  We found that this approximation is valid
to better than 1\%.  

The effects of varying cosmological parameters on $\Delta\Sigma$ are illustrated in 
Fig.~\ref{fig:DSigma}.   Comparing Fig.~\ref{fig:DSigma} and Fig.~\ref{fig:xi}, we see 
that the relative enhancement observed in the halo profiles in $f(R)$ gravity is 
broadened and propagated to larger radial scales by the projection and conversion
to the excess surface mass density.

\section{Observations}\label{sec:observations}

The observations in this paper are derived from the SDSS \citep{2000AJ....120.1579Y}, which imaged roughly $\pi$ steradians
of the sky, and followed up approximately one million of the detected
objects spectroscopically \citep{2001AJ....122.2267E,
  2002AJ....123.2945R,2002AJ....124.1810S}. The imaging was carried
out by drift-scanning the sky in photometric conditions
\citep{2001AJ....122.2129H, 2004AN....325..583I} in five bands
($ugriz$) \citep{1996AJ....111.1748F, 2002AJ....123.2121S} using a
specially-designed wide-field camera
\citep{1998AJ....116.3040G}. These imaging data were used to create
the cluster and source catalogs that we use in this paper.  All of
the data were processed by completely automated pipelines that detect
and measure photometric properties of objects, and astrometrically
calibrate the data \citep{2001ASPC..238..269L,
  2003AJ....125.1559P,2006AN....327..821T}. The SDSS I/II imaging
surveys were completed with a seventh data release
\citep{2009ApJS..182..543A}, though this work relies as well on an
improved data reduction pipeline ({\sc Photo v5\_6}) and updated
photometric calibration (ubercalibration,
\cite{2008ApJ...674.1217P}) that is part of the eighth data
release, from SDSS-III
\citep{2011ApJS..193...29A,2011AJ....142...72E}.

\subsection{Lens cluster sample}

We use cluster-galaxy lensing measurements around a subset of the maxBCG optically detected cluster sample from the SDSS, consisting of $5\,891$ clusters with background sources. The parent sample of clusters from which our lens sample is derived consists of $13\,823$ MaxBCG clusters~\cite{koester:07} that are identified by concentrations of galaxies in color-position space using $7\,500$ square degrees of imaging data from the SDSS. The entire sample is placed into a single redshift slice spanning $0.1<z<0.3$ ($z_{\rm eff}=0.23$), and a redshift-dependent richness cut in $N_{200}$ (the number of red member galaxies above some luminosity threshold) is applied to achieve a redshift-independent number density of $\bar{n}=2\times10^{-5}\hMpcc$. 

The maxBCG sample is particularly well suited for our study on halo profiles 
since the BCG 
is expected to coincide with the center of its host halo, i.e., the minimum of 
the potential well. If this assumption is perfectly satisfied, then our analysis is simplified since no modeling of
the mass distribution around satellite galaxies (including assumptions about their hosts, cf.~\citep{2005MNRAS.362.1451M}) is required. 
 
To ensure that this is the case, and reduce effects from possible ``satellites'' (in reality, clumps of galaxies within some larger cluster that are misidentified as a separate, nearby cluster) contaminating the maxBCG sample, 
we define a cylindrical
region around each cluster with a transverse radius of three virial radii, derived using the mass-richness relation from \cite{mandelbaum:09}, and extent along the line of sight of $\Delta z = \pm 0.045$ (corresponding to $\chi\approx\pm100 \hMpc$, a $3\sigma$ photo-$z$ error).  If there is a lighter cluster candidate in this region, then the lighter cluster is removed from the sample.  
This removes 30\% of the clusters in the sample, resulting in a net observed number density of $\bar{n}=1.4\times10^{-5}\hMpcc$.  As described in \textsection\ref{sec:halomodelpred}, carrying out the same procedure on the halo catalog of the $N$-body simulations removes 20\% of the halos.  This finding suggests that of the 30\% that were removed from the maxBCG sample, 10\% were truly spurious detections and 20\% were removed due to chance projections.  We thus estimate the true parent sample number density to be $\bar{n}=1.8\times10^{-5}\hMpcc$.  This is the value used when abundance-matching the halos from the $f(R)$ and $\Lambda$CDM simulations.  We emphasize that it is not a problem that our procedure is overly conservative; it is better to avoid modeling difficulties at the expense of losing 20\% of the real clusters in the sample.

\subsection{Source catalog}

The catalog of source galaxies ($1.18$ arcmin$^{-2}$) with resolved shape measurements and photometric redshifts is described in detail by Reyes \emph{et al.}~\cite{reyes:11}.  In brief, the correction for the effects of the point-spread function (PSF) uses a method called re-Gaussianization \citep{2003MNRAS.343..459H}, with original systematics tests presented in \cite{2005MNRAS.361.1287M} and an updated treatment by Reyes \emph{et al.}~\cite{reyes:11}.  The effect of errors in the ZEBRA photometric redshifts \citep{2006MNRAS.372..565F} on the lensing signal calibration was studied by \cite{2011arXiv1107.1395N}.

\subsection{Lensing measurements} \label{sec:lensmeas}

\begin{figure}
 \resizebox{1.\hsize}{!}{\includegraphics{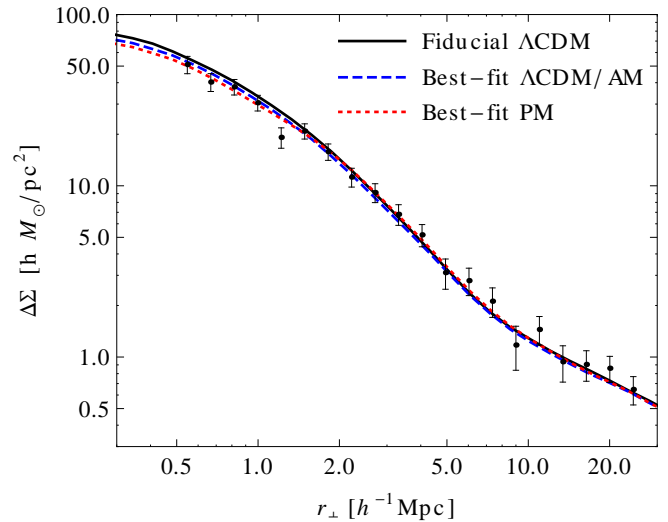}}
\caption{Excess surface mass density, Eq.~(\ref{eq:ds_obs}), as a function of the comoving transverse separation to the cluster center (BCG), $r_{\perp}$, measured in the maxBCG sample (points).  The lines show the predictions from the fiducial and best-fit $\Lambda$CDM models as well as for the best-fit phenomenological scenario (see~\textsection\ref{sec:weak_lensing}).
Note that the best-fit abundance-matched $f(R)$ model is indistinguishable from the best-fit concordance model (see Table~\ref{tab:results}) and is therefore not shown separately.
}
\label{fig:DSigma_obs}
\end{figure}

A description of the procedure for calculating the lensing signal can be found in Reyes \emph{et al.}~\cite{reyes:11}.  In brief, we assign optimal weights $w_\mathrm{ls}$ to each lens-source pair based on the noise in the shape measurement and based on the critical surface mass density $\Sigma_c^{\rm (ls)}=\Sigma_\mathrm{crit}(z_l,z_s)$ estimated using the source photo-$z$.  To estimate the lensing signal $\ds(r_\perp)$, we then compute a weighted average 
\be
\ds(r_\perp) = \frac{\sum_{\mathrm{ls}} w_{\mathrm{ls}} \gamma_t^{(\mathrm{ls})} \Sigma_c^{(\mathrm{ls})}}{2 {\cal
R}\sum_{\mathrm{rs}} w_{\mathrm{rs}}},
\label{eq:ds_obs}
\ee
in logarithmic radial bins.  The denominator includes the sum over weights of {\em random lens}-source pairs $w_{\rm rs}$, to correct for the dilution of the source sample by ``sources'' that are actually associated with the cluster and are not lensed by it.  The factor of $2$ arises due to our ellipticity definition, and ${\cal R}$ is the shear responsivity, which describes how our ellipticity definition responses to a shear \citep{2002AJ....123..583B}.  After computing the signal, we also compute the signal around the random points to check for any systematic shear contamination \citep{2005MNRAS.361.1287M}, and subtract it from the real signal (in practice, for the scales of interest, this correction is only nonzero for $R>10h^{-1}$Mpc and even then, it is well below the statistical errors).
Errors are calculated using jackknife resampling; for this purpose, we divide the survey area and therefore the lens sample into 100 equal-area regions.

We use the same procedures as in Reyes \emph{et al.}~\cite{reyes:11} to assess the impact of various sources of calibration biases on the lensing signal, and we then remove them, assigning an overall $5\%$ calibration uncertainty.  
We therefore divide the theoretical predictions for $\ds$ by the calibration factor $\mathcal{C} = 1.08$, and include a Gaussian scatter of $0.05$ on $\mathcal{C}$ when comparing
to the lensing measurements in the MCMC analysis (see~\textsection\ref{sec:pheno}).

Fig.~\ref{fig:DSigma_obs} shows the measurement of the unbiased excess surface mass density $\ds(r_\perp)$ (multiplied by $\mathcal{C})$ along with the best-fit signals for the $\Lambda$CDM, AM, and phenomenological model (see~\textsection\ref{sec:xigauss}), respectively.

In~\textsection\ref{sec:systematics}, we shall discuss further possible systematics, especially those which have scale-dependence.

\subsection{External priors}\label{sec:priors}

In order to prevent degeneracies of $\absfR$ with other cosmological parameters and combinations thereof, we further employ measurements of the background expansion history and the cosmic microwave background.
For this purpose, we consider the likelihood distribution for the concordance model parameters from \cite{lombriser:10}. This analysis uses the CMB anisotropy data from the five-year Wilkinson Microwave Anisotropy Probe (WMAP)~\cite{WMAP:08}, the Arcminute Cosmology Bolometer Array Receiver (ACBAR)~\cite{ACBAR:07}, the Cosmic Microwave Background Imager (CBI)~\cite{CBI:04}, and the Very Small Sky Array (VSA)~\cite{VSA:03}. It further utilizes data from the Supernova Cosmology Project (SCP) Union~\cite{UNION:08} compilation, the measurement of the Hubble constant from the Supernovae and $H_0$ for the Equation of State (SHOES)~\cite{SHOES:09} program generalized by~\cite{reid:09}, and the BAO distance measurements of~\cite{BAO:09}. For the description of these observables, in particular, for the CMB, a high-redshift parametrization was chosen, constructed from the physical baryon and cold dark matter density $\Omega_{\rm b}h^2$ and $\Omega_{\rm c}h^2$, the ratio of the sound horizon to angular diameter distance at recombination multiplied by 100, $\theta$, the optical depth to reionization $\tau$, the scalar tilt $n_{\rm s}$, and amplitude $A_{\rm s}$ at $k_* = 0.002~\textrm{Mpc}^{-1}$.

For our analysis we restrict to the parameters that are used for predicting the excess surface mass density $\Delta\Sigma$ in~\textsection\ref{sec:DSigma_pre}, i.e., $n_{\rm s}$ and the derived parameters, the total matter density $\Om$ and the power spectrum normalization $\sigma_8^{\Lambda\textrm{CDM}}$. Hence, we marginalize over $\{ \Omega_{\rm b}h^2, \Omega_{\rm c}h^2, \theta, \tau, \ln[10^{10} A_{\rm s}] \}$ to obtain a three-dimensional posterior distribution for $n_{\rm s}$, $\Om$, and $\sigma_8^{\Lambda\textrm{CDM}}$, which serves as our prior within the MCMC analysis.

Note that by construction, at high redshifts, $f(R)$ modifications become negligible, i.e., at large multipoles of the CMB, predictions from $f(R)$ gravity match the predictions from the concordance model. Modifications appear only at low multipoles of the CMB due to the Integrated-Sachs Wolfe effect and lead to constraints on $\absfR$ of around unity~\cite{song:07}. The background expansion history within the Hu-Sawicki $f(R)$ gravity model matches the one of $\Lambda$CDM for $\absfR\ll1$ at the accuracy level of current observations. Since we are interested in constraints on $f(R)$ modifications that originate from the halo profile alone, we restrict to the concordance model predictions for comparison with the data described here.

As a prior on the scatter $\sigma$ we adopt the probability distribution shown in the
top panel of Fig.~3 in~\cite{rozo:08}, obtained from 
comparing cluster richness with X-ray mass measurements.  This constrains the
scatter to be $\lesssim 0.7$ at the 95\% confidence level.  While that analysis
assumed GR, the measurement of the scatter in the 
mass-richness relation only relies on the fact that the X-ray mass proxies trace
true mass with much smaller scatter than richness.  This is expected to hold even in the
modified gravity case, at least when the chameleon mechanism is
not active \cite{schmidt:10} as is the case for the values of $\absfR$ considered here.  

Finally, for the lensing calibration, which we use to scale $\ds$ (see~\textsection\ref{sec:observations}), we use a Gaussian distribution around 1.08 with 5\% standard deviation.

\section{Results}\label{sec:constraints}

\begin{table*}
\centering
\begin{tabular}{|c|cc|cc|cc|}
\hline
Parameter                     & \multicolumn{2}{|c|}{$\Lambda$CDM} & \multicolumn{2}{|c|}{AM}     & \multicolumn{2}{|c|}{PM}     \\
\hline
$\Om$                         & $0.266\pm0.011$ & $0.268$          & $0.251\pm0.013$ & $0.265$    & $0.261\pm0.011$ & $0.258$    \\
$\sigma_8^{\Lambda{\rm CDM}}$ & $0.795\pm0.016$ & $0.791$          & $0.769\pm0.022$ & $0.788$    & $0.785\pm0.017$ & $0.776$    \\
$n_{\rm s}$                   & $0.956\pm0.011$ & $0.951$          & $0.961\pm0.015$ & $0.952$    & $0.956\pm0.012$ & $0.952$    \\
$\sigma$                      & $0.46\pm0.10$   & $0.46$           & $0.53\pm0.13$   & $0.45$     & $0.47\pm0.10$   & $0.42$     \\
$10^{-3}\absfR$               & \multicolumn{2}{|c|}{\ldots}       & $<3.55$         & $0.00$     & \multicolumn{2}{|c|}{\ldots} \\
$F_0$                         & \multicolumn{2}{|c|}{\ldots}       & \multicolumn{2}{|c|}{\ldots} & $0.34\pm0.20$   & $0.34$     \\
$\mathcal{C}$                 & $1.083\pm0.048$ & $1.089$          & $1.114\pm0.052$ & $1.085$    & $1.092\pm0.049$ & $1.084$    \\
\hline
$-2\ln L$                     & \multicolumn{2}{|c|}{14.2}         & \multicolumn{2}{|c|}{14.2}   & \multicolumn{2}{|c|}{11.5}  \\
\hline
\end{tabular}
\caption{Mean, standard deviations, and best-fit values for the concordance model, $f(R)$ gravity in the abundance-matched case, and the phenomenological model, respectively. For $\absfR$
we quote 95\% 1D-marginalized confidence levels. $-2\ln L$ is calculated for the cluster-galaxy lensing data including the priors of~\textsection\ref{sec:priors}.}
\label{tab:results}
\end{table*}

We now move to the MCMC likelihood analysis of the cosmological parameter spaces
\begin{equation}
\mathcal{P}_{\rm AM} = \left\{ \Om, \sigma_8^{\Lambda\textrm{CDM}}, n_{\rm s}, \sigma, \mathcal{C}, \absfR \right\}
\end{equation}
and, in the case of the PM enhancement,
\begin{equation}
\mathcal{P}_{\rm PM} = \left\{ \Omega_{\rm m}, \sigma_8^{\Lambda\textrm{CDM}}, n_{\rm s}, \sigma, \mathcal{C}, F_0 \right\},
\end{equation}
where for the concordance model $\mathcal{P}_{\Lambda\textrm{CDM}} = \mathcal{P}_{\rm AM} \cap \left\{ \absfR = 0 \right\} = \mathcal{P}_{\rm PM} \cap \left\{ F_0 = 0 \right\}$. 
We implement the following flat priors on the parameters in $\mathcal{P}_{\rm AM}\backslash\mathcal{P}_{\Lambda\textrm{CDM}}$ and $\mathcal{P}_{\rm PM}\backslash\mathcal{P}_{\Lambda\textrm{CDM}}$: $\absfR\in (0,10)$
and
$F_0 \in (-5,5)$ for the AM and PM enhancement, respectively. In addition to the priors from the distance and CMB measurements discussed in~\textsection\ref{sec:priors}, we further employ flat priors on top of the priors on the parameters in $\mathcal{P}_{\Lambda\textrm{CDM}}$: $\Om \in (0.05,0.5)$, $\sigma_8^{\Lambda\textrm{CDM}} \in (0.4,1.6)$, $n_{\rm s} \in (0.5,1.5)$, $\sigma \in (0,2)$, and $\mathcal{C} \in (0.5,1.5)$.
Note that these bounds only serve as clear truncations for the parameter exploration in the MCMC code and since the ranges are chosen much wider than the bounds from the external priors in~\textsection\ref{sec:priors} and of $\mathcal{C}$ in~\textsection\ref{sec:lensmeas}, they do not affect the final parameter constraints.

The {\sc cosmomc}~\cite{cosmomc:02} package used for the MCMC likelihood analysis employs the Metropolis-Hastings algorithm~\cite{metropolis:53, hastings:70} for the sampling and the Gelman and Rubin statistic $\mathcal{G}$~\cite{gelman:92} for testing the convergence. We require $\mathcal{G}-1 < 7\times10^{-3}$ for our runs. We summarize our results in Table~\ref{tab:results}.

\subsection{$f(R)$ gravity}

\begin{figure*}
 \resizebox{\hsize}{!}{\includegraphics{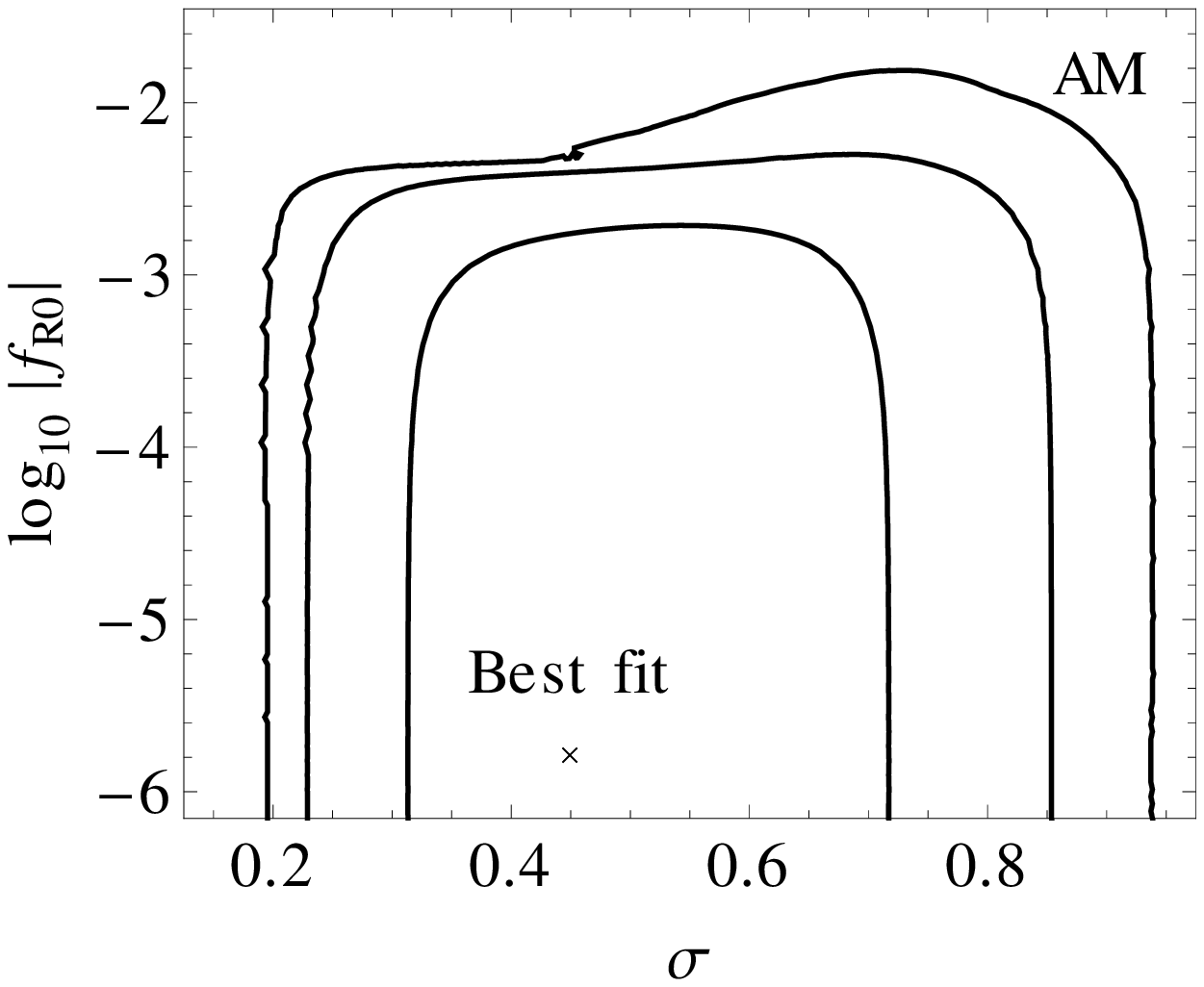}\includegraphics{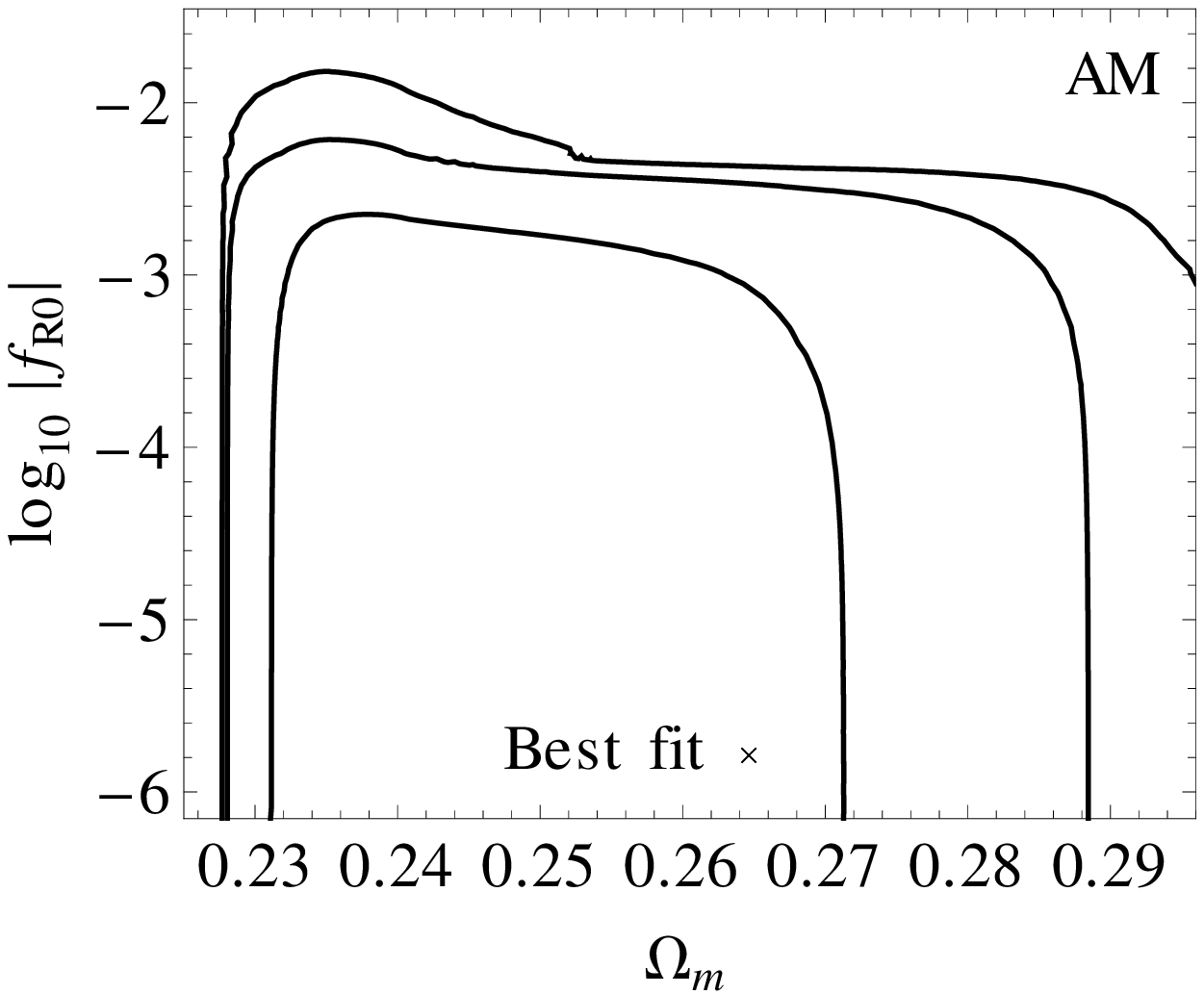}\includegraphics{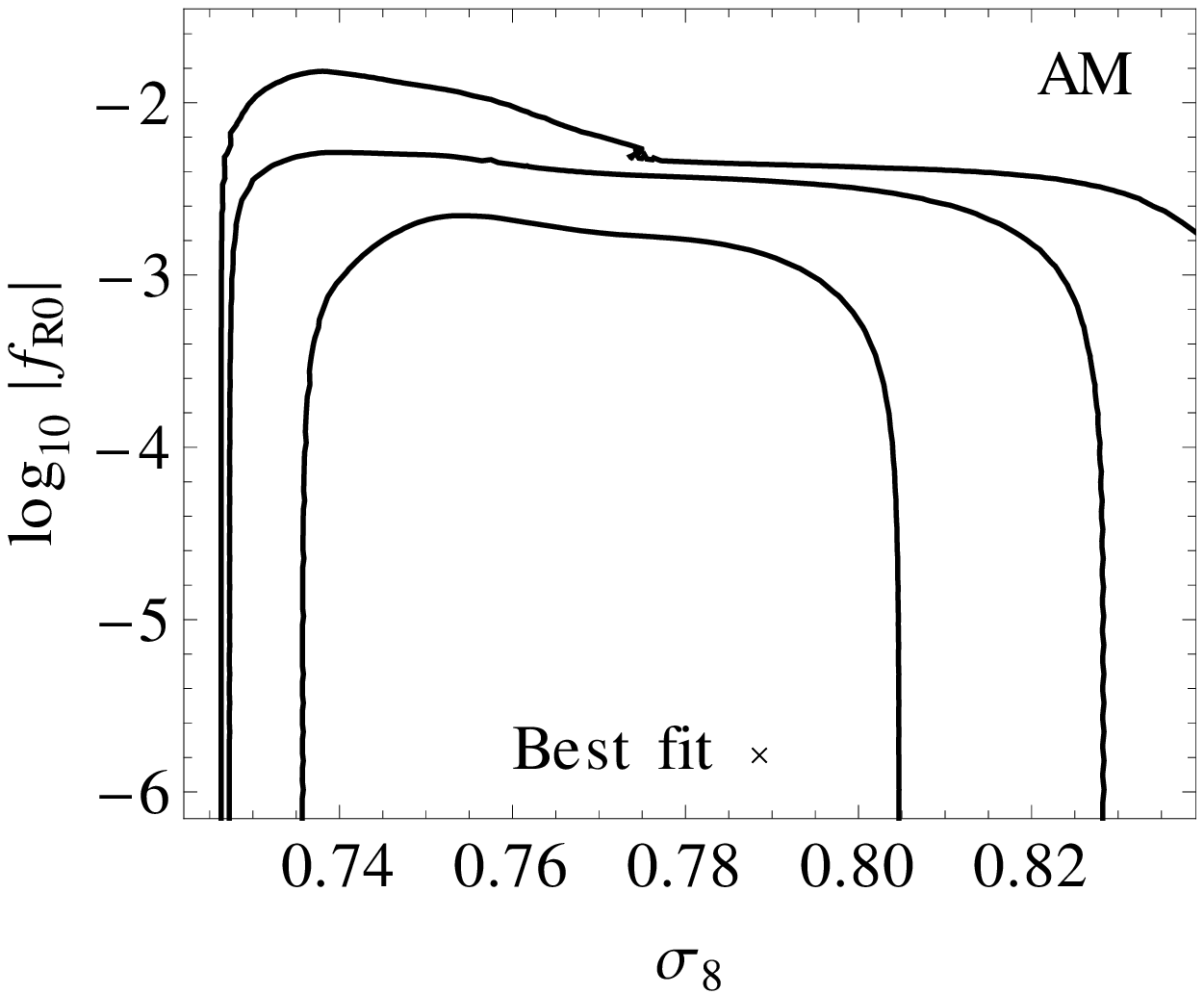}}
 \vspace{0mm} \\
 \resizebox{\hsize}{!}{\includegraphics{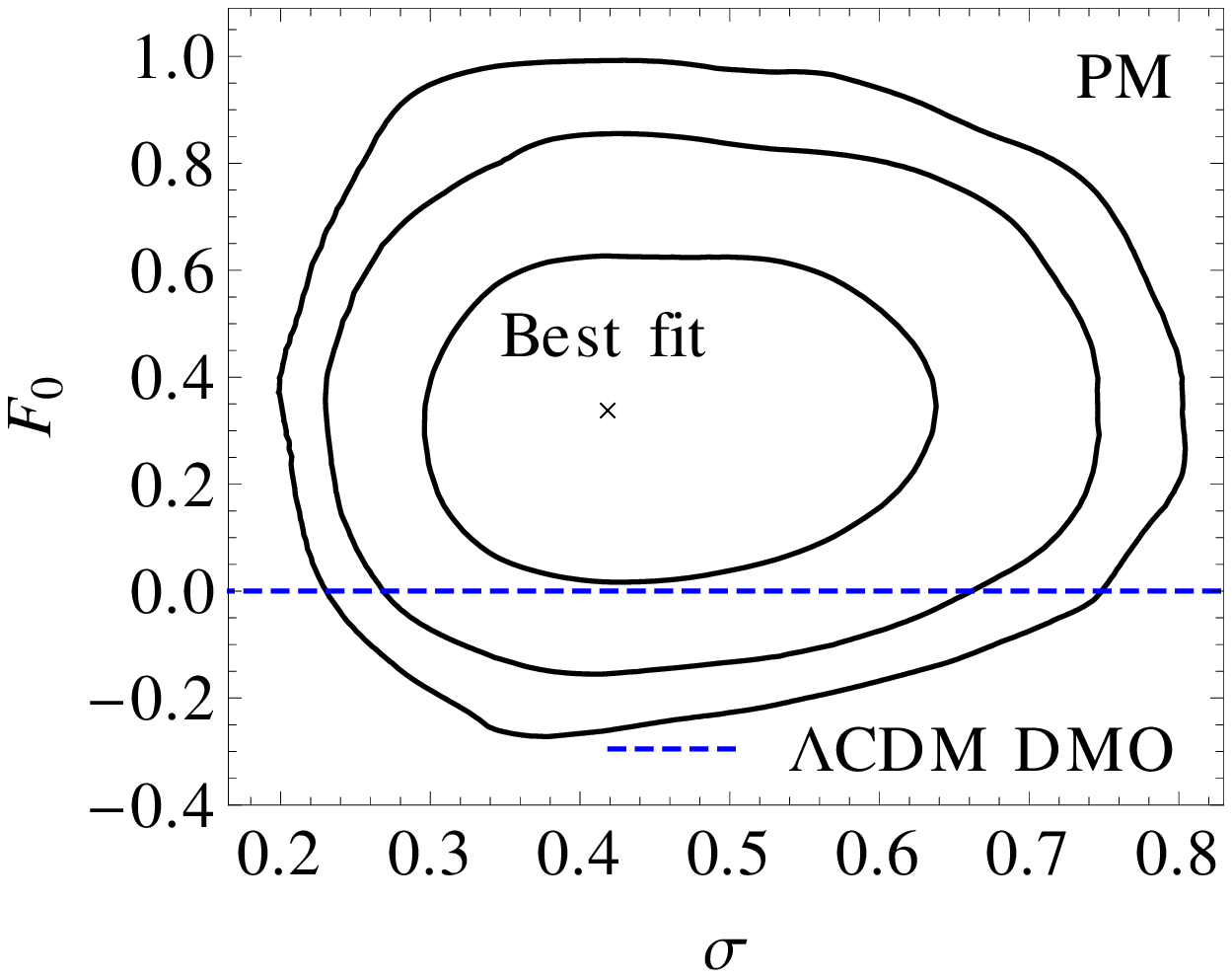}\includegraphics{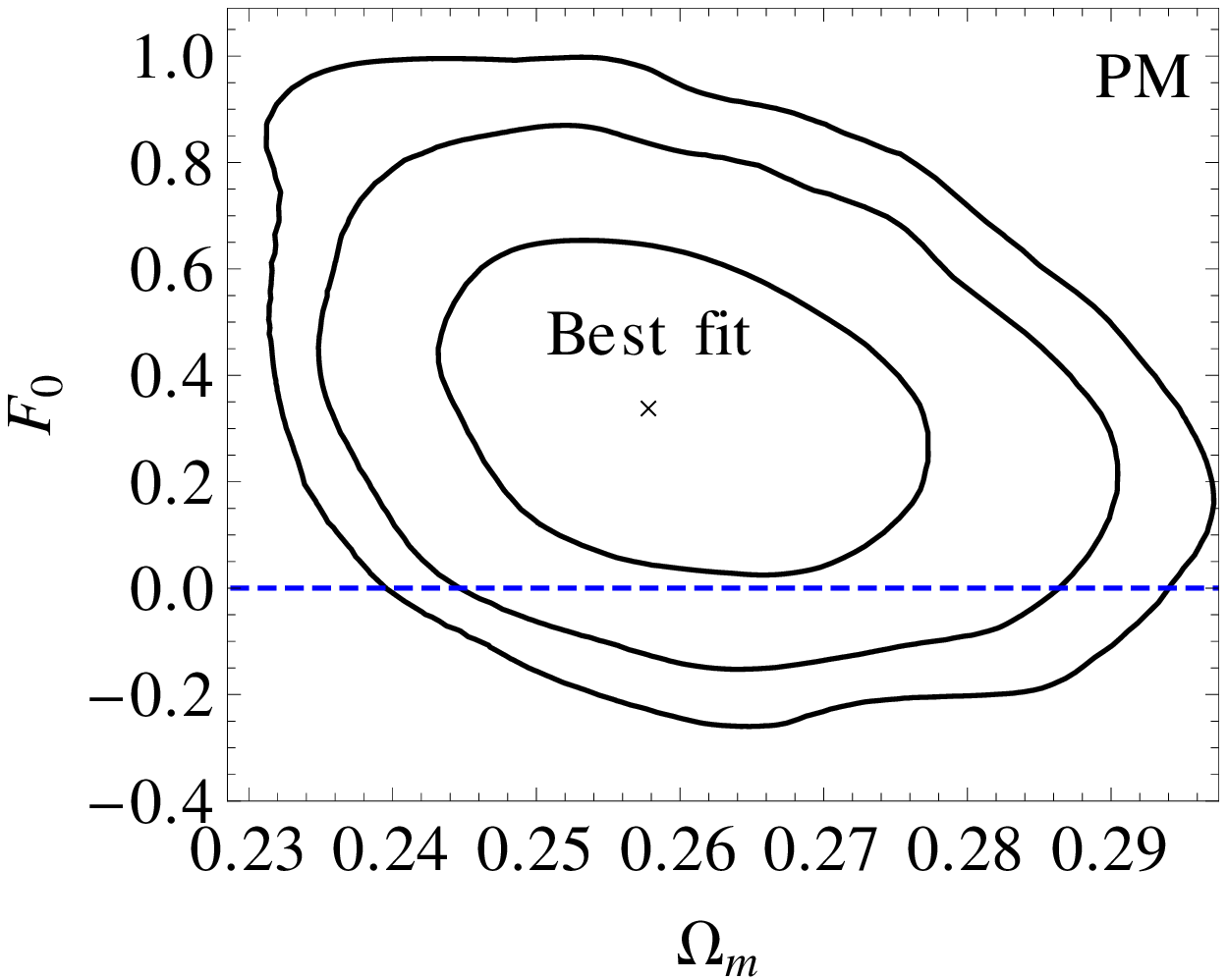}\includegraphics{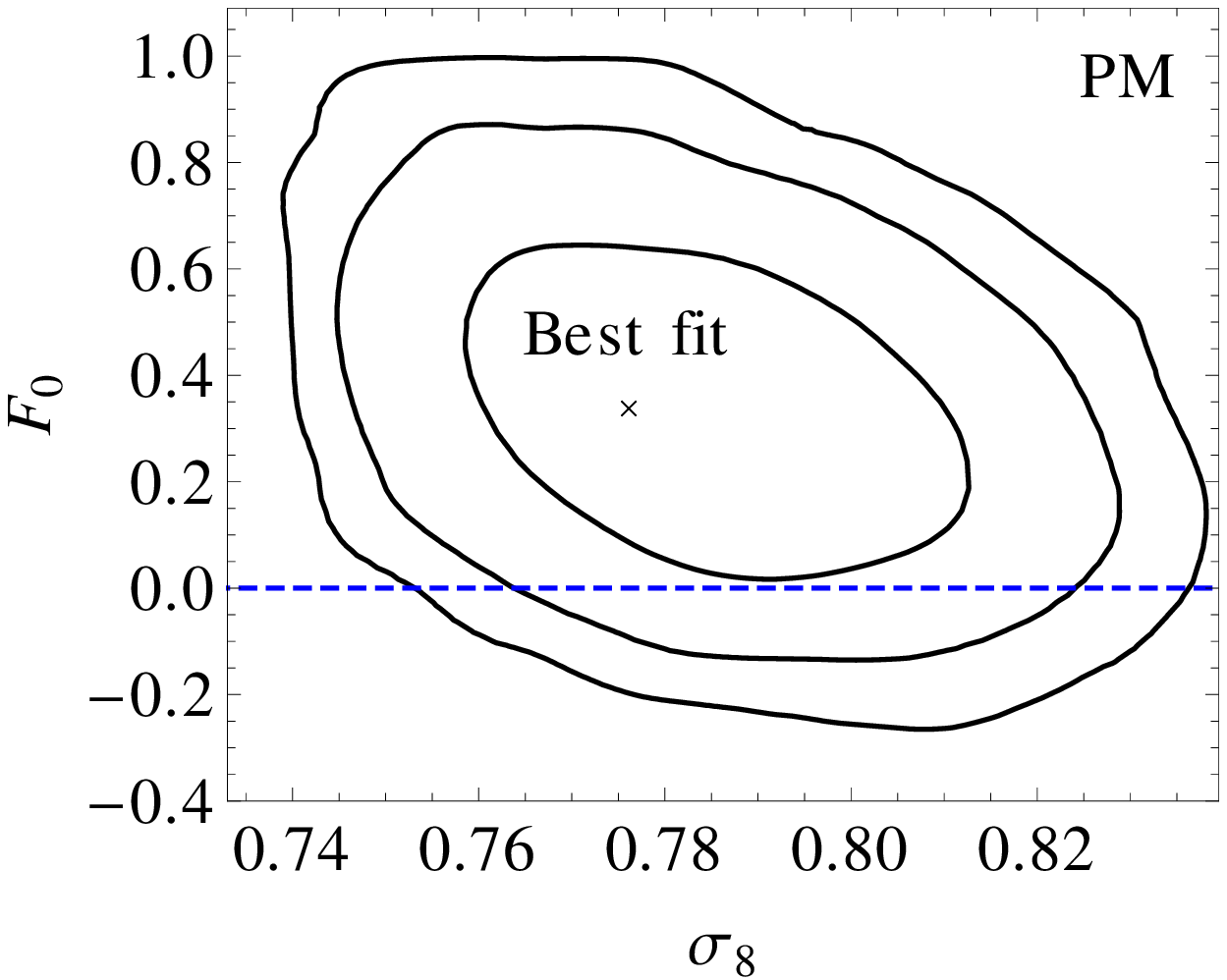}}
\caption{2D-marginalized contour plots for the abundance-matched (top row)
and the phenomenological enhancement case (bottom row), showing 68\%, 95\%, and 99\% confidence levels.
The dashed line corresponds to $\Lambda$CDM predictions from DMO simulations.
}
\label{fig:contours}
\end{figure*}

\begin{figure*}
 \resizebox{\hsize}{!}{\includegraphics{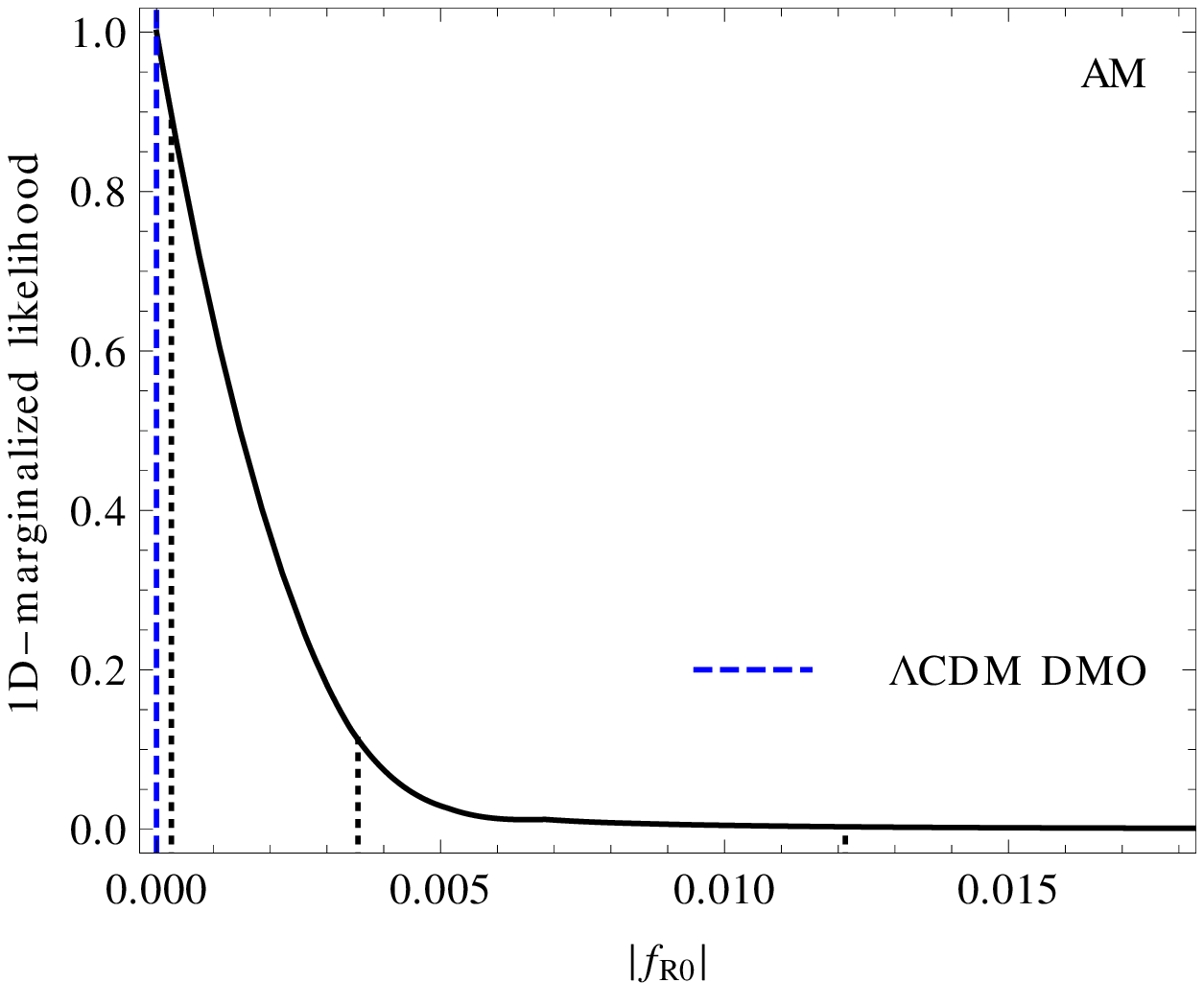}
 \includegraphics{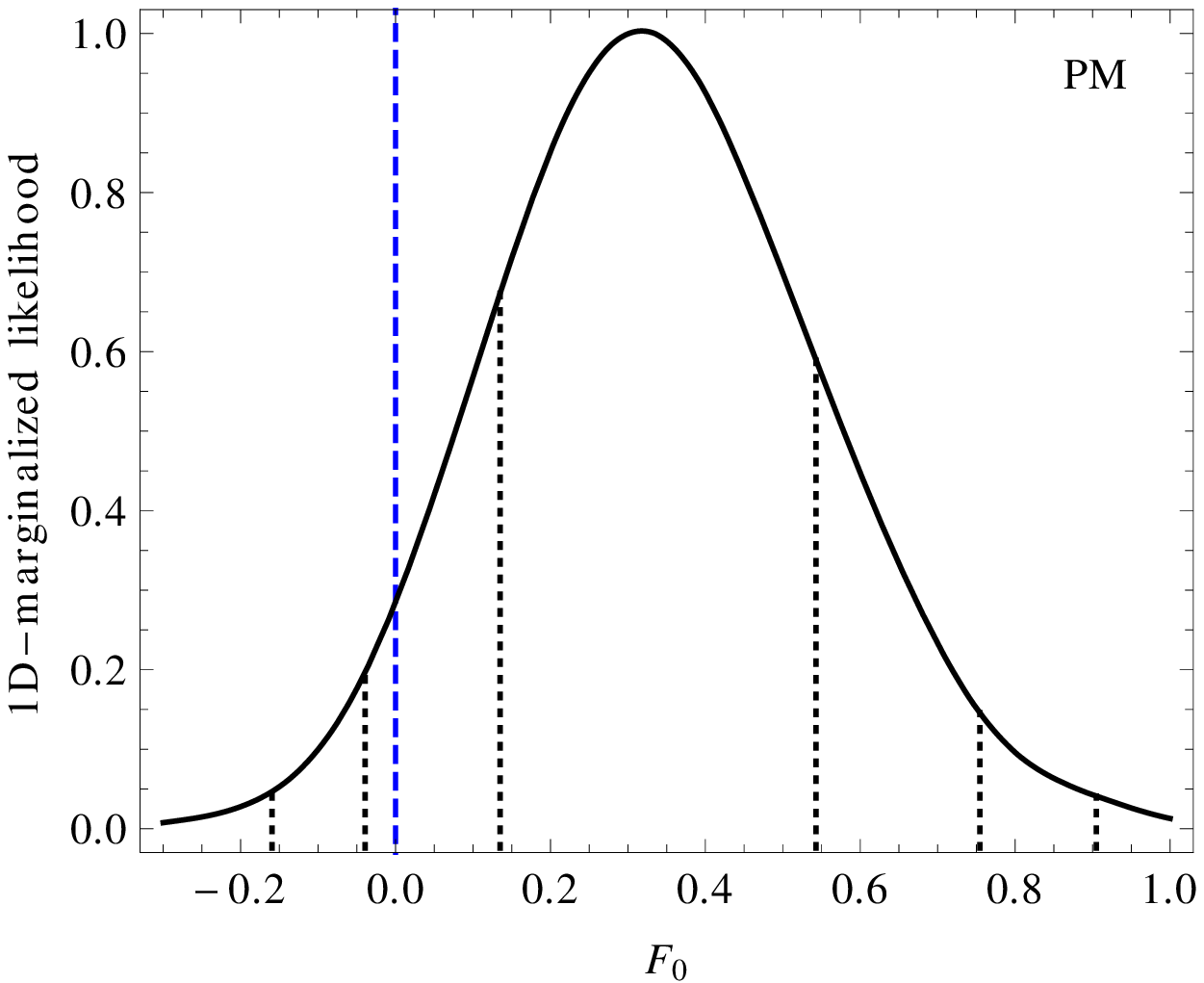}
}
\caption{
One- and two-tail 1D-marginalized likelihood. The dotted lines indicate the 68\%, 95\%, and 99\% confidence levels, the dashed line corresponds to the $\Lambda$CDM prediction from DMO simulations. \textit{Left}: $\absfR$ in the abundance-matched case.
\textit{Right}: $F_0$ in the phenomenological scenario with a Gaussian fit in $\ln r$ to the enhancement in the TM case.
}
\label{fig:constraints}
\end{figure*}

\begin{figure*}
 \resizebox{\hsize}{!}{\includegraphics{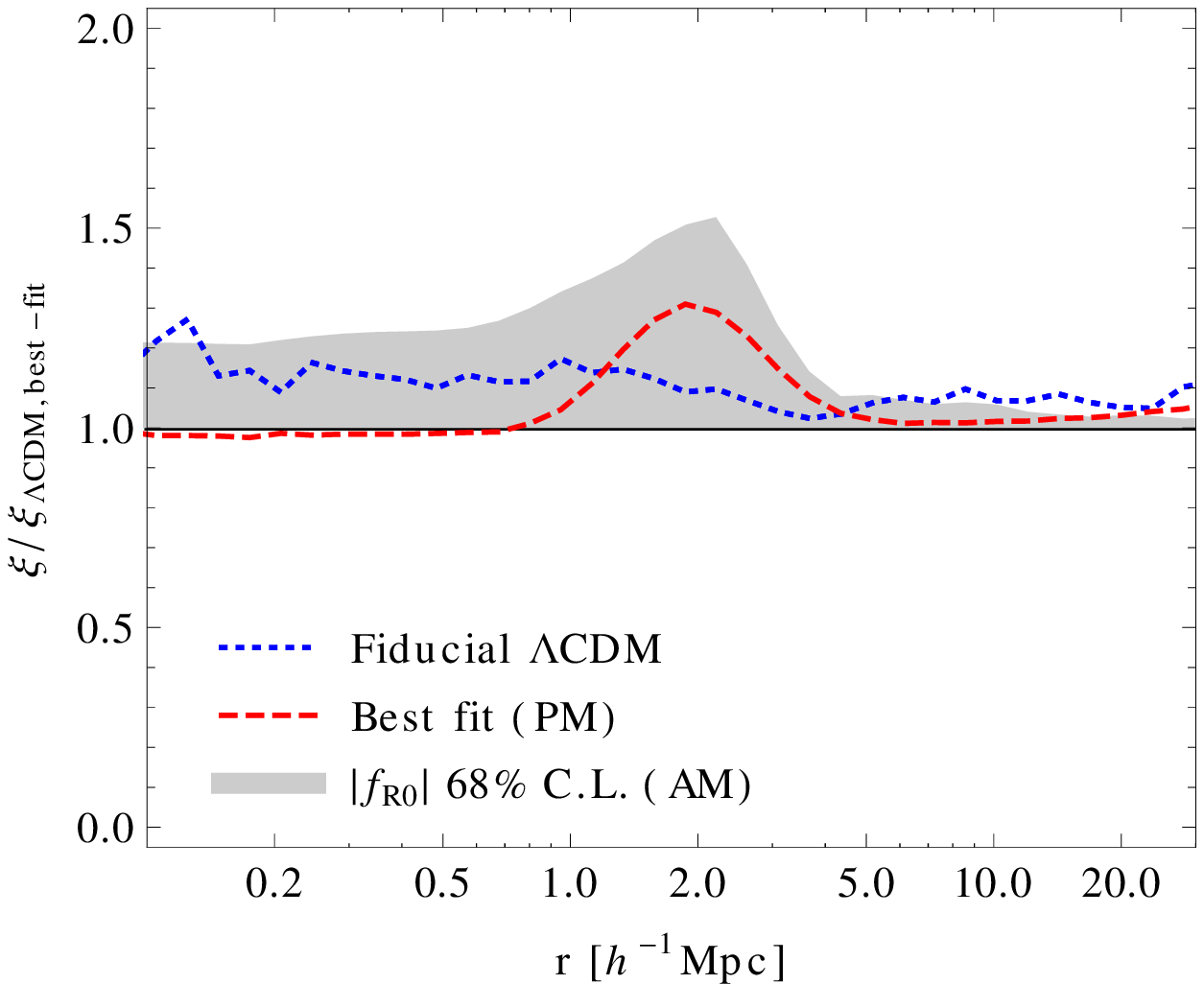}\includegraphics{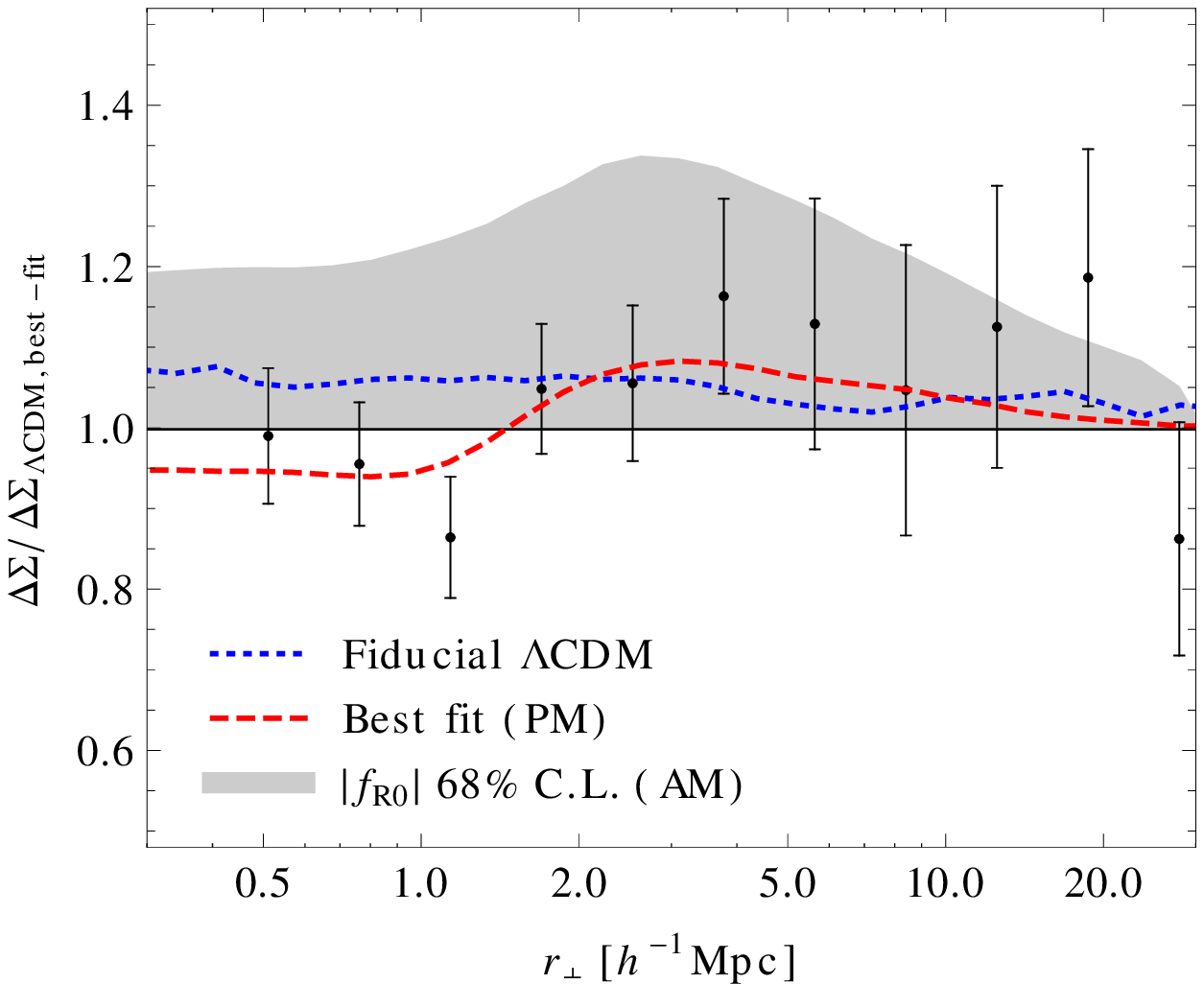}}
 \caption{
\textit{Left:} Best-fit prediction for $\xi_{\rm hm}$ with respect to the best-fit $\Lambda$CDM model prediction for the phenomenological scenario (dashed) and fiducial $\Lambda$CDM cosmology (dotted), respectively.
\textit{Right}: Best-fit prediction for the excess surface mass density $\Delta\Sigma$ in the phenomenological scenario (dashed) and the fiducial $\Lambda$CDM model (dotted) with respect to the best-fit concordance model.
Note that the lensing data has been rebinned for illustrative purposes (cf. Fig.~\ref{fig:DSigma_obs}).  
The shaded areas indicate regions in the abundance-matched case bounded by the best-fit model from below, and $\absfR=3.5\times 10^{-3}$ (corresponding to the 68\% CL bound) with otherwise identical parameter values from above.  
The best-fit model for the abundance-matched case is essentially identical to its $\Lambda$CDM counterpart and is therefore not shown separately.
}
\label{fig:xires}
\end{figure*}

Fig.~\ref{fig:DSigma} shows that the $f(R)$ predictions in the AM case for
$\absfR=10^{-3}$ are in clear tension with the data, at least when excluding scatter.  
Further, due to the strong radial dependence, the  $f(R)$ effects cannot
easily be canceled by varying some of the other cosmological parameters.
This leads to a 1D-marginalized constraint of $\absfR<3.5\times10^{-3}$ at the 95\% confidence level. Note that including a prior on scatter plays an essential role (see Fig.~\ref{fig:contours}), i.e., if we were to remove it from the analysis, very large scatter would make large $\absfR$ models viable (see Fig.~\ref{fig:xi} and Fig.~\ref{fig:DSigma}) and due to a slow increase of the enhancement $A$ as function of $\absfR$ (see Fig.~\ref{fig:amplitude}), there would be a rather loose constraint on $\absfR$. This is what happens
if one wishes to constrain the TM scenario instead.
In contrast to the AM case, when fixing the mass range equally in the $\Lambda$CDM and modified gravity model and therefore lowering the average halo mass within the stacked profiles, the discrepancy in the enhancement on $\Delta\Sigma$ on scales below $r_{\perp}\lesssim 1\hMpc$ and above $r_{\perp}\gtrsim 10\hMpc$ is less severe (see Fig.~\ref{fig:shape} and Fig.~\ref{fig:DSigma}).
Therefore, in that case, smaller values of scatter are already sufficient to make the profile enhancement compatible with the cluster-galaxy lensing data.
This leaves $\absfR$ unconstrained within the region of applicability
of our linearized $f(R)$ equations, $\absfR \lesssim 2\times 10^{-2}$.
Note, however, that the TM case does not consistently take into account the enhanced abundance of clusters of $f(R)$ gravity and we, therefore, restrict our $f(R)$ specific constraints to the AM case.

Fig.~\ref{fig:contours} shows the 2D-marginalized likelihoods for the parameter degeneracies with $\absfR$ and Fig.~\ref{fig:constraints} shows the one-tail 1D-marginalized likelihood for $\absfR$.
In Fig.~\ref{fig:xires}, we illustrate the band of $\xi_{\rm hm}$ and $\ds$ predictions bounded from below by the best-fit AM $f(R)$ gravity model, which is essentially identical to the best-fit concordance model ($\absfR=1.7\times10^{-6}$), and from above by the upper 68\% confidence level value of $\absfR$ with otherwise identical parameter values.

\subsection{Phenomenological scenario}

In the phenomenological case based on the TM halo profile enhancements,
we use a Gaussian function in $\ln r$ with width and position fixed to fit the simulation and only consider the amplitude of the Gaussian function $F_0$ as an additional free parameter (see~\textsection\ref{sec:xigauss}).  We obtain a
mean and standard deviation of $F_0=0.34\pm0.20$. 
Fig.~\ref{fig:contours} illustrates parameter correlations with $F_0$,
Fig.~\ref{fig:constraints} shows the
two-tail 1D-marginalized likelihood for the amplitude $F_0$,
and in Fig.~\ref{fig:xires} we present the best-fit predictions for $\xi_{\rm hm}$ and $\ds$.
The best-fit parameter values, as well as the corresponding $-2\ln L$ are listed in Table~\ref{tab:results}.

$F_0=0$ corresponds to the $\Lambda$CDM model with DMO simulations and is consistent at the 1D-marginalized
95\% confidence level (Fig.~\ref{fig:constraints}). The best fit is obtained for
$F_0=0.34$, achieving 
$-2\Delta \ln L = -2.7$ with respect to the best-fit concordance model.

The data thus slightly prefer an enhancement in halo profiles over $\Lambda$CDM in the phenomenological case.  Future surveys will thus either strengthen the constraints on modified gravity parameters, or even more interestingly, provide additional evidence for $F_0 > 0$.  
Note that for the best-fit $\Lambda$CDM model the reduced $\chi^2$ is roughly unity and that we therefore do not expect our error bars to be significantly underestimated.

\subsection{Systematic effects} \label{sec:systematics}

The shape of the enhancement effect, $g(r)$, on the cluster profile $\xi_{\rm hm}$ and the excess surface mass density $\Delta\Sigma$ observed in $f(R)$ gravity simulations cannot be reproduced by any reasonable deviations in the parameter values of the fiducial cosmology (see Figs.~\ref{fig:xi} and \ref{fig:DSigma}). Our comparison of theoretical predictions to the lensing observable is, however, affected by the following possible scale-dependent systematics.

\begin{itemize}

 \item \emph{Mass scatter}: In~\textsection\ref{sec:halomodelpred}, we include a log-normal scatter in the mass-richness relation in our theoretical modeling.  The actual form of the scatter might differ from log-normal, though it seems unlikely that this would result in an enhancement of $\Delta\Sigma$ localized at $r_{\perp}\simeq(1-10)\hMpc$.  

 \item \emph{Baryons}: In order to understand the formation of galaxies within clusters, it is essential to include the baryonic components. Realistic models comprise mechanisms such as gas cooling, star formation, supernovae feedback, as well as the feedback from supermassive black holes to avoid the overcooling and accumulation of gas in the core of the cluster, the so-called active galactic nucleus (AGN) feedback. AGN outbursts produce shock waves that move the gas from the core to larger radii, i.e., between $r_{\rm v}$ and $2r_{\rm v}$, as was shown in~\cite{teyssier:10} by employing simulations of Virgo-like galaxy clusters. Moreover, due to the AGN feedback, there is a slight adiabatic expansion of the dark matter when compared to DMO simulations (see Fig.~9 of~\cite{teyssier:10}), leading
to a $\lesssim 10\%$ effect on the density profiles.  
These effects have a radial dependence that is qualitatively different from the
modified gravity enhancements considered here.  
Note that the moved mass by modified gravity can be calculated as
\begin{equation}
\Delta M=4\pi \bar{\rho}\int_{0}^\infty \xi_\text{hm}(r)\arr(r) r^2 dr
\end{equation}
and amounts to $\Delta M\approx6 \times 10^{12} \hMs$ for the phenomenological fit with 
$F_0=0.3$.

 \item \emph{Intrinsic
 alignment}: High-precision weak-lensing measurements may be contaminated by the intrinsic alignment of galaxies (see, e.g.,~\cite{blazek:11a}). The correlation of intrinsic alignment and gravitational shear distortion can contribute to the observed ellipticity correlation function and $\Delta\Sigma$ at the $\lesssim 10\%$ level~\cite{hirata:07,blazek:11b}.  

 \item \emph{Miscentering and satellites}: The cluster centers in the MaxBCG 
sample are identified by the brightest cluster galaxy (BCG). The true cluster 
center may, however, be offset from the BCG position 
(see, e.g., discussion in~\cite{mandelbaum:09}). This effect causes a 
suppression of the lensing signal in the inner parts of the halo, which 
subsequently leads to an underestimation of the cluster mass and the concentration. A miscentered $\Delta \Sigma$ can have a bump relative to a correctly centered $\Delta \Sigma$, which is, however, located further inwards than the $f(R)$ gravity enhancement (cf.~\cite{hilbert:09}). A similar enhancement around the virial radius can further be introduced by galaxy satellites. To prevent the contamination of the excess surface mass density through satellites, we apply a cylindrical cut in the projected radius at $r_{\rm cut} = 3 r_{\rm v}$ in the simulations (see~\textsection\ref{sec:LCDMsimulations}) and the observations (see~\textsection\ref{sec:observations}).
Note that we applied this cut only to the {\sc zhorizon} simulations and not to the $f(R)$ gravity simulations. We verified, however, that changing $r_{\rm cut}$ has a negligible impact on the dependence of $\ds$ on cosmological parameters. 
Furthermore, the cut only affects $\ds$ on scales $r_{\perp}\gtrsim5\hMpc$.  
Therefore, we can safely assume that there is no significant impact on the relative $f(R)$ enhancement by the cylindrical cut. 

 \item \emph{Wrong cosmology}: The analysis of lensing as used in this study 
requires the assumption of an \emph{a priori} cosmological model to estimate the critical surface mass density
$\Sigma_\text{crit}$ and to convert angles to distances. Within $\Lambda$CDM, a wrong prior on the cosmological model produces a radial horizontal shift of $\Delta\Sigma$ at the $\lesssim 2\%$ level for $\Omega_m=0.25\pm 0.05$ (see discussion in~\cite{baldauf:09}).  
Note that the Hu-Sawicki $f(R)$ gravity model matches the $\Lambda$CDM
background to order $\absfR$.  Deviations of this magnitude have
a negligible impact on $\Delta\Sigma$.  

 \item \emph{Simulation systematics}: 
In order to test the convergence of
the halo profiles of the large-scale cosmological simulations on the scales
used in this study, we compared the halo profiles from the {\sc zhorizon} simulations to the halo profiles of the {\sc millennium} simulations~\cite{millenium:05}, which employ $N=2\,160^3$ particles in a $500^3\hMpcc$ box. The profiles agree at the $\lesssim 5\%$ level on the scales of interest. We therefore conclude that the {\sc zhorizon} simulations have converged for $r\sim(0.2-100)\hMpc$.  
The halos in the {\sc zhorizon} simulations are identified using an FoF halo finder,
while the $f(R)$ effects were measured on a SO-identified halo
sample.
However, since we only use the 
enhancements from the $f(R)$ simulations relative to $\absfR=0$, we expect the difference to be smaller than the residual statistical error ($\sim 20$\%) on the 
modified gravity effects.  Note that the environmental effects found in 
\cite{schmidt:10,ZhaoEtal:11} are induced by the chameleon mechanism and are 
not relevant for the values of $\absfR$ considered here.
Moreover, halo finders typically agree at the scales relevant to our halo profile measurements~\cite{knebe:11}. 

\item \emph{Survey geometry}: While we are mimicking the selection process as 
closely as possible, including the removal of fake clusters, the simulation 
measurements provide the dark matter and halo positions in a cubic box. 
Furthermore the simulation results are obtained from a single redshift slice at the effective redshift of the sample, 
and are thus not accounting for the redshift evolution of the lens sample.

\end{itemize}

Except for the case of the scatter, we neither model the systematics described above nor include them as additional errors to the measurement when performing the likelihood analysis. In order to consistently include these systematics, they should not only be carefully analyzed within $\Lambda$CDM but also in the context of $f(R)$ gravity, which is beyond the scope of this paper. Note that, when added in quadrature, the described uncertainties sum up to a $\lesssim 15\%$ and $\lesssim 25\%$ error in the predicted $\ds$ for $\Lambda$CDM and the modified gravity  cases, respectively. This work is based on the assumption that the above systematics, except for the mass scatter, can be neglected and that the observations can correctly be described by an average over the DMO simulations.  Note that our
$\Lambda$CDM model indeed provides a good fit to the data (see Fig.~\ref{fig:xires}).

\section{Conclusion}\label{sec:discussion}

\begin{figure}
 \resizebox{\hsize}{!}{\includegraphics{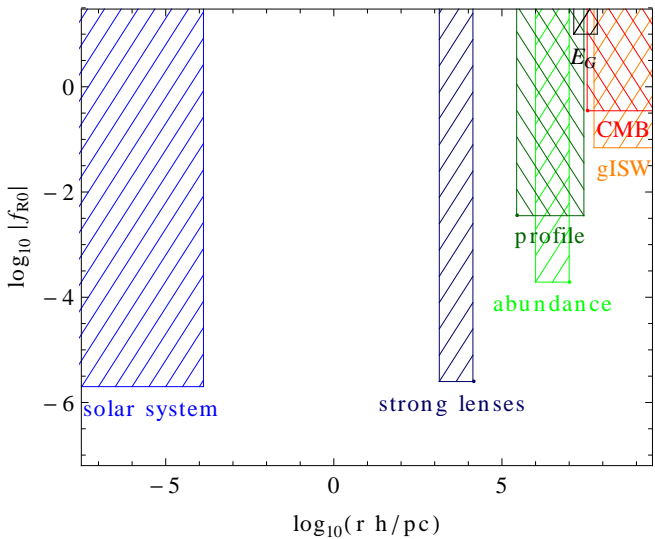}}
 \caption{Current constraints on $f(R)$ gravity. On linear scales, the strongest bound on $\absfR$ is obtained through the comparison of predicted to observed cross correlations of the ISW with foreground galaxies. In the nonlinear regime, enhancements of the abundance of clusters and the cluster density profile due to the $f(R)$ modification are incompatible with observations unless $\absfR$ is smaller than $10^{-4}$ and $10^{-3}$, respectively.
The currently strongest bounds on $\absfR$, however, are inferred from requiring the modification to be suppressed by the chameleon mechanism within the solar system and the dark matter halo as well as from strong gravitational lenses.
}
\label{fig:constsumm}
\end{figure}

Modifications of GR as in the $f(R)$ gravity model under consideration in this paper generically predict departures from the standard growth produced in the concordance model. On the largest, cosmological scales ($r\gtrsim10~{\rm Mpc}$) and on small, solar-system scales ($r\lesssim 20$~AU) such deviations have extensively been instrumentalized to probe gravity.
However, structures on intermediate scales also offer opportunities to test the gravitational interactions.

In this paper, we test modifications of gravity on scales around the virial radius of a cluster, i.e., $r\simeq(0.2-20)~{\rm Mpc}$. 
The modification of the Poisson equation leads to a difference in the accretion of mass onto massive dark matter halos.  
The resulting halos exhibit enhanced density profiles at a few virial radii that offer a unique opportunity for testing gravity. We use the projected mass distribution measured through cluster-galaxy lensing around maxBCG clusters from the SDSS to put constraints on the modifications induced by the Hu-Sawicki $f(R)$ gravity model.
For consistent theoretical predictions we rely on $f(R)$ gravity and concordance model $N$-body DMO simulations.
Matching simulated to observed halos by abundance, we obtain a one-tail upper bound of $\absfR<3.5\times10^{-3}$ at the 1D-marginalized 95\% confidence level.  
This places a new independent constraint on $f(R)$ gravity at intermediate scales, where $\absfR\lesssim \rm{few}\:10^{-4}$ and $\absfR\lesssim(10^{-6}-10^{-5})$ are current bounds inferred from large cosmological and solar-system scales, respectively.
We summarize current constraints on $\absfR$ in Fig.~\ref{fig:constsumm}, showing bounds inferred from measurements in the solar system~\cite{hu:07a}, of strong lenses~\cite{smith:09t}, the abundance of clusters~\cite{schmidt:09,lombriser:10}, galaxy-ISW (gISW) cross correlations~\cite{giannantonio:09,lombriser:10}, and the CMB~\cite{song:07,lombriser:10}, as well as our constraint from halo density profiles measured via weak gravitational lensing. We extrapolate results presented in~\cite{lombriser:10} to estimate an upper bound on $\absfR$ from the $E_G$ measurement of~\cite{reyes:10}, which combines weak lensing measurements around galaxies with their large-scale velocities. Note that Fig.~\ref{fig:constsumm} does not include the measurement of gravitational redshifts of galaxies in clusters at around $(1-6)\hMpc$ of~\cite{wojtak:11} since it was found to be consistent with $f(R)$ gravity and cannot be illustrated in the same manner as the previous measurements.

In order to assess the ability of halo profiles to
provide constraints independently of halo abundances,
we also considered a phenomenological parametrization of the modified gravity
effects on halo profiles at fixed mass.
In this scenario, the concordance model (with amplitude of the modification $F_0=0$) is consistent with the lensing measurement at the
95\% 1D-marginalized confidence level.
Thereby, we considered a Gaussian enhancement of the cluster density profile due to modified gravity located at a few virial radii.
The best-fit value of $F_0 = 0.34\pm 0.20$ indicates that the data slightly prefer an enhancement in halo profiles over $\Lambda$CDM;  future surveys will thus either strengthen the constraints on modified gravity parameters, or even more interestingly, provide additional evidence for $F_0 > 0$.

\section*{Acknowledgments}

We thank Jonathan Blazek, Michael Busha, Vincent Desjacques, Bhuvnesh Jain, Doug Potter, Darren Reed, Ravi Sheth, An\v{z}e Slosar, and Romain Teyssier for useful discussions.
We are also grateful to the anonymous referee for helpful suggestions and comments.
LL thanks the Lawrence Berkeley National Laboratory, the Berkeley Center for Cosmological Physics, and Ewha Womans University for hospitality while parts of this work have been carried out. Computational resources were provided on the Schr\"odinger supercomputer at the University of Zurich and on the supercomputer at the Institute for the Early Universe at Ewha University. This work was supported by the Swiss National Foundation under Contract No. 2000 124835/1 and WCU Grant No. R32-2008-000-10130-0.  FS is supported by the Gordon and Betty Moore Foundation at Caltech.

\appendix

\section{Halo model predictions for the density profiles} \label{app:HM}

\begin{figure*}
\resizebox{\hsize}{!}{\includegraphics{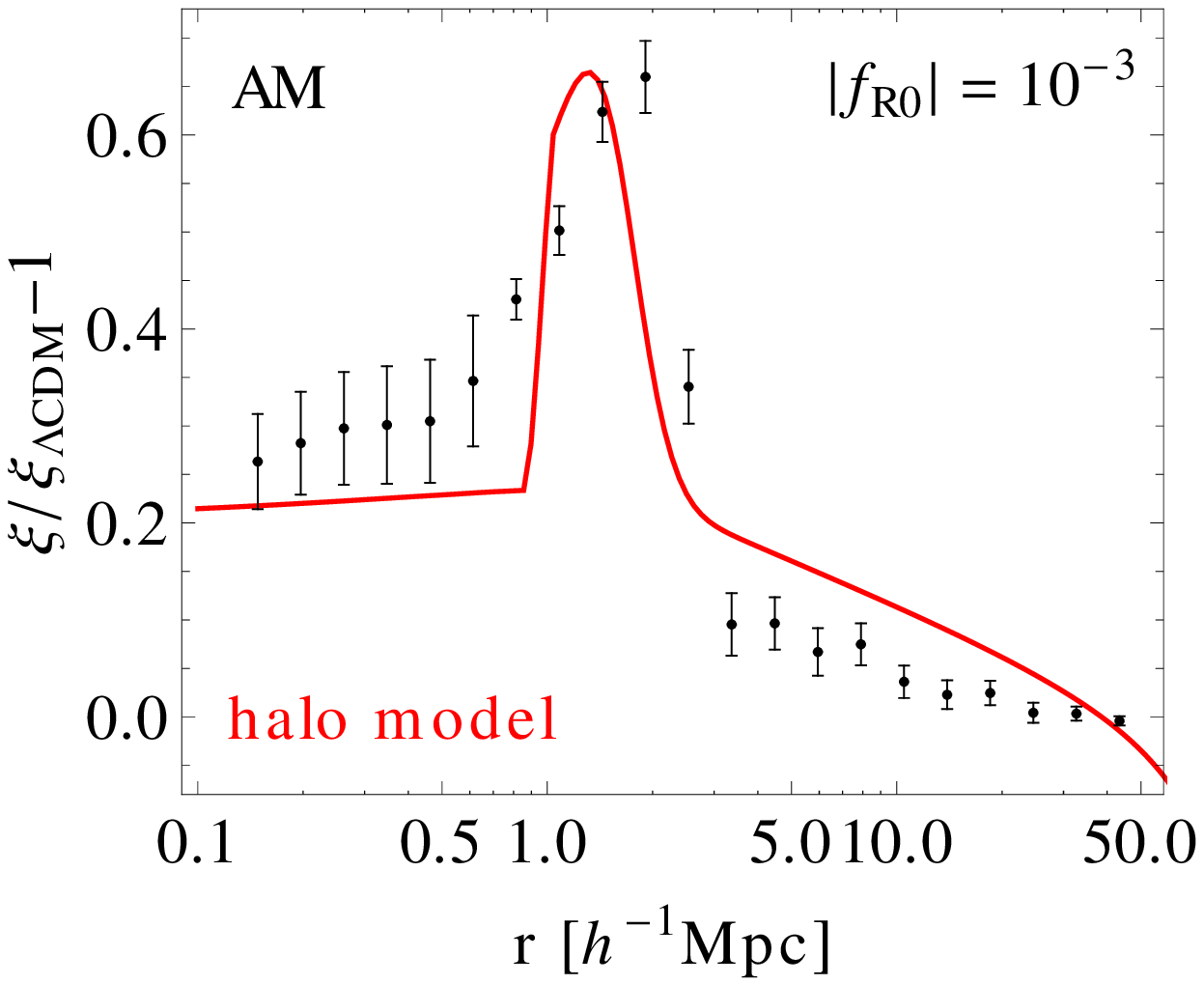}\includegraphics{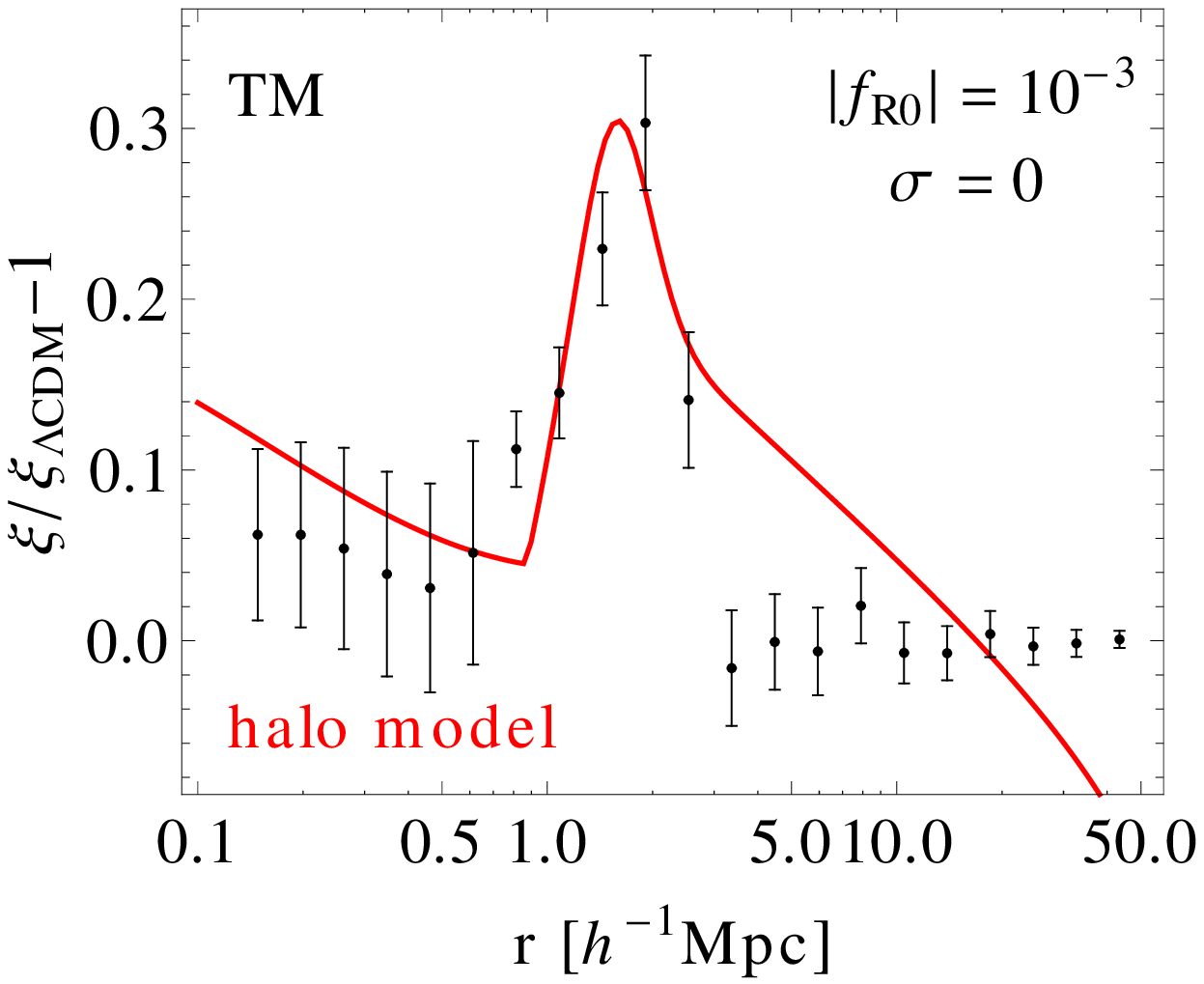}\includegraphics{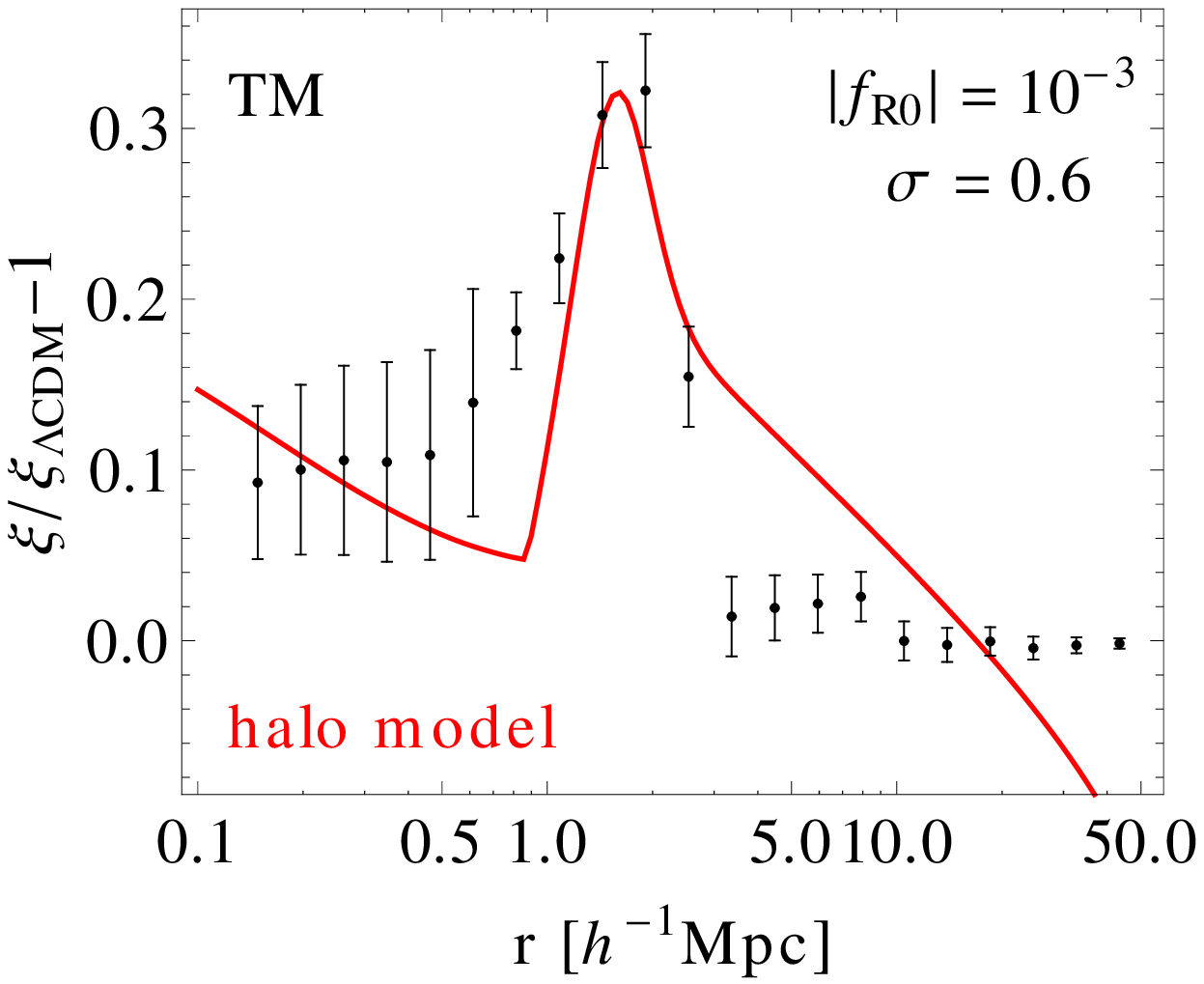}}
\caption{
\textit{Left:}  Halo model prediction scaled by the overall factor $\alpha=0.52$ 
(Sec.~\ref{sec:halomodelpred}) in comparison with simulation measurements for the 
abundance-matched case and $\absfR=10^{-3}$.  
\textit{Middle:} Same as left panel, but for the threshold-matched case without
scatter (halo model scaled by $\alpha=0.73$).
\textit{Right:}  Same as middle panel, but with a scatter of $\sigma=0.6$
(halo model scaled by $\alpha=0.77$).
}
\label{fig:halomodel}
\end{figure*}

In this appendix, we describe the halo model prediction for $\arr(r)$, the 
enhancement in $\xi_{\rm hm}(r)$ induced by $f(R)$ gravity, Eq.~(\ref{eq:xiratio}).  In the halo model,
the halo-mass cross-correlation function is given by a sum of two terms,
\begin{equation}
\xi_{\rm hm}(r) = \xi_{\rm hm}^{1h}(r) + \xi_{\rm hm}^{2h}(r),
\end{equation}
denoting the 1-halo and 2-halo contributions, respectively.  Throughout,
all quantities are evaluated at the
redshift of the $f(R)$ simulation output, $z=0.22$.  For the TM case,
we consider halos with mass $M_\Delta > M_0$, where $\Delta = 300$
is the overdensity in units of the background matter density today,
and
$M_0=10^{13.91}h^{-1}~M_\odot$
is a fixed threshold mass determined by
matching the halo abundance in simulations to the observed abundance, i.e.,
the same for $f(R)$ and GR.    
In the AM case, $M_0$ is determined separately for $f(R)$ and GR through
\be
\int_{\ln M_{\rm v,0}}^\infty n_{\rm v} d\ln M_{\rm v} = \bar n,
\ee
where $n_{\rm v}$ is the mass function of dark matter halos per logarithmic
interval in the virial mass $M_{\rm v} = M_{\Delta_{\rm v}}$.  
We adopt a fixed virial overdensity of $\Delta_{\rm v} = 390$.  
The virial mass threshold is then converted to the threshold $M_0$ 
for $\Delta = 300$ through the rescaling described in \cite{HuKravtsov}.  
For the virial mass function, we adopt the Sheth-Tormen
prescription \cite{sheth:99},
\begin{equation}
 n_{\rm v} \equiv \frac{dn}{d \ln M_{\rm v}} = \frac{\bar{\rho}_{\rm m}}{M_{\rm v}} f(\nu) \frac{d\nu}{d \ln M_{\rm v}},
\end{equation}
where $\nu = \delta_c/ \sigma(M)$, $\sigma(M)$ being the variance of the
density field for a top-hat enclosing mass $M$ at the background density, and
\begin{equation}
 \nu \, f(\nu) = \mathcal{A} \sqrt{\frac{2}{\pi} a \, \nu^2} \left[ 1 + (a \, \nu^2)^{-p} \right] e^{-a \, \nu^2/2}
\end{equation}
with $a = 0.75$, $p = 0.3$, and $\delta_{c} = 1.673$.  $\mathcal{A}$ is fixed so that 
$\int d\nu f(\nu) = 1$.  

The 2-halo contribution to $\xi_{\rm hm}(r)$ 
is most easily written in terms of its Fourier-space counterpart
$P^{2h}_{\rm hm}(k)$, defined through
\be
\xi^{2h}_{\rm hm}(r) = \int \frac{d^3k}{(2\pi)^3} P^{2h}_{\rm hm}(k) e^{i \bm{k\cdot r}}.
\ee 
The two-halo halo-mass power spectrum is given by
\be
P_{\rm hm}^{2h} = b(> M_0) I(k) P_m(k),
\ee
where $P_m$ is the linear matter power spectrum,
\begin{align}
b(> M_0) =\:& \frac{\int_{\ln M_{\rm v,0}}^\infty b(M_{\rm v}) n_{\rm v}(M_{\rm v}) d\ln M_{\rm v}}
{\int_{\ln M_{\rm v,0}}^\infty n_{\rm v}(M_{\rm v}) d\ln M_{\rm v}}\\
I(k) =\:& 
 \int_0^\infty n_{\rm v} \frac{M_{\rm v}}{\bar{\rho}_{\rm m}} y(k,M_{\rm v}) b(M_{\rm v}) d\ln M_{\rm v}\,,
\end{align}
and $b(M_{\rm v})$ is the scale-independent linear peak-background split bias
derived from the Sheth-Tormen mass function:
\begin{eqnarray}
b(M_{\rm v}) & \equiv & b(k=0,M_{\rm v}) \nonumber\\
&=&  1 + {a \nu^2 -1 \over \delta_c}
         + { 2 p \over \delta_c [ 1 + (a \nu^2)^p]}\,.
\label{eqn:bias}
\end{eqnarray}
$y(k, M)$ is the Fourier transform of a Navarro-Frenk-White (NFW)~\cite{NFW} density profile
which is truncated at the virial radius $r_{\rm v}$.  $y$ is normalized so that
$y(k=0, M) = 1.$  We adopt the mass-concentration
relation of \cite{BullockEtal}.  

The one-halo contribution is simply the normalized stacked NFW profile,
\begin{align}
\xi_{\rm hm}^{1h}(r) =\:& {\cal N}^{-1} \int_{\ln M_{\rm v}}^\infty 
\rho_{\rm NFW}(r; M_{\rm v})\: n_{\rm v}\: d\ln M_{\rm v}\\
\cal{N} =\:& \int_{\ln M_{\rm v}}^\infty n_{\rm v} d\ln M_{\rm v}.
\end{align}

The halo model prediction for $\xi_{\rm hm}(r)$ in $f(R)$ gravity 
is then obtained by substituting the linear $f(R)$ matter power spectrum
into the above expressions.  We do not change the concentration relation,
motivated by the fact that the inner profiles of halos seem relatively little
affected by $f(R)$ \cite{schmidt:08, lombriser:12}.  
The result is shown in Fig.~\ref{fig:halomodel}, scaled by the factor $\alpha$
introduced in~\textsection\ref{sec:halomodelpred}, 
together with the simulation results.  
The halo model prediction produces a bump at a few virial radii, because
halos are on average more massive in $f(R)$ gravity, leading to slightly larger
virial radii.  This becomes noticeable because the stacked truncated profile
becomes very steep outside the virial radius corresponding to $M_{\rm v,0}$.  
Clearly, there are discrepancies between the halo model predictions
and the simulation results at both small and large $r$.  Hence, we only
rely on the halo model prediction for the overall amplitude, whose
scaling as function of $\absfR$ is predicted well (Fig.~\ref{fig:amplitude}),
whereas the radial dependence is taken from the simulation measurements.  

As an aside, in \cite{schmidt:08}, modified
spherical collapse parameters were derived for a collapse with enhanced
forces throughout (i.e., the limiting case of infinite reach of the fifth
force).  This set of parameters can be used to estimate the
spread in the halo model predictions induced by modified gravitational
forces.  We found that the halo profile predictions for both sets of 
spherical collapse parameters are very similar, with the unmodified parameters
yielding a somewhat better approximation to the simulation results.

\section{Extrapolation of the halo model predictions} \label{sec:extrapolation}

In order for the MCMC runs to converge, the chains need to cover a large parameter space. At the most extreme values of the cosmological parameters, the halo model approach (see~\textsection\ref{sec:fR_halomodel}) breaks down and we need to rely on a more \emph{ad hoc} extrapolation for $A\left(\absfR,\sgL\right)$.
We design it to fit the simulations and halo model predictions within the AM scenario in the range $\absfR \leq 2 \times 10^{-2}$ and $\sgL \in [0.7,0.9]$ using the functional form
\begin{equation}
 A = a_0(\sigma_8) + a_1(\sigma_8) \, x + a_2(\sigma_8) \, e^x,
 \label{eq:Aext}
\end{equation}
where $x=\log_{10}\absfR$.
The approximation, Eq.~(\ref{eq:Aext}), is accurate at the $\lesssim 0.1\%$ level within the range of simulated values of $\absfR$, i.e., the regime of correspondence to the $f(R)$ gravity model. For the coefficients, we use the fit
\begin{eqnarray}
 a_i(\sigma_8) & = & a_{i0} + a_{i1} \, \sigma_8 + a_{i2} \, \sigma_8^2,
 \label{eq:Aextcoeff}
\end{eqnarray}
where $i=0,1,2$. Note that we used $\sigma_8=\sgL$ in Eqs.~(\ref{eq:Aext}) and (\ref{eq:Aextcoeff}) to simplify notation.

Furthermore, note that the exact form of this extrapolation does not affect our constraints, which are well within the halo model inter-/extrapolation.

\vfill
\bibliographystyle{arxiv_physrev}
\bibliography{fRlens}

\def\eprinttmppp@#1arXiv:@{#1}
\providecommand{\arxivlink[1]}{\href{http://arxiv.org/abs/#1}{arXiv:#1}}
\def\eprinttmp@#1arXiv:#2 [#3]#4@{\ifthenelse{\equal{#3}{x}}{\ifthenelse{
\equal{#1}{}}{\arxivlink{\eprinttmppp@#2@}}{\arxivlink{#1}}}{\arxivlink{#2}
  [#3]}}
\providecommand{\eprintlink}[1]{\eprinttmp@#1arXiv: [x]@}
\renewcommand{\eprint}[1]{\eprintlink{#1}}
\providecommand{\eprintmod}[1][XXXX.XXXX]{\eprintlink{#1}}
\providecommand{\adsurl}[1]{\href{#1}{ADS}}
\renewcommand{\bibinfo}[2]{\ifthenelse{\equal{#1}{isbn}}{\href{http://cosmolog%
ist.info/ISBN/#2}{#2}}{#2}}
\begin{thebibliography}{116}
\expandafter\ifx\csname natexlab\endcsname\relax\def\natexlab#1{#1}\fi
\expandafter\ifx\csname bibnamefont\endcsname\relax
  \def\bibnamefont#1{#1}\fi
\expandafter\ifx\csname bibfnamefont\endcsname\relax
  \def\bibfnamefont#1{#1}\fi
\expandafter\ifx\csname citenamefont\endcsname\relax
  \def\citenamefont#1{#1}\fi
\expandafter\ifx\csname url\endcsname\relax
  \def\url#1{\texttt{#1}}\fi
\expandafter\ifx\csname urlprefix\endcsname\relax\def\urlprefix{URL }\fi

\bibitem{will:05}
C.~M. Will,
\newblock Living Rev. Rel. {\bf 9}, 3 (2005),
  [\eprintmod[arXiv:gr-qc/0510072]].

\bibitem{fang:08a}
W.~Fang {\em et~al.},
\newblock Phys. Rev. {\bf D78}, 103509 (2008), [\eprintmod[arXiv:0808.2208]].

\bibitem{lombriser:09}
L.~Lombriser, W.~Hu, W.~Fang and U.~Seljak,
\newblock Phys. Rev. {\bf D80}, 063536 (2009), [\eprintmod[arXiv:0905.1112]].

\bibitem{reyes:10}
R.~Reyes {\em et~al.},
\newblock Nature {\bf 464}, 256 (2010), [\eprintmod[arXiv:1003.2185]].

\bibitem{song:07}
Y.-S. Song, H.~Peiris and W.~Hu,
\newblock Phys. Rev. {\bf D76}, 063517 (2007), [\eprintmod[arXiv:0706.2399]].

\bibitem{giannantonio:09}
T.~Giannantonio, M.~Martinelli, A.~Silvestri and A.~Melchiorri,
\newblock JCAP {\bf 1004}, 030 (2010), [\eprintmod[arXiv:0909.2045]].

\bibitem{lombriser:10}
L.~Lombriser, A.~Slosar, U.~Seljak and W.~Hu,
\newblock \eprintmod[arXiv:1003.3009].

\bibitem{hojjati:11}
A.~Hojjati, L.~Pogosian and G.-B. Zhao,
\newblock JCAP {\bf 1108}, 005 (2011), [\eprintmod[arXiv:1106.4543]].

\bibitem{schmidt:09}
F.~Schmidt, A.~Vikhlinin and W.~Hu,
\newblock Phys. Rev. {\bf D80}, 083505 (2009), [\eprintmod[arXiv:0908.2457]].

\bibitem{diporto:07}
C.~Di~Porto and L.~Amendola,
\newblock Phys. Rev. {\bf D77}, 083508 (2008), [\eprintmod[arXiv:0707.2686]].

\bibitem{rapetti:08}
D.~Rapetti, S.~W. Allen, A.~Mantz and H.~Ebeling,
\newblock Mon. Not. Roy. Astron. Soc. {\bf 400}, 699 (2009),
  [\eprintmod[arXiv:0812.2259]].

\bibitem{daniel:09}
S.~F. Daniel {\em et~al.},
\newblock Phys.Rev. {\bf D80}, 023532 (2009), [\eprintmod[arXiv:0901.0919]].

\bibitem{rapetti:09}
D.~Rapetti, S.~W. Allen, A.~Mantz and H.~Ebeling,
\newblock Mon. Not. Roy. Astron. Soc. {\bf 406}, 1796 (2010),
  [\eprintmod[arXiv:0911.1787]].

\bibitem{bean:10}
R.~Bean and M.~Tangmatitham,
\newblock Phys.Rev. {\bf D81}, 083534 (2010), [\eprintmod[arXiv:1002.4197]].

\bibitem{daniel:10}
S.~F. Daniel {\em et~al.},
\newblock Phys. Rev. {\bf D81}, 123508 (2010), [\eprintmod[arXiv:1002.1962]].

\bibitem{daniel:10b}
S.~F. Daniel and E.~V. Linder,
\newblock Phys. Rev. {\bf D82}, 103523 (2010), [\eprintmod[arXiv:1008.0397]].

\bibitem{dossett:10}
J.~Dossett, M.~Ishak, J.~Moldenhauer, Y.~Gong and A.~Wang,
\newblock JCAP {\bf 1004}, 022 (2010), [\eprintmod[arXiv:1004.3086]].

\bibitem{tereno:10}
I.~Tereno, E.~Semboloni and T.~Schrabback,
\newblock Astron. Astrophys. {\bf 530}, A68 (2011),
  [\eprintmod[arXiv:1012.5854]].

\bibitem{zhao:10}
G.-B. Zhao {\em et~al.},
\newblock Phys. Rev. {\bf D81}, 103510 (2010), [\eprintmod[arXiv:1003.0001]].

\bibitem{dossett:11}
J.~Dossett, J.~Moldenhauer and M.~Ishak,
\newblock Phys. Rev. {\bf D84}, 023012 (2011), [\eprintmod[arXiv:1103.1195]].

\bibitem{dossett:11b}
J.~N. Dossett, M.~Ishak and J.~Moldenhauer,
\newblock Phys.Rev. {\bf D84}, 123001 (2011), [\eprintmod[arXiv:1109.4583]].

\bibitem{lombriser:11}
L.~Lombriser,
\newblock Phys. Rev. {\bf D83}, 063519 (2011), [\eprintmod[arXiv:1101.0594]].

\bibitem{zhao:11}
G.-B. Zhao {\em et~al.},
\newblock \eprintmod[arXiv:1109.1846].

\bibitem{smith:09t}
T.~L. Smith,
\newblock \eprintmod[arXiv:0907.4829].

\bibitem{wojtak:11}
R.~Wojtak, S.~H. Hansen and J.~Hjorth,
\newblock Nature {\bf 477}, 567 (2011), [\eprintmod[arXiv:1109.6571]].

\bibitem{carroll:03}
S.~M. Carroll, V.~Duvvuri, M.~Trodden and M.~S. Turner,
\newblock Phys. Rev. {\bf D70}, 043528 (2004),
  [\eprintmod[arXiv:astro-ph/0306438]].

\bibitem{nojiri:03}
S.~Nojiri and S.~D. Odintsov,
\newblock Phys. Rev. {\bf D68}, 123512 (2003),
  [\eprintmod[arXiv:hep-th/0307288]].

\bibitem{capozziello:03}
S.~Capozziello, S.~Carloni and A.~Troisi,
\newblock Recent Res. Dev. Astron. Astrophys. {\bf 1}, 625 (2003),
  [\eprintmod[arXiv:astro-ph/0303041]].

\bibitem{starobinsky:79}
A.~A. Starobinsky,
\newblock JETP Lett. {\bf 30}, 682 (1979).

\bibitem{starobinsky:80}
A.~A. Starobinsky,
\newblock Phys. Lett. {\bf B91}, 99 (1980).

\bibitem{khoury:03}
J.~Khoury and A.~Weltman,
\newblock Phys. Rev. {\bf D69}, 044026 (2004),
  [\eprintmod[arXiv:astro-ph/0309411]].

\bibitem{navarro:06}
I.~Navarro and K.~Van~Acoleyen,
\newblock JCAP {\bf 0702}, 022 (2007), [\eprintmod[arXiv:gr-qc/0611127]].

\bibitem{faulkner:06}
T.~Faulkner, M.~Tegmark, E.~F. Bunn and Y.~Mao,
\newblock Phys. Rev. {\bf D76}, 063505 (2007),
  [\eprintmod[arXiv:astro-ph/0612569]].

\bibitem{hu:07a}
W.~Hu and I.~Sawicki,
\newblock Phys. Rev. {\bf D76}, 064004 (2007), [\eprintmod[arXiv:0705.1158]].

\bibitem{song:06}
Y.-S. Song, W.~Hu and I.~Sawicki,
\newblock Phys. Rev. {\bf D75}, 044004 (2007),
  [\eprintmod[arXiv:astro-ph/0610532]].

\bibitem{schmidt:08}
F.~{Schmidt}, M.~{Lima}, H.~{Oyaizu} and W.~{Hu},
\newblock \prd {\bf 79}, 083518 (2009), [\eprintmod[arXiv:0812.0545]].

\bibitem{schmidt:09b}
F.~Schmidt,
\newblock Phys. Rev. {\bf D80}, 123003 (2009), [\eprintmod[arXiv:0910.0235]].

\bibitem{koester:07}
SDSS, B.~Koester {\em et~al.},
\newblock Astrophys. J. {\bf 660}, 239 (2007),
  [\eprintmod[arXiv:astro-ph/0701265]].

\bibitem{SDSS:09}
SDSS, K.~N. Abazajian {\em et~al.},
\newblock Astrophys. J. Suppl. {\bf 182}, 543 (2009),
  [\eprintmod[arXiv:0812.0649]].

\bibitem{KnoxSongTyson}
L.~{Knox}, Y.-S. {Song} and J.~A. {Tyson},
\newblock \prd {\bf 74}, 023512 (2006).

\bibitem{Schmidt:08a}
F.~{Schmidt},
\newblock \prd {\bf 78}, 043002 (2008), [\eprintmod[arXiv:0805.4812]].

\bibitem{JainZhang}
B.~Jain and P.~Zhang,
\newblock Phys. Rev. {\bf D78}, 063503 (2008), [\eprintmod[arXiv:0709.2375]].

\bibitem{Tsujikawa2008}
S.~Tsujikawa and T.~Tatekawa,
\newblock Phys. Lett. {\bf B665}, 325 (2008), [\eprintmod[arXiv:0804.4343]].

\bibitem{borisov:11}
A.~Borisov, B.~Jain and P.~Zhang,
\newblock Phys.Rev. {\bf D85}, 063518 (2012), [\eprintmod[arXiv:1102.4839]].

\bibitem{martino:08}
M.~C. Martino, H.~F. Stabenau and R.~K. Sheth,
\newblock Phys. Rev. {\bf D79}, 084013 (2009), [\eprintmod[arXiv:0812.0200]].

\bibitem{schmidt:09a}
F.~Schmidt,
\newblock Phys. Rev. {\bf D80}, 043001 (2009), [\eprintmod[arXiv:0905.0858]].

\bibitem{narikawa:12}
T.~Narikawa and K.~Yamamoto,
\newblock JCAP {\bf 1205}, 016 (2012), [\eprintmod[arXiv:1201.4037]].

\bibitem{oyaizu:08a}
H.~Oyaizu,
\newblock Phys. Rev. {\bf D78}, 123523 (2008), [\eprintmod[arXiv:0807.2449]].

\bibitem{oyaizu:08b}
H.~Oyaizu, M.~Lima and W.~Hu,
\newblock Phys. Rev. {\bf D78}, 123524 (2008), [\eprintmod[arXiv:0807.2462]].

\bibitem{zhao:10b}
G.-B. Zhao, B.~Li and K.~Koyama,
\newblock Phys. Rev. {\bf D83}, 044007 (2011), [\eprintmod[arXiv:1011.1257]].

\bibitem{li:11}
B.~Li, G.-B. Zhao, R.~Teyssier and K.~Koyama,
\newblock JCAP {\bf 1201}, 051 (2012), [\eprintmod[arXiv:1110.1379]].

\bibitem{FerraroEtal}
S.~{Ferraro}, F.~{Schmidt} and W.~{Hu},
\newblock \prd {\bf 83}, 063503 (2011), [\eprintmod[arXiv:1011.0992]].

\bibitem{squires:95}
G.~Squires and N.~Kaiser,
\newblock Astrophys. J. {\bf 473}, 65 (1996),
  [\eprintmod[arXiv:astro-ph/9512094]].

\bibitem{smith:08}
R.~E. Smith,
\newblock Mon. Not. Roy. Astron. Soc. {\bf 400}, 851 (2009),
  [\eprintmod[arXiv:0810.1960]].

\bibitem{jenkins:00}
A.~Jenkins {\em et~al.},
\newblock Mon. Not. Roy. Astron. Soc. {\bf 321}, 372 (2001),
  [\eprintmod[arXiv:astro-ph/0005260]].

\bibitem{spergel:03}
WMAP, D.~N. Spergel {\em et~al.},
\newblock Astrophys. J. Suppl. {\bf 148}, 175 (2003),
  [\eprintmod[arXiv:astro-ph/0302209]].

\bibitem{spergel:06}
WMAP, D.~N. Spergel {\em et~al.},
\newblock Astrophys. J. Suppl. {\bf 170}, 377 (2007),
  [\eprintmod[arXiv:astro-ph/0603449]].

\bibitem{smith:09}
R.~E. Smith, C.~Hernandez-Monteagudo and U.~Seljak,
\newblock Phys. Rev. {\bf D80}, 063528 (2009), [\eprintmod[arXiv:0905.2408]].

\bibitem{springel:05}
V.~Springel,
\newblock Mon. Not. Roy. Astron. Soc. {\bf 364}, 1105 (2005),
  [\eprintmod[arXiv:astro-ph/0505010]].

\bibitem{seljak:96}
U.~Seljak and M.~Zaldarriaga,
\newblock Astrophys. J. {\bf 469}, 437 (1996),
  [\eprintmod[arXiv:astro-ph/9603033]].

\bibitem{scoccimarro:97}
R.~Scoccimarro,
\newblock Mon. Not. Roy. Astron. Soc. {\bf 299}, 1097 (1998),
  [\eprintmod[arXiv:astro-ph/9711187]].

\bibitem{crocce:06}
M.~Crocce, S.~Pueblas and R.~Scoccimarro,
\newblock Mon. Not. Roy. Astron. Soc. {\bf 373}, 369 (2006),
  [\eprintmod[arXiv:astro-ph/0606505]].

\bibitem{davis:85}
M.~Davis, G.~Efstathiou, C.~S. Frenk and S.~D.~M. White,
\newblock Astrophys. J. {\bf 292}, 371 (1985).

\bibitem{rozo:08}
E.~Rozo {\em et~al.},
\newblock Astrophys. J. {\bf 699}, 768 (2009), [\eprintmod[arXiv:0809.2794]].

\bibitem{2000AJ....120.1579Y}
D.~G. {York} {\em et~al.},
\newblock \aj {\bf 120}, 1579 (2000).

\bibitem{2001AJ....122.2267E}
D.~J. {Eisenstein} {\em et~al.},
\newblock \aj {\bf 122}, 2267 (2001).

\bibitem{2002AJ....123.2945R}
G.~T. {Richards} {\em et~al.},
\newblock \aj {\bf 123}, 2945 (2002).

\bibitem{2002AJ....124.1810S}
M.~A. {Strauss} {\em et~al.},
\newblock \aj {\bf 124}, 1810 (2002).

\bibitem{2001AJ....122.2129H}
D.~W. {Hogg}, D.~P. {Finkbeiner}, D.~J. {Schlegel} and J.~E. {Gunn},
\newblock \aj {\bf 122}, 2129 (2001).

\bibitem{2004AN....325..583I}
{\v Z}.~{Ivezi{\' c}} {\em et~al.},
\newblock Astronomische Nachrichten {\bf 325}, 583 (2004).

\bibitem{1996AJ....111.1748F}
M.~{Fukugita} {\em et~al.},
\newblock \aj {\bf 111}, 1748 (1996).

\bibitem{2002AJ....123.2121S}
J.~A. {Smith} {\em et~al.},
\newblock \aj {\bf 123}, 2121 (2002).

\bibitem{1998AJ....116.3040G}
J.~E. {Gunn} {\em et~al.},
\newblock \aj {\bf 116}, 3040 (1998).

\bibitem{2001ASPC..238..269L}
R.~H. {Lupton} {\em et~al.},
\newblock {The SDSS Imaging Pipelines},
\newblock in {\em ASP Conf. Ser. 238: Astronomical Data Analysis Software and
  Systems X}, pp. 269--+, 2001.

\bibitem{2003AJ....125.1559P}
J.~R. {Pier} {\em et~al.},
\newblock \aj {\bf 125}, 1559 (2003).

\bibitem{2006AN....327..821T}
D.~L. {Tucker} {\em et~al.},
\newblock Astronomische Nachrichten {\bf 327}, 821 (2006),
  [\eprintmod[arXiv:astro-ph/0608575]].

\bibitem{2009ApJS..182..543A}
K.~N. {Abazajian} {\em et~al.},
\newblock \apjs {\bf 182}, 543 (2009), [\eprintmod[arXiv:0812.0649]].

\bibitem{2008ApJ...674.1217P}
N.~{Padmanabhan} {\em et~al.},
\newblock \apj {\bf 674}, 1217 (2008), [\eprintmod[arXiv:astro-ph/0703454]].

\bibitem{2011ApJS..193...29A}
H.~{Aihara} {\em et~al.},
\newblock \apjs {\bf 193}, 29 (2011), [\eprintmod[arXiv:1101.1559]].

\bibitem{2011AJ....142...72E}
D.~J. {Eisenstein} {\em et~al.},
\newblock \aj {\bf 142}, 72 (2011), [\eprintmod[arXiv:1101.1529]].

\bibitem{2005MNRAS.362.1451M}
R.~{Mandelbaum}, A.~{Tasitsiomi}, U.~{Seljak}, A.~V. {Kravtsov} and R.~H.
  {Wechsler},
\newblock Mon. Not. Roy. Astron. Soc. {\bf 362}, 1451 (2005),
  [\eprintmod[arXiv:astro-ph/0410711]].

\bibitem{mandelbaum:09}
R.~Mandelbaum, U.~Seljak, T.~Baldauf and R.~E. Smith,
\newblock Mon. Not. Roy. Astron. Soc. {\bf 405}, 2078 (2010),
  [\eprintmod[arXiv:0911.4972]].

\bibitem{reyes:11}
R.~Reyes {\em et~al.},
\newblock \eprintmod[arXiv:1110.4107].

\bibitem{2003MNRAS.343..459H}
C.~{Hirata} and U.~{Seljak},
\newblock Mon. Not. Roy. Astron. Soc. {\bf 343}, 459 (2003).

\bibitem{2005MNRAS.361.1287M}
R.~{Mandelbaum} {\em et~al.},
\newblock Mon. Not. Roy. Astron. Soc. {\bf 361}, 1287 (2005),
  [\eprintmod[arXiv:astro-ph/0501201]].

\bibitem{2006MNRAS.372..565F}
R.~{Feldmann} {\em et~al.},
\newblock Mon. Not. Roy. Astron. Soc. {\bf 372}, 565 (2006),
  [\eprintmod[arXiv:astro-ph/0609044]].

\bibitem{2011arXiv1107.1395N}
R.~{Nakajima} {\em et~al.},
\newblock Mon. Not. Roy. Astron. Soc. {\bf 420}, 3240 (2012),
  [\eprintmod[arXiv:1107.1395]].

\bibitem{2002AJ....123..583B}
G.~M. {Bernstein} and M.~{Jarvis},
\newblock \aj {\bf 123}, 583 (2002).

\bibitem{WMAP:08}
WMAP, J.~Dunkley {\em et~al.},
\newblock Astrophys. J. Suppl. {\bf 180}, 306 (2009),
  [\eprintmod[arXiv:0803.0586]].

\bibitem{ACBAR:07}
C.-L. Kuo {\em et~al.},
\newblock Astrophys. J. {\bf 664}, 687 (2007),
  [\eprintmod[arXiv:astro-ph/0611198]].

\bibitem{CBI:04}
A.~C.~S. Readhead {\em et~al.},
\newblock Astrophys. J. {\bf 609}, 498 (2004),
  [\eprintmod[arXiv:astro-ph/0402359]].

\bibitem{VSA:03}
K.~Grainge {\em et~al.},
\newblock Mon. Not. Roy. Astron. Soc. {\bf 341}, L23 (2003),
  [\eprintmod[arXiv:astro-ph/0212495]].

\bibitem{UNION:08}
Supernova Cosmology Project, M.~Kowalski {\em et~al.},
\newblock Astrophys. J. {\bf 686}, 749 (2008), [\eprintmod[arXiv:0804.4142]].

\bibitem{SHOES:09}
A.~G. Riess {\em et~al.},
\newblock Astrophys. J. {\bf 699}, 539 (2009), [\eprintmod[arXiv:0905.0695]].

\bibitem{reid:09}
B.~A. Reid, L.~Verde, R.~Jimenez and O.~Mena,
\newblock JCAP {\bf 1001}, 003 (2010), [\eprintmod[arXiv:0910.0008]].

\bibitem{BAO:09}
W.~J. Percival {\em et~al.},
\newblock Mon. Not. Roy. Astron. Soc. {\bf 401}, 2148 (2010),
  [\eprintmod[arXiv:0907.1660]].

\bibitem{schmidt:10}
F.~Schmidt,
\newblock Phys. Rev. {\bf D81}, 103002 (2010), [\eprintmod[arXiv:1003.0409]].

\bibitem{cosmomc:02}
A.~Lewis and S.~Bridle,
\newblock Phys. Rev. {\bf D66}, 103511 (2002),
  [\eprintmod[arXiv:astro-ph/0205436]].

\bibitem{metropolis:53}
N.~Metropolis, A.~W. Rosenbluth, M.~N. Rosenbluth, A.~H. Teller and E.~Teller,
\newblock J. Chem. Phys. {\bf 21}, 1087 (1953).

\bibitem{hastings:70}
W.~K. Hastings,
\newblock Biometrika {\bf 57}, 97 (1970).

\bibitem{gelman:92}
A.~Gelman and D.~B. Rubin,
\newblock Statist. Sci. {\bf 7}, 457 (1992).

\bibitem{teyssier:10}
R.~Teyssier, B.~Moore, D.~Martizzi, Y.~Dubois and L.~Mayer,
\newblock Mon. Not. Roy. Astron. Soc. {\bf 414}, 195 (2011),
  [\eprintmod[arXiv:1003.4744]].

\bibitem{blazek:11a}
J.~Blazek, M.~McQuinn and U.~Seljak,
\newblock JCAP {\bf 1105}, 010 (2011), [\eprintmod[arXiv:1101.4017]].

\bibitem{hirata:07}
C.~M. Hirata {\em et~al.},
\newblock Mon. Not. Roy. Astron. Soc. {\bf 381}, 1197 (2007),
  [\eprintmod[arXiv:astro-ph/0701671]].

\bibitem{blazek:11b}
J.~Blazek, R.~Mandelbaum, U.~Seljak and R.~Nakajima,
\newblock \eprintmod[arXiv:1204.2264].

\bibitem{hilbert:09}
S.~Hilbert and S.~D.~M. White,
\newblock Mon. Not. Roy. Astron. Soc. {\bf 404}, 486 (2010),
  [\eprintmod[arXiv:0907.4371]].

\bibitem{baldauf:09}
T.~Baldauf, R.~E. Smith, U.~Seljak and R.~Mandelbaum,
\newblock Phys. Rev. {\bf D81}, 063531 (2010), [\eprintmod[arXiv:0911.4973]].

\bibitem{millenium:05}
V.~Springel {\em et~al.},
\newblock Nature {\bf 435}, 629 (2005), [\eprintmod[arXiv:astro-ph/0504097]].

\bibitem{ZhaoEtal:11}
G.-B. {Zhao}, B.~{Li} and K.~{Koyama},
\newblock Physical Review Letters {\bf 107}, 071303 (2011),
  [\eprintmod[arXiv:1105.0922]].

\bibitem{knebe:11}
A.~{Knebe} {\em et~al.},
\newblock Mon. Not. Roy. Astron. Soc. {\bf 415}, 2293 (2011),
  [\eprintmod[arXiv:1104.0949]].

\bibitem{HuKravtsov}
W.~{Hu} and A.~V. {Kravtsov},
\newblock \apj {\bf 584}, 702 (2003), [\eprintmod[arXiv:astro-ph/0203169]].

\bibitem{sheth:99}
R.~K. Sheth and G.~Tormen,
\newblock Mon. Not. Roy. Astron. Soc. {\bf 308}, 119 (1999),
  [\eprintmod[arXiv:astro-ph/9901122]].

\bibitem{NFW}
J.~F. Navarro, C.~S. Frenk and S.~D.~M. White,
\newblock Astrophys. J. {\bf 490}, 493 (1997),
  [\eprintmod[arXiv:astro-ph/9611107]].

\bibitem{BullockEtal}
J.~S. {Bullock} {\em et~al.},
\newblock Mon. Not. Roy. Astron. Soc. {\bf 321}, 559 (2001),
  [\eprintmod[arXiv:astro-ph/9908159]].

\bibitem{lombriser:12}
L.~Lombriser, K.~Koyama, G.-B. Zhao and B.~Li,
\newblock \eprintmod[arXiv:1203.5125].

\end{thebibliography}

\end{document}